\pdfoutput=1
\documentclass[12pt]{article}
\usepackage[utf8]{inputenc}      % input font encoding
\usepackage[colorlinks=true,citecolor=blue,urlcolor=blue,linktocpage=true,linkcolor=blue]{hyperref}
\usepackage{amsmath,amssymb,mathtools}
\usepackage[T1]{fontenc}          % output font encoding
\usepackage{booktabs,tabularx}
\usepackage{graphicx}
\usepackage{xspace}
\usepackage{lmodern}
\usepackage[usenames]{xcolor}\definecolor{fscolor}{RGB}{44,118,255}
\usepackage{geometry}
\usepackage{todonotes}
\usepackage{listings}
\usepackage[absolute]{textpos}
\usepackage[many]{tcolorbox}
\usepackage{xparse}
\usepackage[font=small,labelfont=bf,format=plain,margin=0.05\textwidth]{caption}
\usepackage{bbm}
\usepackage{tabularx}
\usepackage{soul}
\usepackage[capitalize]{cleveref}
\usepackage{slashed}
\usepackage{pifont}
\usepackage[shortlabels]{enumitem}
\usepackage{nicefrac}
\usepackage[sorting=none,%
  citestyle=numeric-comp,%
  bibstyle=mynumeric,%
  giveninits=true]{biblatex}
\usepackage{subfig}
\usepackage[super]{nth}
\usepackage{lineno}
\usepackage{authblk}

\allowdisplaybreaks

\evensidemargin \oddsidemargin
\renewcommand{\arraystretch}{1.5}

\newcommand{\cp}{\ensuremath{{\cal CP}}\xspace}
\newcommand{\SM}{{\text{SM}}}

\newcommand{\ttH}{\ensuremath{t\bar t H}\xspace}
\newcommand{\ttbar}{\ensuremath{t\bar t}\xspace}
\newcommand{\tH}{\ensuremath{tHq}\xspace}
\newcommand{\tWH}{\ensuremath{tWH}\xspace}

\newcommand{\pT}{\ensuremath{p_\text{T}}\xspace}
\newcommand{\pTx}[1]{\ensuremath{p_{T,{#1}}}\xspace}
\newcommand{\tstar}{\ensuremath{\theta^*}\xspace}

\newcommand{\tev}{\,\, \mathrm{TeV}}
\newcommand{\gev}{\,\, \mathrm{GeV}}

\newcommand{\invfb}{\,\, \mathrm{fb}^{-1}}

\definecolor{Darkgreen}{rgb}{0.,.7,0.2}
\definecolor{Darkblue}{rgb}{0.,.2,0.7}
\definecolor{Magenta}{rgb}{0.7,0.,0.7}

\newcommand{\gt}{\ensuremath{g_t}\xspace}
\newcommand{\at}{\ensuremath{\alpha_t}\xspace}

\newcommand{\ct}{\ensuremath{c_t}\xspace}

\newcommand{\ctt}{\ensuremath{\tilde c_{t}}\xspace}

\def\be{\begin{equation}}
\def\ee{\end{equation}}

\NewBibliographyString{refname}
\NewBibliographyString{refsname}
\DefineBibliographyStrings{english}{%
  refname = {Ref\adddot},
  refsname = {Refs\adddot}
}
\DeclareCiteCommand{\ccite}
  {%
  \ifnum\thecitetotal=1
    \bibstring{refname}%
  \else%
    \bibstring{refsname}%
  \fi%
  \addnbspace\bibopenbracket%
  \usebibmacro{cite:init}%
   \usebibmacro{prenote}}
  {\usebibmacro{citeindex}%
   \usebibmacro{cite:comp}}
  {}
  {\usebibmacro{cite:dump}%
   \usebibmacro{postnote}%
   \bibclosebracket}
\newrobustcmd*{\Ccite}{\bibsentence\ccite}

\def\thefootnote{\fnsymbol{footnote}}

\begin{document}
\title{
\vspace{2em}
\begin{large}
\textbf{\cp-sensitive simplified template cross-sections for \ttH}
\end{large}
}
\author[1]{Henning Bahl\thanks{bahl@thphys.uni-heidelberg.de}}
\author[2]{Alberto Carnelli\thanks{alberto.carnelli@cea.fr}}
\author[2]{Fr\'ed\'eric D\'eliot \thanks{frederic.deliot@cern.ch}}
\author[3,4]{Elina Fuchs\thanks{elina.fuchs@cern.ch}}
\author[2]{Anastasia Kotsokechagia\thanks{anastasia.kotsokechagia@cern.ch}}
\author[2]{Tanguy Lucas Marsault\thanks{marsault.tanguy@cea.fr}}
\author[3,4]{Marco Menen\thanks{marco.menen@itp.uni-hannover.de}}
\author[2]{Laurent Schoeffel\thanks{laurent.schoeffel@cea.fr}}
\author[2]{Matthias Saimpert\thanks{matthias.saimpert@cea.fr}}

\affil[1]{Institut für Theoretische Physik, Universität Heidelberg, Philosophenweg 16, 61920 Heidelberg, Germany} 
\affil[2]{Université Paris-Saclay, CEA, Département de Physique des Particules, 91191, Gif-sur-Yvette, France}
\affil[3]{Institut für Theoretische Physik, Leibniz Universität Hannover, Appelstraße 2, 30167 Hannover, Germany}
\affil[4]{Physikalisch-Technische Bundesanstalt, Bundesallee 100, 38116 Braunschweig, Germany}

\maketitle

\vspace{-1.5ex}
\begin{abstract}

The \cp structure of the Higgs boson is a fundamental property which has not yet been constrained with high precision. \cp violation in the Yukawa coupling between the Higgs boson and top quark pair can be probed directly at the Large Hadron Collider by measuring top-quark-associated Higgs production. Multivariate analysis techniques commonly developed so far by the experiments are designed for a specific signal model and, therefore, complicate reinterpretations and statistical combinations. With this motivation in mind, we propose a \cp-sensitive extension of the simplified template cross-section (STXS) framework. Considering multiple Higgs decay channels, we perform an in-depth comparison of \cp-sensitive observables and combinations thereof. Our resulting proposal is to extend the existing binning in the transverse momentum of the Higgs boson \pTx{H} by either the pseudorapidity difference of the two top-quarks $\Delta \eta_{t\bar t}$, or a variable that is based on the top quark momenta, namely $b_2$ or the Collins-Soper angle $|\cos\theta^*|$. We demonstrate that this variable selection provides close to optimal sensitivity to the \cp mixture in the top Yukawa coupling for an integrated luminosity of $300\invfb$,  by comparing it to the results of a multivariate analysis. Our results also suggest a benefit of the two-dimensional STXS extension at 3000$\invfb$.
\end{abstract}

\newpage
\tableofcontents
\newpage
\def\thefootnote{\arabic{footnote}}

%%%%%%%%%%%%%%%%%%%%%%%%%%%%%%%%%%

 \fontsize{11.5}{16}\selectfont

%%%%%%%%%%%%%%%%%%%%%%%%%%%%%%%%%%
%%%%%%%%%%%%%%%%%%%%%%%%%%%%%%%%%%
%%%%%%%%%%%%%%%%%%%%%%%%%%%%%%%%%%

\section{Introduction}
\label{sec:intro}

The Standard Model (SM) of particle physics does not contain sufficient \cp violation to explain the observed baryon asymmetry of the Universe (BAU)~\cite{Gavela:1993ts,Huet:1994jb}. Therefore, new sources of \cp violation are needed and the search for corresponding \cp-violating interactions is an important target for searches beyond the SM (BSM) at the LHC.

The investigation of the Higgs sector is especially important in this context. First, the \cp structure of the Higgs couplings is still comparably unexplored. Second, the BAU can be linked to the Higgs sector via the mechanism of electroweak baryogenesis~\cite{Cohen:1993nk} (for reviews, see e.g.\ \ccite{Trodden:1998ym, Cline:2006ts,Morrissey:2012db,White:2016nbo,Bodeker:2020ghk}). While the \cp structure of the Higgs couplings to massive gauge bosons is already relatively tightly constrained~\cite{ATLAS:2015zhl,ATLAS:2016ifi,CMS:2017len,CMS:2019jdw,CMS:2019ekd,ATLAS:2020evk,ATLAS:2021pkb,CMS:2022mlq,ATLAS:2022tan,ATLAS:2023mqy}, the test of the \cp structure of the Higgs--fermion interactions at the LHC only started recently~\cite{CMS:2020cga,ATLAS:2020ior,CMS:2021sdq,CMS:2021nnc,CMS:2022dbt,CMS:2022mlq,ATLAS:2022akr,ATLAS:2023cbt}. Moreover, most BSM theories predict a larger amount of \cp violation in the Yukawa couplings than in the interactions with massive vector bosons, since the latter are loop-suppressed.

The top-Yukawa coupling has a special relevance because of its large size. It can be parameterized in the following way in the Higgs Characterization Model~\cite{Artoisenet:2013puc},
\begin{align}
    \mathcal{L}_\text{top-Yuk} = \frac{y_t^\text{SM}\gt}{\sqrt{2}}\bar t\left(\cos\at + i \gamma_5 \sin\at\right)t H\,,
\label{eq:top-Yukawa}
\end{align}
where $y_t^\text{SM}$ is the SM top-Yukawa coupling, 
$g_t$ is the (real) modifier of the strength of the top-Yukawa coupling, and $\alpha_t$ is the \cp-mixing angle. The SM corresponds to $\gt=1$, $\at=0$. Equivalently, this interaction can be parameterized in terms of a \cp-even and a \cp-odd coupling, $\ct \equiv \gt\cos\at$ and $\ctt \equiv \gt\sin\at$, respectively. 

In an EFT framework (e.g.\ the SM effective field theory), this gauge-invariant modification is generated by the operator $(\Phi^\dagger\Phi)Q_L u_R \tilde\Phi$ (where $\Phi$ is the Higgs doublet, $Q_L$ is the left-handed quark doublet, and $u_R$ is the right-handed up-type quark singlet) and the coefficients $y_t^\text{SM} g_t \cos\at$ and $y_t^\text{SM} g_t \sin\at$ in \cref{eq:top-Yukawa} can be related to the real and imaginary parts of the corresponding Wilson coefficient (see e.g.~\ccite{Fuchs:2020uoc,AharonyShapira:2021ize}).\footnote{In the SM effective field theory, the operator $(\Phi^\dagger\Phi)Q_L u_R \tilde\Phi$ also generates additional multi-Higgs and Goldstone boson interactions with top quarks, which are in general needed to guarantee gauge invariance. These are, however, not relevant to top-associated Higgs production studied in this work.}

This general coupling structure can be probed at the LHC but also via measurements of electric dipole moments (EDMs) of particles like the electron or the neutron, for which so far only upper bounds exist~\cite{Roussy:2022cmp,Abel:2020pzs}.
If one assumes that the first-generation Yukawa couplings do not deviate from their SM predictions, these EDM bounds put strong constraints on the \cp-odd part of the top-Yukawa coupling, constraining $\ctt\lesssim 10^{-3}$ at the 90\% CL~\cite{Brod:2013cka,Fuchs:2020uoc,Brod:2022bww}; this translates into $\at \lesssim 0.06^\circ$ for $\gt=1$.
Since the first-generation Yukawa couplings are only very weakly constrained~\cite{Altmannshofer:2015qra,ATLAS:2019old,Zhou:2015wra,Soreq:2016rae,Bonner:2016sdg,Yu:2016rvv,Alasfar:2019pmn,Aguilar-Saavedra:2020rgo,Falkowski:2020znk,Aguilar-Saavedra:2020rgo,Vignaroli:2022fqh,Balzani:2023jas}, this assumption, however, does not necessarily hold, resulting in relaxed EDM bounds. Therefore, it is crucial to measure the \cp properties of the top-Yukawa coupling directly, i.e.\ without a strong dependence on the other Yukawa couplings. This can be achieved at the LHC where the measurement of distinct production and decay channels provides access to specific couplings, whereas the measured value of an EDM results from the sum over all \cp-violating contributions.

Bounds on the \cp structure of the top-Yukawa interaction at the LHC can be derived from rate measurements~\cite{Freitas:2012kw,Agrawal:2012ga,Djouadi:2013qya,Ellis:2013yxa,Chang:2014rfa,He:2014xla,Boudjema:2015nda,Demartin:2015uha,Demartin:2016axk,Kobakhidze:2016mfx,Hou:2018uvr,Cao:2019ygh,Fuchs:2020uoc,Bahl:2020wee,Bahl:2022yrs,Brod:2022bww,Bahl:2023qwk}. In this case, the most stringent constraints arise from Higgs production via gluon fusion when the Higgs boson decays to two photons (constraining $\at\lesssim 28^\circ$ at the 95\% CL assuming that $g_t=1$~\cite{Bahl:2023qwk}). These constraints are, however, highly model-dependent since other particles also contribute to the respective loop diagrams. A more direct probe is top-associated Higgs production (consisting of the three sub-channels \ttH, \tH, and \tWH). Top-associated Higgs production can be used to constrain the \cp properties of the top-Yukawa coupling via rate measurements (constraining $\at\lesssim 72^\circ$ at the 95\% CL~\cite{Bahl:2020wee} assuming that $g_t=1$), kinematic distributions, and \cp-odd observables. While rate measurements are again model-dependent (e.g., the Higgs-gluon coupling contributes to top-associated Higgs production at next-to-leading order), \cp-odd observables in principle offer a way to unambiguously probe \cp. For top-associated Higgs production, they are, however, very difficult to measure due to the necessity of reconstructing the polarization of the top quark~\cite{Mileo:2016mxg,Faroughy:2019ird,Bortolato:2020zcg,Goncalves:2021dcu,Barman:2021yfh,Azevedo:2022jnd,Ackerschott:2023nax}. Specifically, in hadronic decays, it requires to differentiate between up- and down-type quarks. While this might change during the high-luminosity phase of the LHC, at the moment they certainly only provide little sensitivity for testing the \cp character of the top-Yukawa interaction. On the other hand, kinematic distributions of \cp-even observables offer a good compromise between the \cp-odd observables and rate measurements. After the observation of top-associated Higgs production in 2018~\cite{CMS:2018uxb,ATLAS:2018mme}, the additional data collected and to be collected at the LHC allows now for differential measurements even for this comparably rare process, making it a promising way to probe \cp during over the next few years.

Kinematic information can be best exploited using multivariate analysis techniques (see e.g.~\ccite{Ren:2019xhp,Barman:2021yfh,Bahl:2021dnc,Esmail:2024gdc}) or the matrix element method (see e.g.~\ccite{Martini:2021uey,Kraus:2019myc,Gritsan:2016hjl,Goncalves:2018agy,Butter:2022vkj}). These methods have been used by the experimental collaborations to put constraints on \at, constraining at the 95\% CL $\at\lesssim 43^\circ$ (ATLAS using the $H\to\gamma\gamma$ channel~\cite{ATLAS:2020ior}; in addition, see also the currently less sensitive $H\to b\bar b$ channel in~\ccite{ATLAS:2023cbt}) and $\at\lesssim 45^\circ$ (CMS using $H\to\gamma\gamma,4\ell,\tau\tau$~\cite{CMS:2020cga,CMS:2021nnc,CMS:2022mlq,CMS:2022dbt})\footnote{In the CMS papers, the limits are set on $|f_\cp^{Htt}|\equiv\nicefrac{\ctt^2}{(\ct^2+\ctt^2)} = \sin^2\at$.} assuming that $g_t=1$. For an overview of future prospects, see \ccite{Gritsan:2022php}. These techniques are typically tailored to specific Higgs decay channels, detector environments, and background assumptions. Moreover, they are designed for specific signal models and intrinsically depend on the assumptions about the other couplings involved. As a consequence, the combination of different decay channels (see e.g.~\ccite{CMS:2021nnc,CMS:2022dbt}) and across experiments is non-trivial. The usage of such approaches also makes the reinterpretation of the analyses difficult, as it usually requires the signal efficiency as a function of the final discriminating algorithm output score. To mitigate such issues --- before providing likelihoods encoding all relevant dependencies becomes feasible ---, the simplified template cross-section (STXS) framework~\cite{deFlorian:2016spz,Badger:2016bpw,Berger:2019wnu,Amoroso:2020lgh} has been established. This study, focusing on \cp-sensitive but \cp-even observables, aims to enhance the \cp sensitivity of this framework for \ttH production by taking the interplay of two variables into account.
As a complementary example, also for Drell-Yan production, double- and triple-differential distributions have recently been found to yield a higher sensitivity to SMEFT operators compared to a finer binning in a single variable~\cite{Grossi:2024tou}. 

We start in \cref{sec:obs} with a comparison of the kinematic shapes of various \cp-sensitive observables evaluated in different reference frames building upon existing proposals in the literature~\cite{Gunion:1996xu,Ellis:2013yxa,Yue:2014tya,Demartin:2014fia,He:2014xla,Demartin:2015uha,Buckley:2015vsa,Demartin:2016axk,Gritsan:2016hjl,Azevedo:2017qiz,Goncalves:2018agy,Cao:2020hhb,Martini:2021uey,Barman:2021yfh,Bahl:2021dnc,Azevedo:2022jnd,Ackerschott:2023nax}. In \cref{sec:mostsensitive}, we evaluate how these observables could be impacted by detector effects using a simplified model and describe how their performance is evaluated. In \cref{sec:results}, we perform a sensitivity study based on one- and two-dimensional distributions, optimize the binning of the most sensitive observables assuming $300\invfb$ of data (which should correspond to the available LHC dataset by the end of 2025), and compare the outcome to a multivariate analysis based on a boosted decision tree (BDT). Based on these findings, we make a proposal for a \cp-sensitive extension of the latest version of the STXS framework (v1.2) in \cref{sec:mostsensitive_STXS}. It also includes a qualitative discussion about the Higgs and non-Higgs backgrounds and a comparison between the expected \cp sensitivity of the STXS framework v1.2 and our proposed extension. We present conclusions in \cref{sec:conclusion}.

%%%%%%%%%%%%%%%%%%%%%%%%%%%%%%%%%%
%%%%%%%%%%%%%%%%%%%%%%%%%%%%%%%%%%

\section{\cp-sensitive observables in \texorpdfstring{\ttH}{ttH} events}
\label{sec:obs}

The $t{\bar t}H$ production process possesses many final state particles and therefore also numerous kinematic observables. In experimental analyses targeting \cp violation in the top-Yukawa coupling, a number of such observables have been used already (see e.g.~\ccite{ATLAS:2023cbt}). These are, however, only optimized for a specific channel and selection. In this work, we aim to find observables that are optimal to use in the context of an STXS extension and should therefore be independent of the analysis details.

We describe the generation of the $t{\bar t}H$ events used in this study in~\cref{subsec:event_generation}. We then introduce an extended number of kinematic observables in various rest frames in~\cref{subsec:observables}. The parton level distributions of these observables are shown in~\cref{subsec:obs_distr}. 

%%%%%%%%%%%%%%%%%%%%%%%%%%%%%%%%%%

\subsection{Event generation}
\label{subsec:event_generation}

We generate parton-level events for the $pp \rightarrow t{\bar t}H$ process at a center-of-mass energy of $\sqrt{s}=13\tev$ using the event generator \texttt{MadGraph5\_aMC@NLO}~\cite{Alwall:2014hca} (version 3.3.2) and the \texttt{NNPDF23\_lo\_\\as\_0130\_qed} PDF set~\cite{Ball:2012cx}. The renormalization and factorization scales are set to $1/2 \cdot (2 m_t + m_H)$. Events are generated using the ``Higgs Characterization'' (HC) \texttt{UFO} model~\cite{Artoisenet:2013puc} for $\gt = 1$ and two different values for the \cp phase \at: $0^\circ$ (pure scalar, corresponding to the SM) and $90^\circ$ (pure \cp-odd case). All simulated samples were generated at leading order (LO) and include one million events each. 

In principle, the HC model allows generating events at next-to-leading order (NLO) in QCD. Within this model, if the top-quark mass is treated as finite (as needed for $t\bar t H$ production), it, however, does not allow considering a non-zero effective Higgs--gluon interaction (generated by BSM particles), which would affect $t\bar t H$ production at NLO in $\alpha_s$. As discussed in \cref{sec:intro}, for a zero BSM Higgs--gluon coupling, Higgs production via gluon fusion allows setting much stronger (indirect) constraints on the \cp character of the top-Yukawa coupling with respect to top-associated Higgs production. Since we are interested in a scenario in which $t\bar t H$ provides the leading constraints on the \cp nature of the top-Yukawa coupling and this scenario cannot be generated with the HC model at NLO, we, therefore, generate all events at LO (see also the discussion in~\ccite{Bahl:2020wee}). Moreover, as discussed in~\cite{Demartin:2014fia,Bahl:2020wee}, the NLO correction to the \ttH total cross-section depends only very weakly on the \cp character of the top-Yukawa coupling. Similarly, this observation still holds for kinematic distributions.\footnote{In \ccite{Bahl:2020wee}, the NLO $N_\text{jet}$ distribution for \tH production was found to significantly differ from the LO distribution. This is not relevant to the discussion in this paper which focuses on \ttH production.} Hence, a simple scaling factor of 1.14~\cite{Demartin:2014fia} is used as NLO correction in the following.  

Since only \cp-even observables are considered in this study, signal yields for mixed-\cp\ scenarios are computed by combining the yields from the SM and the pure \cp-odd samples. The following formula is used : 
\begin{align}
    N(\gt,\at) = \gt^2\left[N_{\mathrm{SM}}\cos^2\at  + N_{\mathrm{odd}}\sin^2\at\right],
\label{eq:ttH_yield}
\end{align}
where $N(\gt,\at)$ is the expected yield for the signal corresponding to the parameters $(\gt,\at)$, with $N_{\mathrm{SM}}=N(1,0)$ and $N_{\mathrm{odd}}=N(1,90^\circ)$. The yields for $\alpha_t=35^\circ$ and $\alpha_t=45^\circ$ obtained from~\cref{eq:ttH_yield} were validated against the yields obtained from dedicated samples generated with these $\alpha_t$ values and an agreement within 1\% was obtained.  

The SM background processes discussed in \cref{subsec:bkg}, i.e. $t\bar{t}\gamma\gamma$, $t\bar tW$ and $t\bar{t} b\bar{b}$, were generated using a similar setup. These samples are used at the parton level only and without any additional selection nor corrections applied. We checked that showering the events with \texttt{Pythia8} \cite{Sjostrand:2014zea} has only a small impact on the resulting distributions shown in~\cref{subsec:bkg}. Half a million events are generated at LO for each process using the \texttt{NNPDF23\_lo\_as\_0130\_qed} PDF set~\cite{Ball:2012cx} and the renormalization and factorization scales were set to the sum of the transverse mass of all the final state particles divided by a factor of two (corresponding to \texttt{dynamical\_scale\_choice} $=3$ in \texttt{MadGraph5\_aMC@NLO}).

%%%%%%%%%%%%%%%%%%%%%%%%%%%%%%%%%%

\subsection{Observable definition}
\label{subsec:observables}

\begin{figure}[!htpb]
\centering
\includegraphics[width=0.7\textwidth]{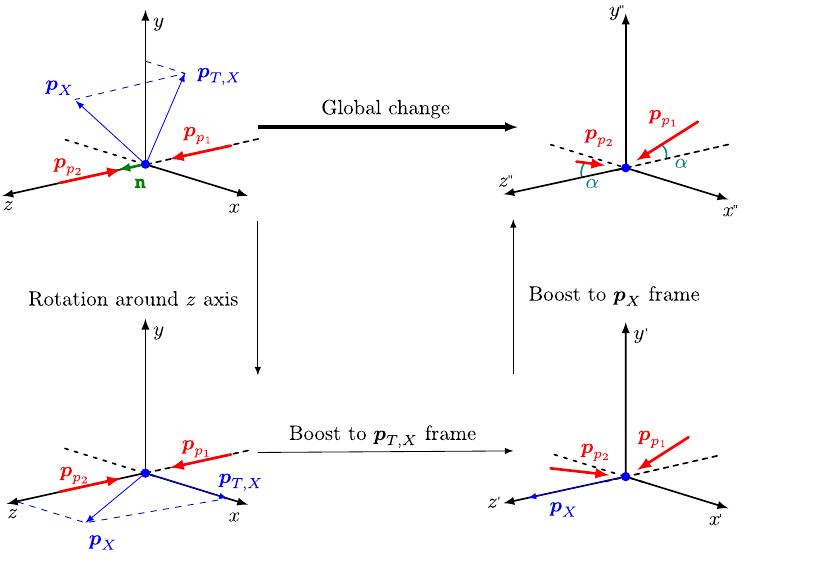}
\caption{Sketch illustrating the rest frame definition adopted in this work. The rest frame $X$ is shown here, defined by $\boldsymbol{p}_X=\boldsymbol{0}$, where $X= H$, $t\bar t$, $t\bar t H$.}
\label{fig:lab-frames}
\end{figure}

We consider a right-handed coordinate system with its origin at the nominal interaction point (IP). In the laboratory frame, see~\cref{fig:lab-frames}, top-left, the \(z\)-axis is defined along the beam axis, and $\boldsymbol{n}$ denotes the associated unit vector. The \(x\)-axis points from the IP to the centre of the LHC ring, and the \(y\)-axis points upwards. Cylindrical coordinates \((r,\phi)\) are used in the transverse ($x,y$) plane, \(\phi\) being the azimuthal angle around the \(z\)-axis. The pseudorapidity is defined in terms of the polar angle \(\theta\) as \(\eta = -\ln \tan(\theta/2)\). Four-momenta are denoted by $p$ and their corresponding three-momenta are denoted by $\boldsymbol{p}$. The magnitudes of the components of the three-momenta in the transverse plane and along the $x$ and $z$-axes are labeled as $\pT$, $p^x$ and $p^z$, respectively. The subscripts $t$, $\bar{t}$, $H$, $p_1$ and $p_2$ in the labels indicate to which particle they refer to, i.e. top quark ($t$), anti-top quark ($\bar{t}$), Higgs boson ($H$) or the colliding protons ($p_1$, $p_2$). We consider four possible reference frames for the computation of the observables:
\begin{itemize}
    \item the laboratory frame (lab frame),
    \item the $t\bar t$ rest frame, where $\boldsymbol{p}_t+\boldsymbol{p}_{\bar t}= \boldsymbol{0}$ ($t\bar t$ frame),
    \item the $H$ rest frame, where $\boldsymbol{p}_H=\boldsymbol{0}$ ($H$ frame),
    \item the $t\bar tH$ rest frame, where $\boldsymbol{p}_t+\boldsymbol{p}_{\bar t}+\boldsymbol{p}_H= \boldsymbol{0}$ ($t\bar tH$ frame).
\end{itemize}
To unambiguously define these frames, we start in the lab frame. To move to the frame $X$ (defined by $\boldsymbol{p}_X=\boldsymbol{0}$, where $X= H$, $t\bar t$, $t\bar t H$), we first apply a rotation around the $z$-axis such that $\boldsymbol{p}_{T,X}$ is parallel to the $x$-axis, see~\cref{fig:lab-frames}, left. Then, a boost along the transverse direction is performed such that $\boldsymbol{p}_{T,X}=0$. Finally, a boost along the longitudinal direction ensures that $\boldsymbol{p}_X=0$. These last two steps are illustrated in~\cref{fig:lab-frames}, right. In the resulting coordinate system, the proton momenta lie in the $x-z$ plane and both form the same angle with the $z$-axis.

The \cp-sensitive observables considered in this work are summarised in~\cref{tab:obs}. All of these are \cp-even observables. We analyse them in all applicable rest frames listed above. Any new observable candidate for the extended STXS framework should be defined for all \ttH\ events, regardless of the Higgs boson and top quark pair decay mode. Therefore, in the following, we consider that the Higgs boson, the top quark and anti-top quark are reconstructed experimentally such that their momenta are accessible. In practice, we compute the observables listed in~\cref{tab:obs} at the parton level and apply specific acceptance and smearing factors to mimic detector effects, as described in~\cref{subsec:mostsensitive_detector}. Note that, given the considered rest frames, none of the observables listed in~\cref{tab:obs} requires a distinction between the top and the anti-top quark, which would be very challenging experimentally. 

%%% table %%%
\setlength{\tabcolsep}{18pt}
\renewcommand{\arraystretch}{1.8}
\begin{table}[!htbp]
    \centering
    \begin{tabular}{|c|c|c|c|}
        \hline
        observable & definition & frame & reference \\
        \hline
        % $\pTx{t}$ & - & all & - \\
        % $\pTx{\bar t}$ & - & all & - \\
        $\pTx{H}$ & - & lab, $t\bar t$, \ttH & - \\
        $\Delta\eta_{t\bar t}$ & $|\eta_{t} - \eta_{\bar t}|$ & lab, $H$, \ttH & - \\
        $\Delta\phi_{t\bar t}$ & $|\phi_{t} - \phi_{\bar t}|$ & lab, $H$, \ttH & - \\
        $m_{t\bar t}$ & $(p_t + p_{\bar t})^2$ & frame-invariant & - \\
        $m_{\ttH}$ & $(p_t + p_{\bar t} + p_H)^2$ & frame-invariant & - \\
        $|\cos\tstar|$ & $\frac{|\boldsymbol{p}_t \cdot \boldsymbol{n}|}{|\boldsymbol{p}_t| \cdot |\boldsymbol{n}|}$ & $t\bar t$ &\cite{Collins:1977iv,Goncalves:2018agy} \\
        $b_1$ & $\frac{(\boldsymbol{p}_t \times \boldsymbol{n}) \cdot (\boldsymbol{p}_{\bar t} \times \boldsymbol{n})}{\pTx{t}  \pTx{\bar t}}$ & all & \cite{Gunion:1996xu} \\
        $b_2$ & $\frac{(\boldsymbol{p}_t \times \boldsymbol{n}) \cdot (\boldsymbol{p}_{\bar t} \times \boldsymbol{n})}{\left| \boldsymbol{p}_t \right| \ \left| \boldsymbol{p}_{\bar t} \right|}$ & all & \cite{Gunion:1996xu} \\
        $b_3$ & $\frac{p_t^x \ p_{\bar t}^x}{\pTx{t}\pTx{\bar t}}$ & all & \cite{Gunion:1996xu} \\
         $b_4$ & $\frac{p_t^z \ p_{\bar t}^z}{\left| \boldsymbol{p}_t \right| \ \left| \boldsymbol{p}_{\bar t} \right|}$ & all & \cite{Gunion:1996xu} \\
        $\phi_C$ & $\arccos\bigg(\frac{|(\boldsymbol{p}_{p_1} \times \boldsymbol{p}_{p_2}) \cdot (\boldsymbol{p}_{t} \times \boldsymbol{p}_{\bar t})|}{\left|\boldsymbol{p}_{p_1} \times \boldsymbol{p}_{p_2} \right| \ \left| \boldsymbol{p}_{t} \times \boldsymbol{p}_{\bar t} \right|}\bigg)$ & $H$ & \cite{Cao:2020hhb} \\
        %$\mathcal{A}(\phi_C)$ & $\frac{N(0 < \phi_c < \nicefrac{\pi}{4}) - N(\nicefrac{\pi}{4}< \phi_c < \nicefrac{\pi}{2})}{N(0 < \phi_c < \nicefrac{\pi}{4}) + N(\nicefrac{\pi}{4}< \phi_c < \nicefrac{\pi}{2})})$ & $H$ & - \\
        \hline
    \end{tabular}
    \caption{Overview of the \cp-sensitive observables considered in this work, including their definition, the rest frames in which they are analysed, and references where they are discussed in more detail.}
    \label{tab:obs}
\end{table}
%%% table %%%

%%%%%%%%%%%%%%%%%%%%%%%%%%%%%%%%%%

\clearpage

\subsection{Distributions at the parton-level}
\label{subsec:obs_distr}

The distributions of the observables defined in~\cref{subsec:observables} which will be discussed further in this work are shown in this section, while the others, which were found to have a lower sensitivity, are shown in~\cref{app:distributions}. As a baseline cut to mimic the ATLAS and CMS tracker acceptance~\cite{PERF-2007-01,CMS-CMS-00-001}, all particles are required to satisfy $|\eta|<2.0$.\footnote{the ATLAS and CMS tracker acceptance is $|\eta|<2.5$, however a tighter cut of $|\eta|<2.0$ is applied to the Higgs boson, the top and anti-top quarks to account for the angular distribution of their decay products.} The Higgs boson, top and anti-top quarks being massive particles, no minimum \pT\ cut is applied.

\begin{figure}[!htpb]
    \centering
    \includegraphics[trim=0.cm 0.5cm 0.cm 0.5cm, width=.8\textwidth]{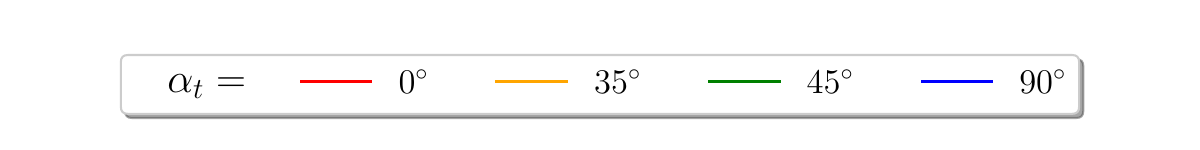}
    \includegraphics[width=.4\textwidth]{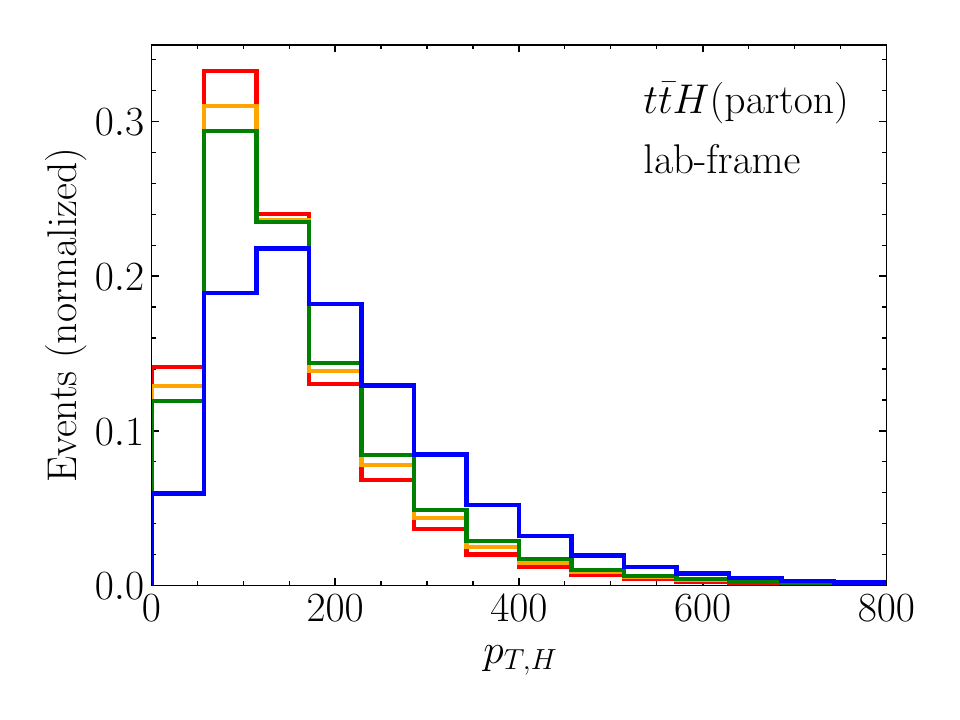}
    \includegraphics[width=.4\textwidth]{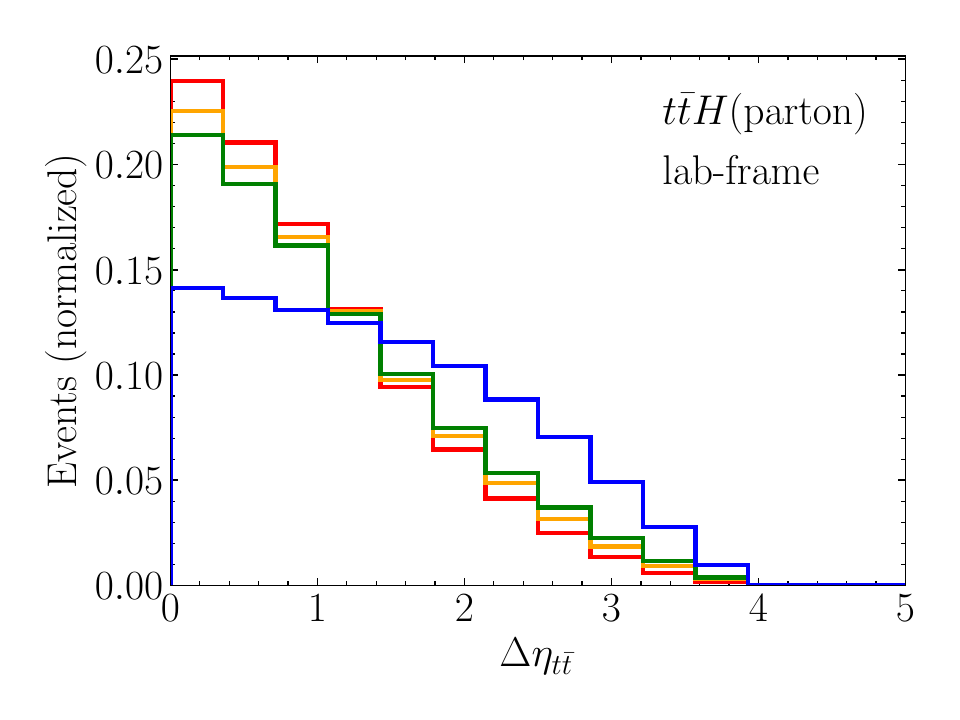}
    \includegraphics[width=.4\textwidth]{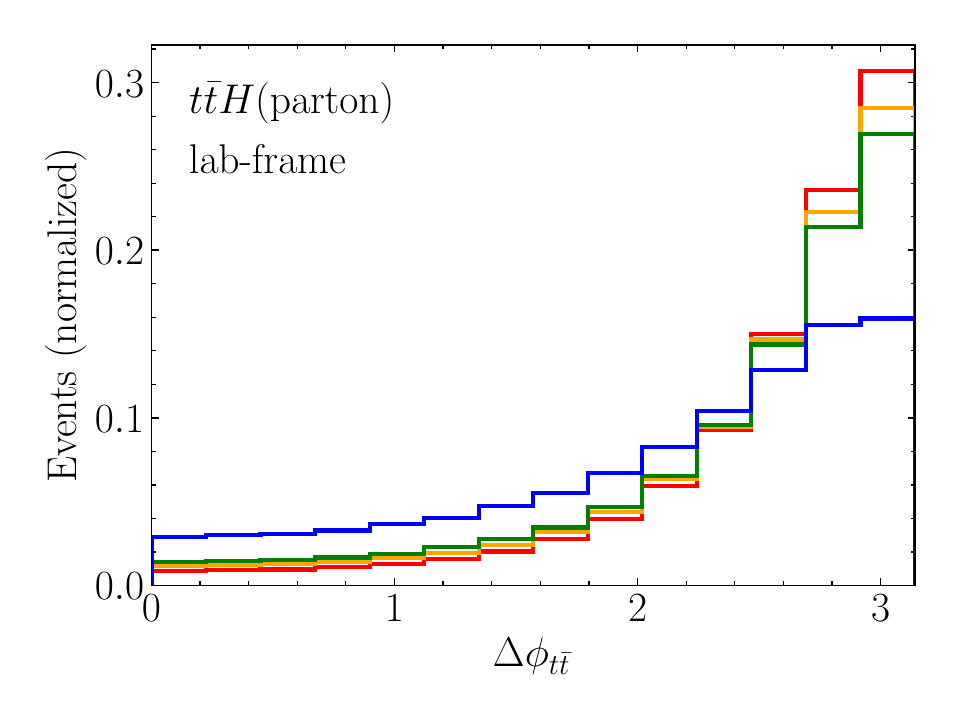}
    \includegraphics[width=.4\textwidth]{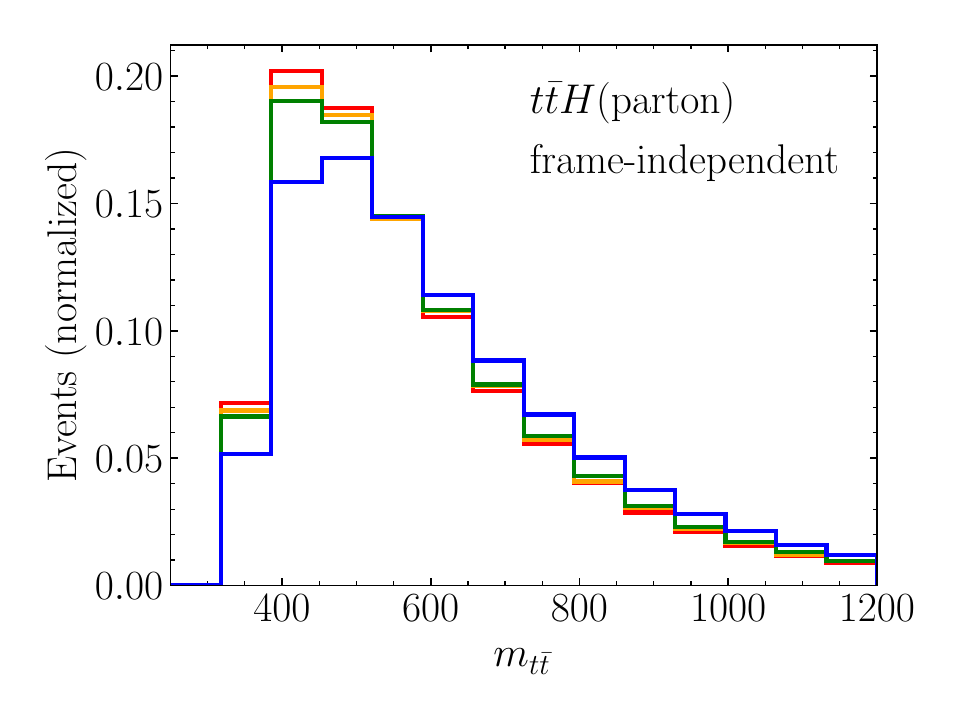}
    \caption{Distributions of the (top, left) Higgs \pT, (top, right) pseudorapidity difference and (bottom, left) azimuthal angle difference between the $t\bar{t}$ pair in the laboratory frame, as well as (bottom, right) the invariant mass of the $t\bar{t}$ pair, which is frame-invariant. $t\bar{t}H$ events generated at parton-level with $\gt=1$ and various values of \cp phase \at are considered following the event generation described in~\cref{subsec:event_generation}. All distributions are normalised to unity. 
    }
    \label{fig:parton_lab}
\end{figure}

\begin{figure}[!htpb]
    \centering
    \includegraphics[trim=0.cm 0.5cm 0.cm 0.5cm, width=.8\textwidth]{figs/example_distributions/legend.pdf}
    \includegraphics[width=.4\textwidth]{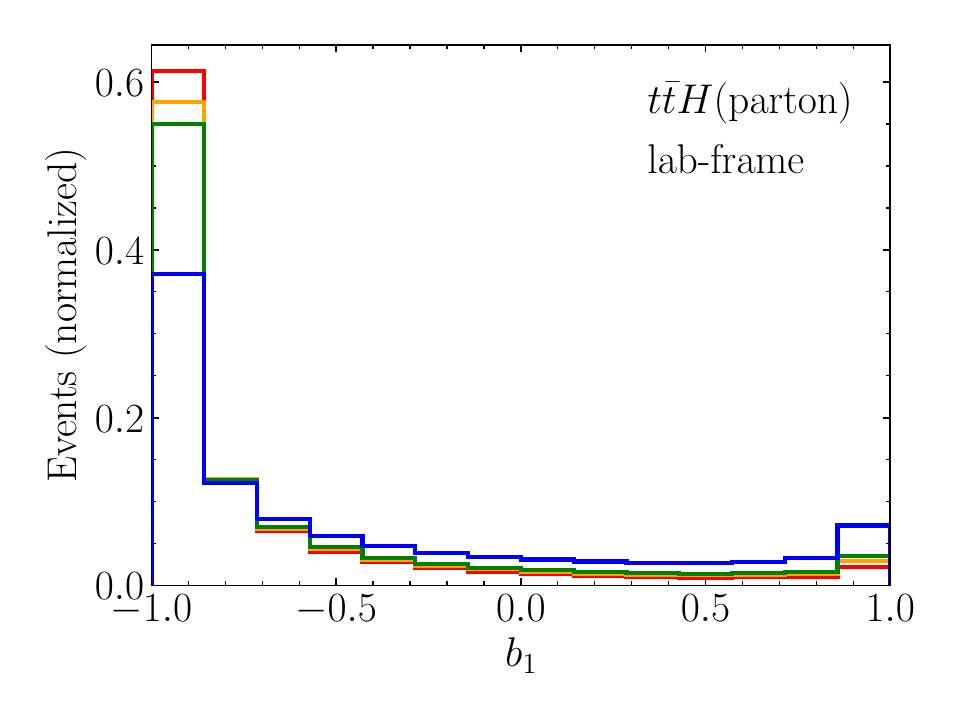}
    \includegraphics[width=.4\textwidth]{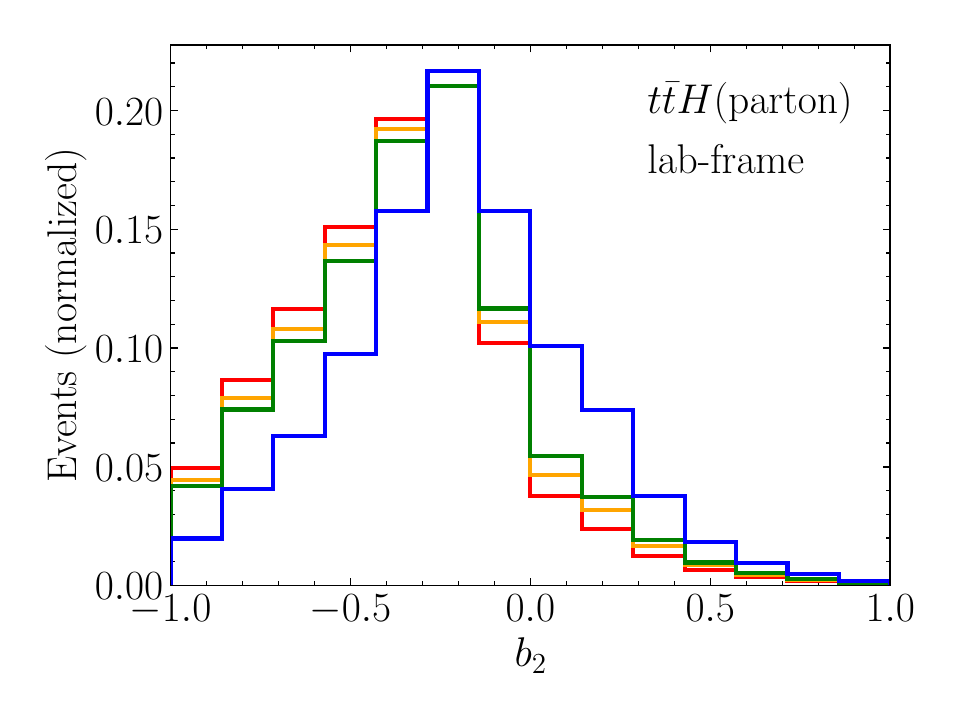}
    \includegraphics[width=.4\textwidth]{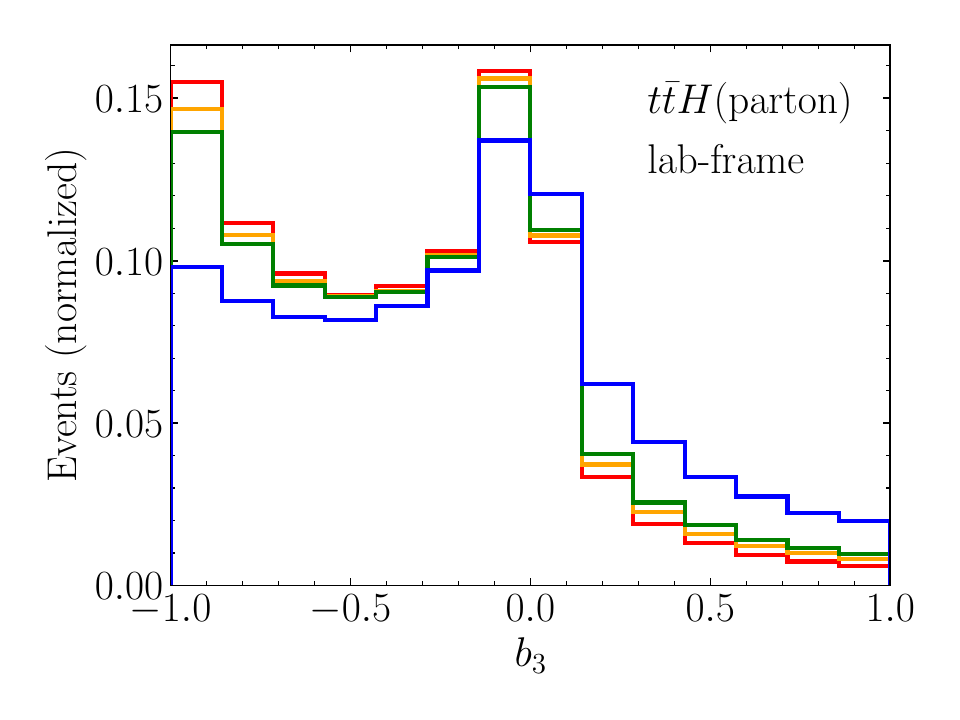}
    \includegraphics[width=.4\textwidth]{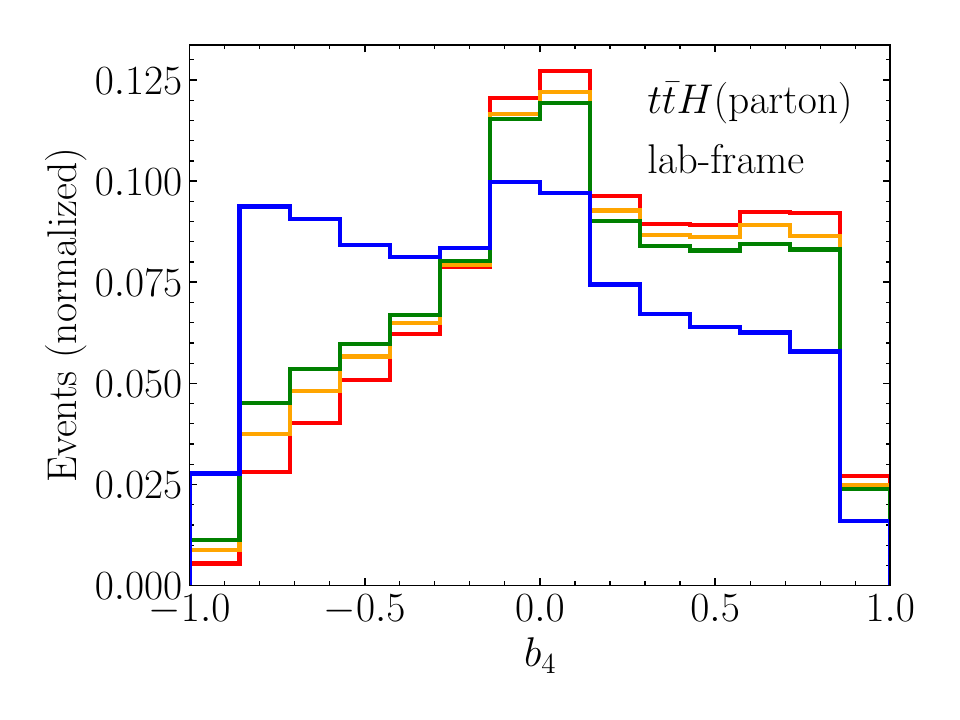}
    \caption{Distributions of the $b_1$, $b_2$, $b_3$ and $b_4$ observables in the laboratory frame. $t\bar{t}H$ events generated at parton-level with $\gt=1$ and various values of \cp phase \at are considered following the event generation described in~\cref{subsec:event_generation}. All distributions are normalised to unity.} 
    \label{fig:parton_blab}
\end{figure}

\begin{figure}[!htpb]
    \centering
    \includegraphics[trim=0.cm 0.5cm 0.cm 0.5cm, width=.8\textwidth]{figs/example_distributions/legend.pdf}
    \includegraphics[width=.4\textwidth]{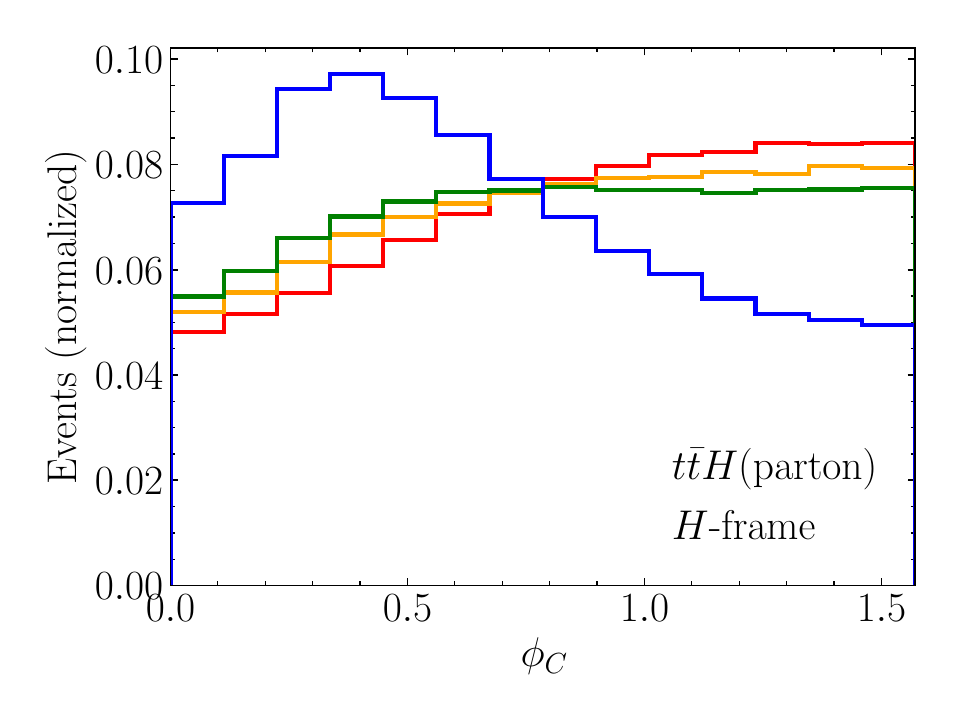}
    \includegraphics[width=.4\textwidth]{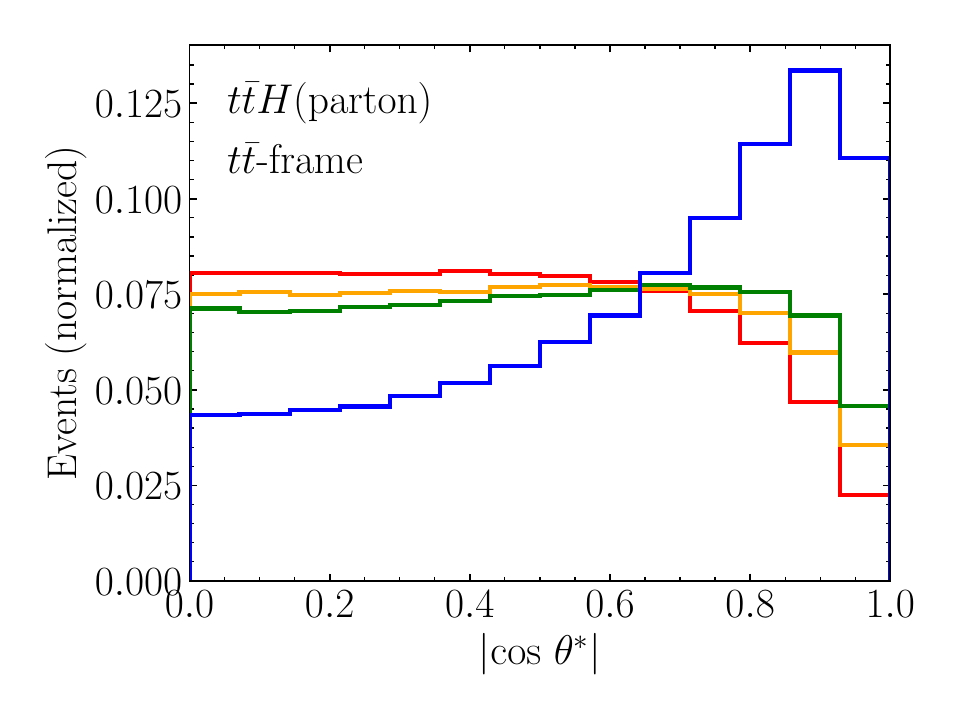}
    \caption{Distributions of (left) the $\phi_C$ and (right) the $|\cos\tstar|$ observables in the Higgs and $t\bar{t}$ frame. $t\bar{t}H$ events generated at parton-level with $\gt=1$ and various values of \cp phase \at are considered following the event generation described in~\cref{subsec:event_generation}. All distributions are normalised to unity.} 
    \label{fig:parton_other}
\end{figure}

The distributions of the Higgs \pT, azimuthal angle difference and pseudorapidity difference between the top and anti-top quarks in the laboratory frame are shown in~\cref{fig:parton_lab}, together with the invariant mass of the $t\bar{t}$ pair, which is frame-invariant. The distributions of the $b_1$, $b_2$, $b_3$ and $b_4$ observables in the laboratory frame are shown in~\cref{fig:parton_blab}, while the distributions of the $\phi_C$ angle in the Higgs boson frame and the cosine of the scattering angle with respect to the direction of the proton beams in the $t\bar{t}$ frame ($|\cos\tstar|$, the so-called Collins-Soper angle) are shown in~\cref{fig:parton_other}.  

Large shape differences are observed between the SM and the pure \cp-odd scenario, whereas the shape differences between the SM and \cp models closer to the current experimental limit ($\at=45^\circ$ or $35^\circ$) are much reduced, yet visible. Larger tails at high (low) value are observed in case of the presence of a \cp-odd component for $\pTx{H}$, $\Delta\eta_{t\bar t}$, $m_{t\bar t}$, $b_1$, $b_2$, $b_3$ and $|\cos\tstar|$ ($b_4$, $\phi_C$ and $\Delta\phi_{t\bar t}$).

The effect from the $|\eta|<2.0$ cut on the top and anti-top quarks is particularly visible in the $\Delta\eta_{t\bar t}$ distribution, which consequently has a sharp cut-off at $\Delta\eta_{t\bar t}=4$. The impact of this cut is also clearly visible on the $\phi_C$ angle and $\cos\tstar$ distributions, as it breaks the approximate, accidental flat distribution observed for these two observables in the SM by depleting the $\phi_C<\pi/4$ and $|\cos\tstar|> \cos(\pi/4)$ regions. A comparison of the distributions of these variables before and after the cut on $\eta$ can be found in \cref{app:eta_cut}. Let us also note here the large peak in the $b_1$ observable distribution in the laboratory frame between the lower bound value of $-1$ and $-0.9$ that occurs irrespective of the \cp hypothesis, but is more pronounced for lower $\at$.

%%% figure %%%
%%% figure %%%

%%% figure %%%
%%% figure %%%

%%% figure %%%
%%% figure %%%

%%%%%%%%%%%%%%%%%%%%%%%%%%%%%%%%%%
%%%%%%%%%%%%%%%%%%%%%%%%%%%%%%%%%%
\clearpage
\section{Evaluation of the observable performance}
\label{sec:mostsensitive}

This section focuses on quantifying the discriminating power of each observable to distinguish between a mixed \cp state of the top-Yukawa coupling and the SM case. With the goal of extending the STXS framework in mind, we examine specifically the three channels explored so far in the experimental $t\bar{t}H$ analyses at the LHC. We discuss in~\cref{subsec:mostsensitive_detector} the limitations in resolution and reconstruction of the top quarks and Higgs boson in each channel, while the formula used for performance evaluation is explained in \cref{subsec:sig_computation}. 

\subsection{Detector effects and selection efficiency}
\label{subsec:mostsensitive_detector}

The $t\bar{t}H$ process at the LHC has been studied so far through three main experimental channels. The first two channels target $H\to\gamma\gamma$ and $H\to b\bar{b}$ decays and will be labeled $t\bar{t}H(\to\gamma\gamma)$ and $t\bar{t}H(\to bb)$ in the following, while the third channel selects events with multiple leptons in the final state and will be labeled $t\bar{t}H$(multilep.). The latest experimental results from the ATLAS and CMS collaborations for each of these channels are available in~\ccite{ATLAS:2020ior,ATLAS:2023cbt,CMS:2020cga,CMS:2021nnc,CMS:2022mlq,CMS:2022dbt}. In order to compare the performance of the various observables defined in~\cref{sec:obs} in discriminating the top-Yukawa \cp properties in a meaningful way, it is necessary to account for the limited efficiency of the event selection performed in experimental data analyses as well as the finite resolution and acceptance of the detectors. The $t\bar{t}H(\to\gamma\gamma)$, $t\bar{t}H(\to bb)$ and $t\bar{t}H$(multilep.) channels differ widely in terms of final state objects and cross-section times branching ratio; taking into account these detector and acceptance effects is highly non-trivial, hence a simplified approach is adopted in this work. We use the $t\bar{t}H$ samples described in~\cref{subsec:event_generation} as a starting point. Relevant decay branching ratios (Higgs, top quark) are taken from the SM, while the limited efficiency of the event selection and acceptance of the detectors are extracted per channel from ATLAS and/or CMS published results (as quoted in the text) and applied as event weights. Finite resolution effects are also extracted per channel from ATLAS and/or CMS published results and accounted for by smearing the four momenta of the Higgs boson and the top and anti-top quarks. All the input parameters to our model are shown in~\cref{tab:yields}, while more details are given below.

The $t\bar{t}H(\to\gamma\gamma)$ channel provides the smallest cross-section times branching ratio, but the best signal-over-background ratio due to the excellent detector resolution on the reconstructed Higgs mass. If any, all high-\pT neutrinos can be assumed to originate from the top or anti-top decays, which eases the $t\bar{t}$ pair reconstruction. Following the $t\bar{t}H$ event yields in signal-enriched regions provided in~\ccite{ATLAS:2020ior}, we scale the $t\bar{t}H$ yields at the parton-level by the branching ratio of $H\to\gamma\gamma$ in the SM times the acceptance and selection efficiency of the $t\bar{t}H$ sample for different \at values. For $H\to\gamma\gamma$, it ranges between 0.25 and 0.32, corresponding to $\at=0^\circ$ and $\at=90^\circ$, respectively. The Higgs boson and top/anti-top quark transverse momenta are smeared using a Gaussian distribution with a width of 4 and $40\gev$, respectively, while the pseudo-rapidity of the top/anti-top quark is smeared by $0.5$. These last numbers were tuned to match the ATLAS photon energy resolution~\cite{PERF-2013-05} and the top quark mass distribution published in~\ccite{ATLAS:2020ior}.

The $t\bar{t}H(\to bb)$ channel provides the highest cross-section times branching ratio, but the lowest signal-over-background ratio due to the large background originating from $t\bar{t}$ pairs produced in association with $b$-quarks and the comparatively poor detector energy resolution on jet measurements with respect to photons. Similarly to $t\bar{t}H(\to\gamma\gamma)$, if any, all high-\pT neutrinos shall originate from the top or anti-top decays. However, jets originating from $b$-hadrons can come from either the Higgs boson decay or the top/anti-top decays, making the top/anti-top reconstruction more challenging with respect to $t\bar{t}H(\to\gamma\gamma)$. Following the $t\bar{t}H$ event yields in signal-enriched regions provided in~\ccite{ATLAS:2023cbt}, we apply on top of the $H(\to bb)$ SM branching ratio a scaling factor ranging from 0.005 to 0.0065 depending on the \cp hypothesis (the extreme values corresponding to $\at=0^\circ$ and $\at=90^\circ$, respectively) to the $t\bar{t}H$ yields at the parton-level. The Higgs boson and top/anti-top quark transverse momenta are smeared using a Gaussian distribution with a width of 80 and $70\gev$, respectively, while the pseudo-rapidity of the top/anti-top quark is smeared by $0.8$ and the azimuth angle by $20^\circ$. The smearing parameters were tuned to match the Higgs boson candidate mass distribution shown in~\ccite{HIGG-2020-23} and the top reconstruction performance described in~\cite{TOPQ-2011-08,TOPQ-2018-15}.

Finally, the $t\bar{t}H$(multilep.) channel lies in between the $t\bar{t}H(\to\gamma\gamma)$ and $t\bar{t}H(\to bb)$ channels in terms of cross-section times branching ratio and signal-over-background ratio. A large number of combinations of top, anti-top quarks and Higgs boson decays lead to multi-lepton final states. In this work, we adopt a simplified approach by considering only the two most sensitive experimental sub-channels, which include the final states with two electrons or muons with the same sign or at least three electrons or muons. The resulting branching ratio is $6.79 \cdot 10^{-2}$ and we obtain the \ttH event yields from signal-enriched regions provided in~\ccite{CMS:2022dbt}. Based on these yields, we apply to the branching ratio an additional acceptance and efficiency factor ranging from 0.036 to 0.042 depending on the \cp hypothesis (the extreme values corresponding to $\at=0^\circ$ and $\at=90^\circ$, respectively). The smearing parameters are the same as the ones considered for the $t\bar{t}H(\to bb)$ channel, except for the Higgs \pT which is smeared by a larger value of $120\gev$ instead of $80\gev$. Indeed, despite the lack of published experimental data, we anticipate lower performance in the Higgs boson reconstruction due to the multiple high-\pT neutrinos in these events, which can typically originate from either the Higgs or the top/anti-top decays.

\setlength{\tabcolsep}{14pt}
\begin{table}[!htpb]
\centering
\begin{tabular}{ |c||c|c|c| } \hline
 & $t\bar{t}H(\to\gamma\gamma)$ & $t\bar{t}H(\mathrm{multilep.})$ & $t\bar{t}H(\to b\bar{b})$ \\
 \hline
  BR & $2.27 \cdot 10^{-3}$ & $6.79 \cdot 10^{-2}$ & $5.81 \cdot 10^{-1}$ \\
 \hline
 \multicolumn{4}{|l|}{Acceptance/efficiency scaling factors} \\
 \hline
 $\at=0^\circ$ & $2.5 \cdot 10^{-1}$ & $3.6 \cdot 10^{-2}$ & $5.0 \cdot 10^{-3}$ \\
 \hline
 $\at=35^\circ$ & $2.5 \cdot 10^{-1}$ & $3.6 \cdot 10^{-2}$ & $5.2 \cdot 10^{-3}$ \\
 \hline
 $\at=45^\circ$ & $2.7 \cdot 10^{-1}$ & $3.8 \cdot 10^{-2}$ & $5.4 \cdot 10^{-3}$ \\
 \hline
 $\at=90^\circ$ & $3.2 \cdot 10^{-1}$ & $4.2 \cdot 10^{-2}$ & $6.5 \cdot 10^{-3}$ \\
  \hline
  \multicolumn{4}{|l|}{Smearing factors} \\
 \hline
 $\Delta \pTx{H}$ [GeV] & 4 & 120 & 80 \\
 \hline
 $\Delta p_{T,t}$ [GeV]& 40 & 70 & 70 \\
 \hline
 $\Delta \eta_t$ & $0.5$ & $0.8$ & $0.8$ \\
 \hline
 $\Delta \phi_t$ [$^\circ$] & -- & 20 & 20 \\
 \hline
  \multicolumn{4}{|l|}{Final \ttH\ event yields at 300 fb$^{-1}$} \\
 \hline
 $\at=0^\circ$ & $86$ & $372$ & $442$ \\
 \hline
 $\at=35^\circ$ & $70$ & $302$ & $373$ \\
 \hline
 $\at=45^\circ$ & $67$ & $281$ & $341$ \\
 \hline
 $\at=90^\circ$ & $47$ & $185$ & $245$ \\
 \hline
\end{tabular}
\caption{Summary of the input parameters to the simplified model used to emulate the detector effects and the analysis selection efficiency in each experimental channel. The first block (from top to bottom) shows the scaling factors accounting for the branching ratio, the second block shows the scaling factors accounting for the limited acceptance of the detector and the limited efficiency of the analysis selection, while the third and fourth blocks show the Gaussian smearing factors applied on the Higgs boson and top/anti-top momenta and the final \ttH event yields obtained for an integrated luminosity of $300\invfb$, respectively. As a reminder, all cross-sections are obtained directly from \texttt{MadGraph5\_aMC@NLO} to which a k-factor of $1.14$ is applied to correct for NLO effects.}
\label{tab:yields}
\end{table}

%%%%%%%%%%%%%%%%%%%%%%%%%%%%%%%%%%

\subsection{Significance computation}
\label{subsec:sig_computation}

In order to quantify the sensitivity of the various observables discussed in~\cref{sec:obs} to \cp violation in the top-Yukawa coupling, we assume to have at our disposal the measurement of the corresponding $t\bar{t}H$ distributions in each experimental channel. The scaling and smearing factors described in the previous section are applied to $t\bar{t}H$ events generated at the parton level to emulate detector effects and obtain realistic yields. 

The sensitivity of an observable to a given BSM model is evaluated by computing a significance $S$. This significance reflects the discriminating power to reject the BSM hypothesis (parameterized by $g_t,\alpha_t$) against the null hypothesis of $g_t=1$, $\at=0$ (corresponding to the SM). Formally, it is defined as the sum in quadrature of the significance computed in each bin of the observable, $(S_i)_{i=1..N_\text{bins}}$. The observed $t\bar{t}H$ event yields are assumed to match the central SM predictions. If we make the additional assumption that they are Poisson distributed and not impacted by additional sources of uncertainty, the $S_i$ can then be evaluated using the Wilk's theorem:
\begin{align}
S_i = \sqrt{-2\left( n_i \ln\frac{m_i}{n_i} + n_i-m_i \right)}\,,
\label{eq:significance_i}
\end{align}
where $n_i$ and $m_i$ are the numbers of events in the $i$th bin of the observable as predicted by the SM and the BSM model under consideration, respectively. 
Accordingly, the significance $S$ results as
\begin{align}
S = \sqrt{\sum_{i=1}^{N_\text{bins}} S_i^2} =\sqrt{-2 \sum_{i=1}^{N_\text{bins}} \left( n_i \ln\frac{m_i}{n_i} + n_i-m_i \right)}\,.
\label{eq:significance}
\end{align}
In the following, all bins with $n_i<2$ are merged with their neighbouring bins until the merged bin contains more than two events, starting with left neighbours, to ensure the validity of Wilk's theorem for significance evaluations. We checked that the signifiance values are stable within a few percent if $n_i<5$ is used instead.

The previous equations rely on two important assumptions, first that the measured $t\bar{t}H$ event yields are Poisson distributed and second that they are not impacted by additional sources of uncertainty. In reality, non-$t\bar{t}H$ events are expected to enter the measurement selection such that 
\begin{align}
n_i = n^{\mathrm{data}}_i - n^{\mathrm{bkg}}_i,
\label{eq:bkg}
\end{align}
where $n^{\mathrm{data}}_i$ and $n^{\mathrm{bkg}}_i$ are the number of observed data events  and the number of predicted non-$t\bar{t}H$ events in the $i$th bin, respectively. The evaluation of $n^{\mathrm{bkg}}_i$ is part of the experimental analysis work. It is impacted by systematic uncertainties related to the detector calibration and physics modelling. Therefore the assumptions used to derive~\cref{eq:significance_i,eq:significance} are valid only in the limit of high signal over background ratio ($S/B$), i.e. $n_i/n^{\mathrm{bkg}}_i\gg 1$. To take account of these deviations from our initial assumptions, the~\cref{eq:significance_i,eq:significance} are corrected in the following way~\cite{ATL-PHYS-PUB-2020-025}:
\begin{align}
S_i &= \sqrt{-2 \bigg( n_i \ln\frac{m_i (n_i + \sigma_i^2)}{n_i^2+m_i \sigma_i^2} - \frac{n_i^2}{\sigma_i^2} \ln\bigg[ 1 + \frac{\sigma_i^2 (m_i - n_i)}{n_i (n_i + \sigma_i^2)} \bigg] \bigg)}, \label{eq:significance_i2} \\
S &= \sqrt{-2 \sum_{i=1}^{N_\text{bins}} \bigg( n_i \ln\bigg[\frac{m_i (n_i + \sigma_i^2)}{n_i^2+m_i\sigma_i^2}\bigg] - \frac{n_i^2}{\sigma_i^2} \ln\bigg[ 1 + \frac{\sigma_i^2 (m_i - n_i)}{n_i (n_i + \sigma_i^2)} \bigg] \bigg) },
\label{eq:significance_uncertainty}
\end{align}
where $\sigma_i$ is a newly introduced term corresponding to an extra uncertainty in $n_i$. It is computed as the sum in quadrature of two terms,
\begin{align}
\sigma_i^2 = (\sigma_i^{\mathrm{stat}})^2 + (\sigma_i^{\mathrm{syst}})^2.
\end{align}
The first term, $\sigma_i^{\mathrm{stat}}$, accounts for the additional Poisson-like uncertainty originating from the fact that $n^{\mathrm{bkg}}_i \ne 0$ and can be approximated by $\sqrt{n^{\mathrm{bkg}}_i}$. The $(n^{\mathrm{bkg}}_i)_{i=1..N_\text{bins}}$ largely depend on the analysis design and the observable considered and therefore are in general not trivial to evaluate in the context of this work. Though we note that they are connected to the $S/B$ in each bin such that $\sigma_i^{\mathrm{stat}}~=~\sqrt{\frac{n_i}{S/B}}$. The $S / B$ typically varies bin-by-bin in the experimental data analyses, however, a single value per channel is considered here for simplification, taken from the typical values observed in the \ttH-enriched regions defined in the corresponding ATLAS and CMS analyses~\cite{CMS:2022dbt,ATLAS:2023cbt,ATLAS:2020ior}. $S / B = 1$, 0.4, and 0.1 is assumed in the $t\bar{t}H(\to\gamma\gamma)$, $t\bar{t}H$(multilep.), and $t\bar{t}H(\to bb)$ channels, respectively. The validity of this simplifying assumption depends on the shape differences between signal and background for the observables under consideration, which will be discussed further in~\cref{subsec:bkg}. 

The second term, $\sigma_i^{\mathrm{syst}}$, accounts for the systematic uncertainty in $n^{\mathrm{bkg}}_i$. We define it so that it contains the uncertainty in the shape of the non-$t\bar{t}H$ events. We take $\sigma_i^{\mathrm{syst}}= n_i~\cdot 0.2$ $[0.5]$ for the $t\bar{t}H$(multilep.) [$t\bar{t}H(\to bb)$] channel, following~\ccite{CMS:2022dbt,ATLAS:2023cbt}, while $\sigma_i^{\mathrm{syst}}=0$ is assumed for the $t\bar{t}H(\to\gamma\gamma)$ channel as this channel is typically dominated by statistical uncertainties~\cite{ATLAS:2020ior}. 
The last source of uncertainty to take into account is the uncertainty in the overall background normalisation. It is significant in the $t\bar{t}H(\to bb)$ and $t\bar{t}H$(multilep.) channels only. For these two channels, we decided to account for it by rescaling the total BSM \ttH event yield to match the total SM \ttH event yields prior to the computation of the significance. This way no discrimination power is obtained from differences in the total rate between SM and BSM in these two channels. In practice, the $m_i$ are replaced in~\cref{eq:significance_i2,eq:significance_uncertainty} by $m'_i$, which corresponds to the number of events in the $i$th bin of the observable as predicted by the BSM model under consideration rescaled so that $\sum_{i=1}^{N_\text{bins}} m'_i = \sum_{i=1}^{N_\text{bins}} n_i$.

\section{Results}
\label{sec:results}

We present results for a luminosity of $300\invfb$, targeting the end of LHC Run-3. Since the current experimental limits exclude values of $g_t=1,~\at\gtrsim 43^\circ$ at the~$95\%~\mathrm{CL}$ using $139\invfb$~\cite{ATLAS:2020ior}, we choose $g_t=1,~\at = 35^\circ$ as a benchmark. In~\cref{subsec:default_binning}, we calculate the significance for all the observables mentioned in~\cref{subsec:observables} as well as for the associated two-dimensional observable combinations. This is done with a fixed number of evenly spaced bins. In~\cref{subsec:mostsensitive_binning}, we optimize the binning individually for the most promising variables based on the combined significance of the three experimental channels. We compare our findings to a multivariate approach in~\cref{subsec:BDT}. Finally, we show in~\cref{subsec:sensitivity} the expected sensitivity in the full two-dimensional $(\gt,\at)$ parameter plane.

\subsection{Results with default binning}
\label{subsec:default_binning}

We first evaluate the sensitivity of each observable separately starting from a default binning of $14$ evenly distributed bins. In addition, we evaluate the sensitivity of the combination of two observables starting from a two-dimensional binning of $6 \times 6$ evenly distributed bins. These (not yet optimised) binnings were chosen to maximise the number of bins to avoid losing valuable kinematic shape information while populating most of the bins with at least a few events. The bin merging method described in the previous section is then applied to guarantee at least two SM $t\bar{t}H$ events per bin. 
In total, $31$ different observables (and their combinations) are considered across the four different rest frames defined in~\cref{subsec:observables}. While the results for all 465 different combinations can be found in~\cref{app:significance_tables}, we focus in the following on the most promising observables and related combinations. 

\begin{figure}[!htbp]
    \centering
    \includegraphics[width=\textwidth]{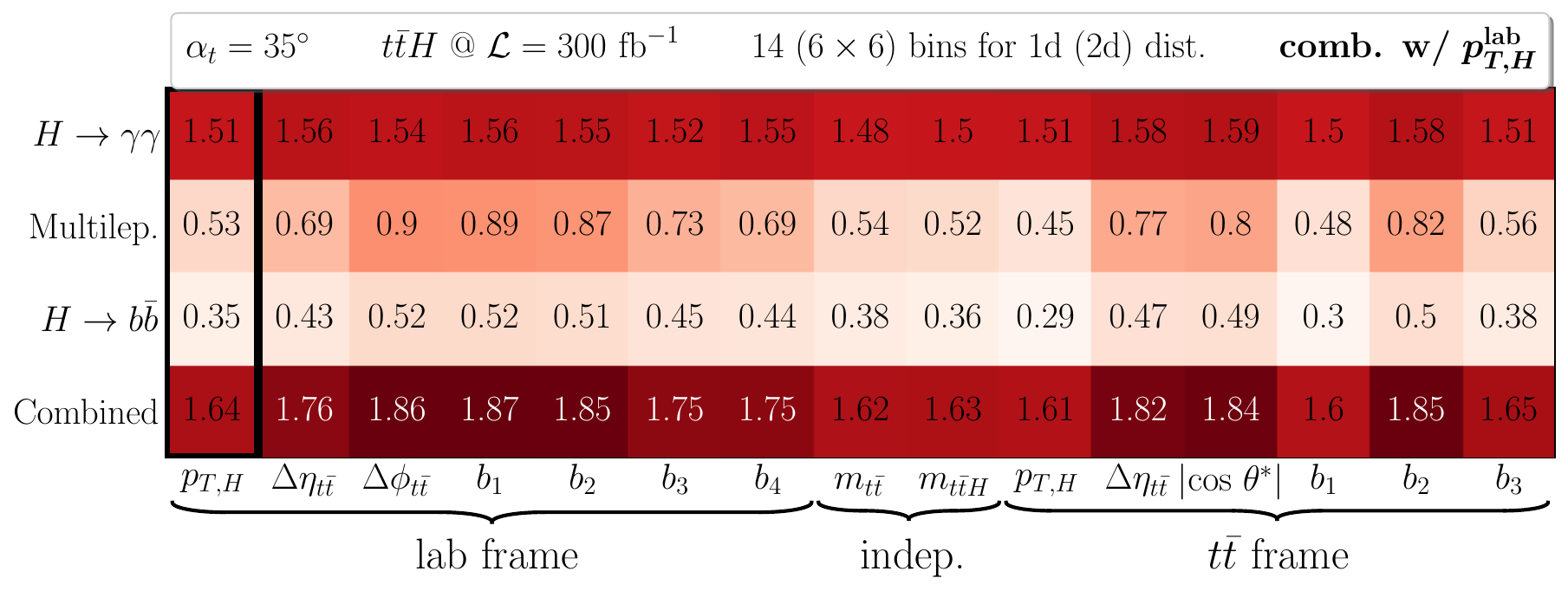}
    \includegraphics[width=\textwidth]{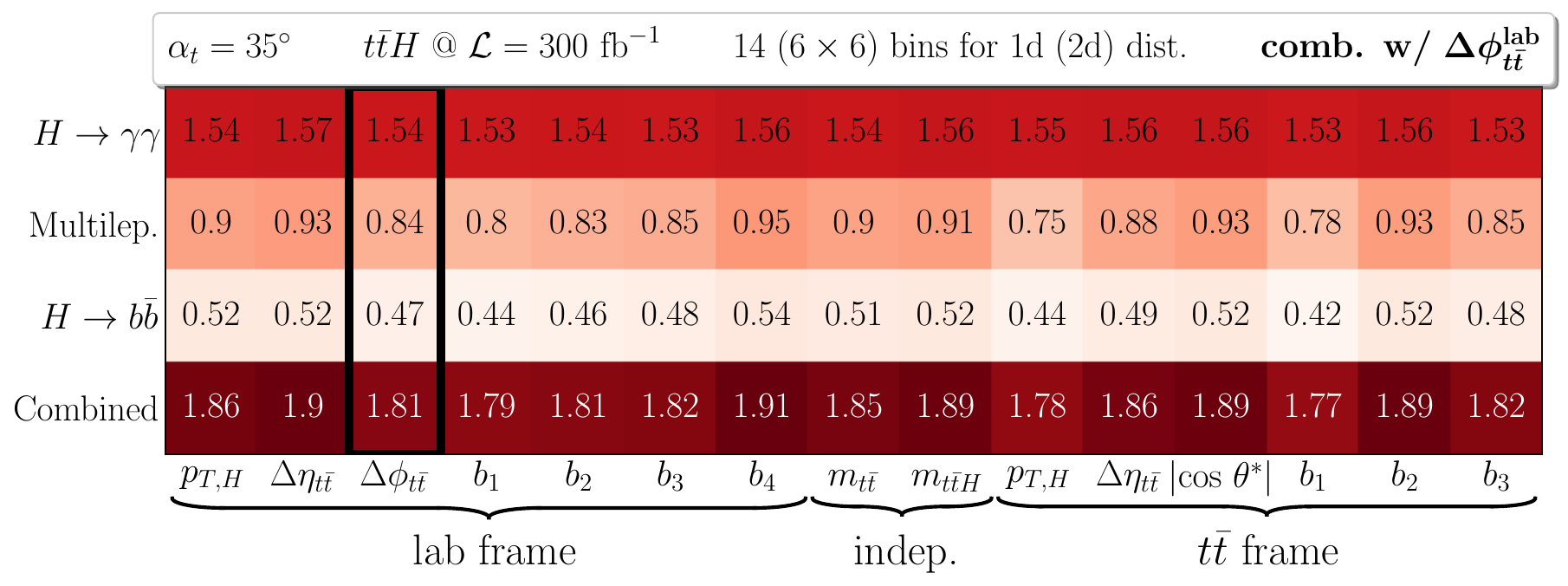}
    \caption{Significances evaluated from the (top) $p_{\mathrm{T},H}$ and (bottom) $\Delta\phi_{t\bar{t}}$ distributions in the laboratory frame for a BSM signal corresponding to $g_t=1,~\at=35^{\circ}$. The emphasised column shows the significance obtained from the one-dimensional distribution while the others show the significances obtained when combining with a second observable. The significances are shown per experimental channel and after their statistical combinations.}
    \label{fig:significance_rows}
\end{figure}

The sensitivity of $p_{\mathrm{T},H}$ and $\Delta\phi_{t\bar{t}}$ in the laboratory frame to a BSM signal corresponding to $g_t=1,~\at=35^{\circ}$ is shown in~\cref{fig:significance_rows} for an integrated luminosity of 300 fb$^{-1}$, separately (emphasized column) and in combination with a second observable (other columns). The significance in each channel as well as the combined significance are shown. The $p_{\mathrm{T},H}$ observable was chosen to be displayed here as it shows high sensitivity and corresponds to the observable used in the STXS binning v1.2~\cite{deFlorian:2016spz,Badger:2016bpw,Berger:2019wnu,Amoroso:2020lgh}, while $\Delta\phi_{t\bar{t}}$ was chosen to be displayed here as it provides the best significance both when considered alone and when combined with certain observables.

The sensitivity of the $t\bar{t}H(\to\gamma\gamma)$ channel largely comes from the difference in terms of total event rate between the SM and BSM hypotheses, and therefore does not depend strongly on the choice of observable. On the other hand, the other channels suffer from a larger background and a less precise event reconstruction, such that no discrimination from the total event rate was considered (see~\cref{subsec:sig_computation}). Hence, the choice of the best observable for the combined three channels is mostly driven by the expected performance in the $t\bar{t}H(\to bb)$ and $t\bar{t}H$(multilep.) channels. The highest significance is obtained when combining $\Delta\phi_{t\bar{t}}$ with $b_4$ in the lab frame and reaches $S=1.91$. The best combinations involving $\pTx{H}$ reach significances ranging from $1.82$ to $1.87$, and are obtained with $\Delta\phi_{t\bar{t}}$, $b_1$, and $b_2$ in the lab frame, as well as $\Delta\eta_{t\bar{t}}$, $|\cos\theta^*|$, and $b_2$ in the $\ttbar$ frame.

While $\pTx{H}$ alone yields a significance considerably weaker than the best combinations, the difference between the 1D $\Delta\phi_{t\bar{t}}$ result, the best 2D combinations involving $\pTx{H}$ and any 2D combination involving $\Delta\phi_{t\bar{t}}$ is smaller. This result brings some flexibility in the final choice of observables. Since the \ttH\ process is split into \pTx{H} bins in the current STXS binning, we favour for practical reasons a reduced set of observables which reach the highest significances when paired with \pTx{H}. For comparison, we also show in \cref{fig:significance_rows} the best combinations involving $\Delta\phi_{t\bar{t}}$. As we will see, given the small difference to the combinations with $\Delta\phi_{t\bar{t}}$, we understand our chosen combinations with \pTx{H} as close to optimal.

%%% figure %%%
%%% figure %%%

\clearpage

%%%%%%%%%%%%%%%%%%%%%%%%%%%%%%%%%%

\subsection{Final results with optimised binning}
\label{subsec:mostsensitive_binning}

We now optimize the binning of the observables preselected in the previous section for an integrated luminosity of $300\invfb$. In combination with $\pTx{H}$, the $b_2$ variable yields the same significance in the lab and $\ttbar$ frame. Taking into account that $\pTx{H}$ is given in the lab frame, we will consider the $b_2$ observable only in the lab frame in the following. For practical reasons, we follow the STXS v1.2 framework for the Higgs transverse momentum, which specifies the following six bins:

\begin{itemize}
    \item $\pTx{H}$: $[0,60,120,200,300,450,+\infty]$~(GeV)
\end{itemize}

\begin{figure}[!htpb]
\centering
\includegraphics[trim=0.2cm 0.2cm 1.5cm 0.2cm, width=.99\textwidth]{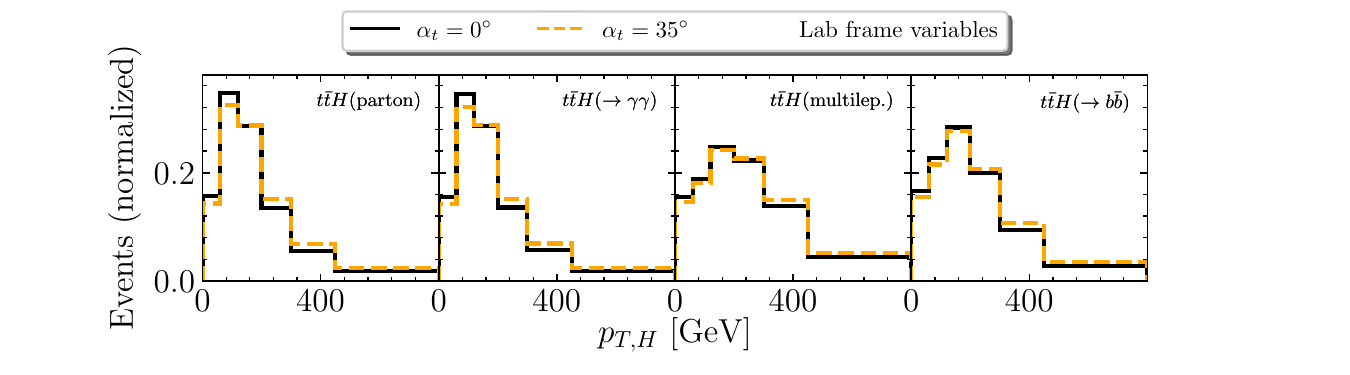}
\includegraphics[trim=0.2cm 0.2cm 1.5cm 0.2cm, width=.99\textwidth]{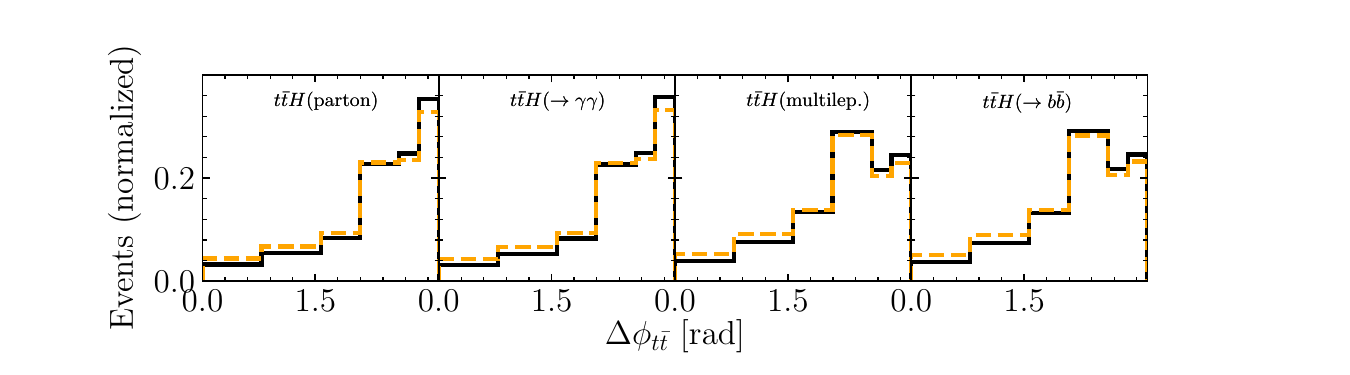}
\includegraphics[trim=0.2cm 0.2cm 1.5cm 0.2cm, width=.99\textwidth]{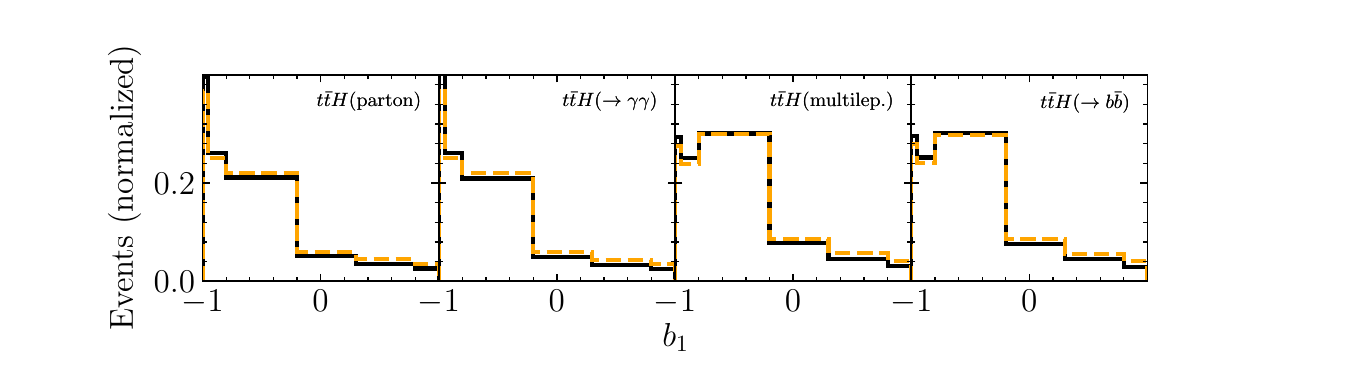}
\includegraphics[trim=0.2cm 0.2cm 1.5cm 0.2cm, width=.99\textwidth]{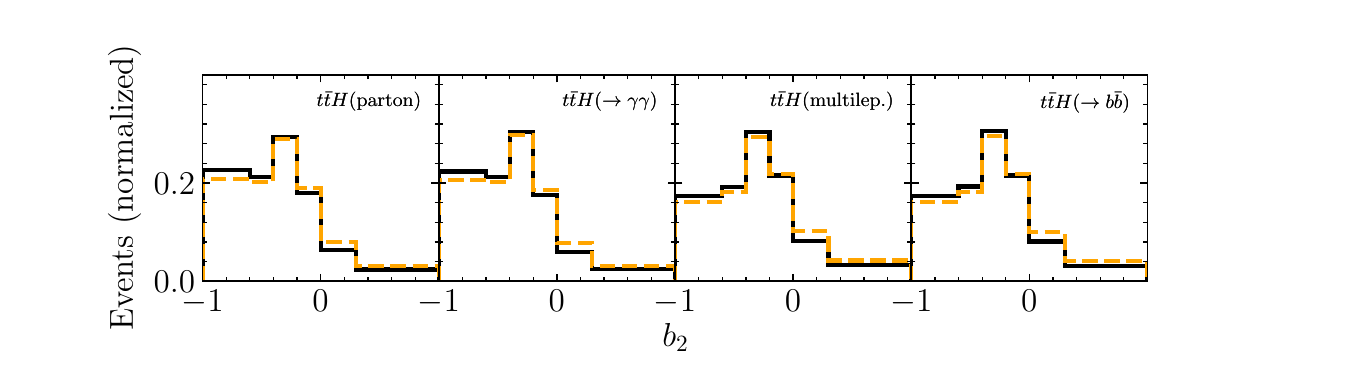}
\caption{Distributions (from top to bottom) of $\pTx{H}$, $\Delta\phi_{t\bar{t}}$, $b_1$ and $b_2$ in the lab frame with optimised binning. Each canvas is split into four panels, from left to right the following distributions are shown: parton-level and smeared distributions for the $t\bar{t}H(\to\gamma\gamma)$, $t\bar{t}H(\mathrm{multilep.})$, and $t\bar{t}H(\to b\bar{b})$ channels. The SM distributions are represented as solid black lines, while the distributions for ${\at = 35^\circ}$ are shown as dashed orange lines.
}
\label{fig:smearings_1}
\end{figure}

\begin{figure}[!htpb]
\centering
\includegraphics[trim=0.2cm 0.2cm 1.5cm 0.2cm, width=.99\textwidth]{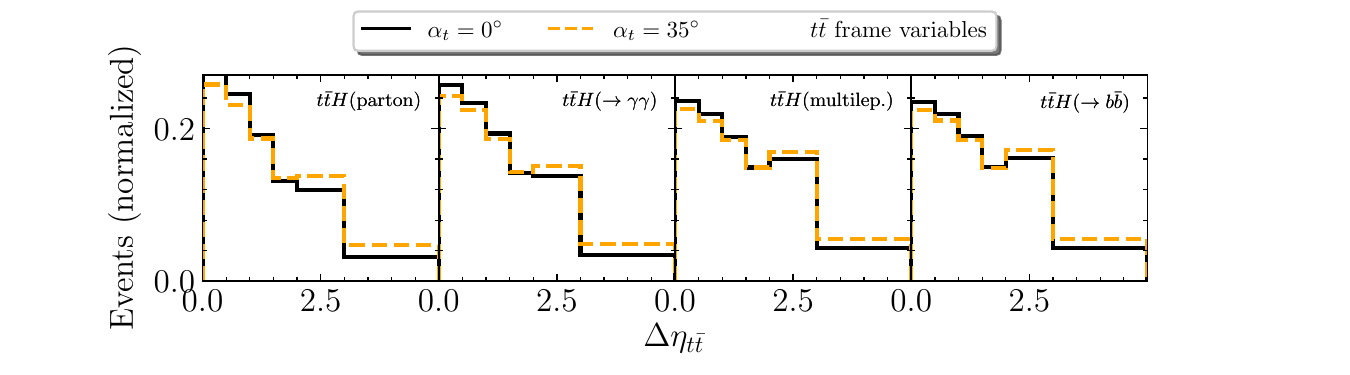}
\includegraphics[trim=0.2cm 0.2cm 1.5cm 0.2cm, width=.99\textwidth]{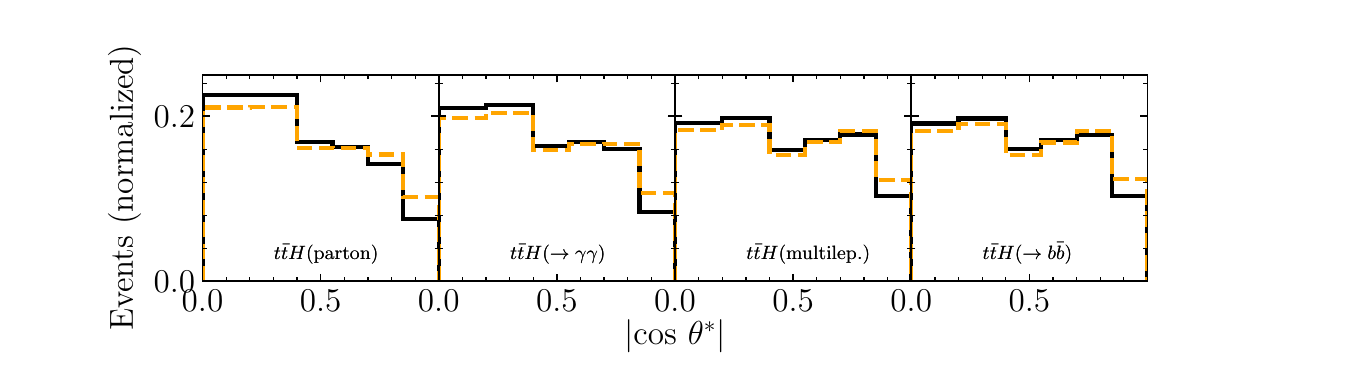}
\caption{Distributions of (top) $\Delta\eta_{t\bar{t}}$ in the $\ttbar$ frame and (bottom) $|\cos\theta^*|$ with optimized binning. Each canvas is split into four panels, from left to right the following distributions are shown: parton level and smeared distributions for the $t\bar{t}H(\to\gamma\gamma)$, $t\bar{t}H(\mathrm{multilep.})$, and $t\bar{t}H(\to b\bar{b})$ channels, respectively. The SM distributions are represented as solid black lines, while the distributions for ${\at = 35^\circ}$ are shown as dashed orange lines.}
\label{fig:smearings_2}
\end{figure}

\begin{figure}[!htpb]
\centering
\includegraphics[trim=0cm 0.2cm 0cm 0.2cm, width=.48\textwidth]{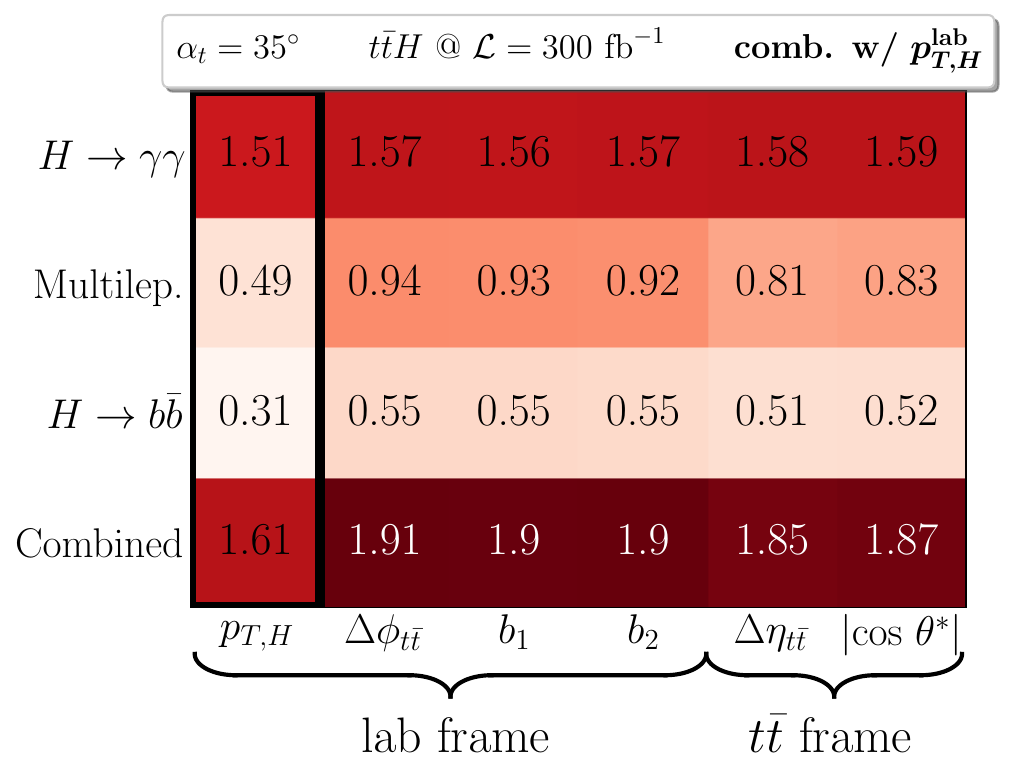}
\includegraphics[trim=0cm 0.2cm 0cm 0.2cm, width=.48\textwidth]{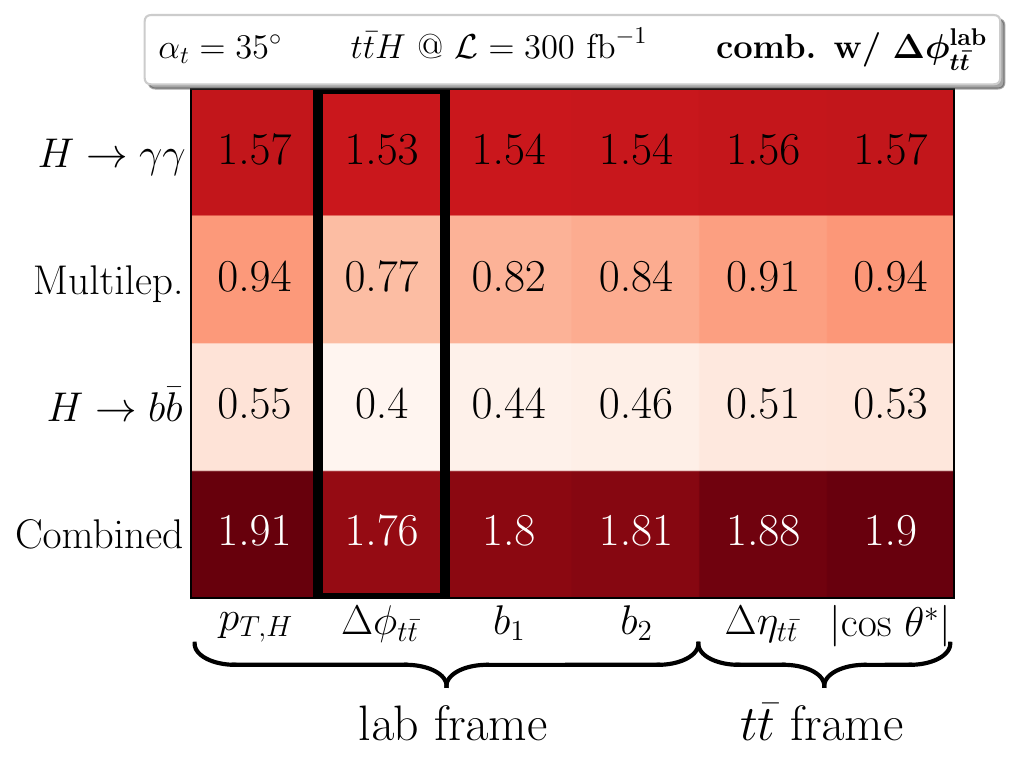}
\caption{Significances evaluated from the (left) $p_{\mathrm{T},H}$ and (right) $\Delta\phi_{t\bar{t}}$ distributions in the lab frame for a BSM signal corresponding to $g_t=1,~\at=35^{\circ}$. The emphasised column shows the significance obtained from the one-dimensional distribution while the others show the significance obtained when combining with a second observable. All observables are considered in their optimised binning. The significances are shown per experimental channel and after their statistical combinations.}
\label{fig:significances_optimized}
\end{figure}

For the remaining five observables, i.e. $\Delta\phi_{t\bar{t}}$, $b_1$, and $b_2$ in the lab frame, as well as $\Delta\eta_{t\bar{t}}$ and $|\cos\theta^*|$ in the $t\bar{t}$ frame, we define six bins as well such that the less populated regions in the tails of the distributions get wider bins\footnote{This adjustment of the bins is what we refer to as optimised binning without strictly employing an optimization algorithm.}: 

\begin{itemize}
    \item $\Delta\phi_{t\bar t}^{\mathrm{lab}}$: [0, $\pi/4$, $\pi/2$, $2\pi/3$, $5\pi/6$, $11\pi/12$, $\pi$]~(rad.),
    \item $b_1^{\mathrm{lab}}$: [-1, -0.95, -0.8, -0.2, 0.3, 0.8, 1.0],
    \item $b_2^{\mathrm{lab}}$: [-1, -0.6, -0.4, -0.2, 0., 0.3, 1.0],
    \item $\Delta\eta_{t\bar t}^{t\bar t}$: [0, 0.5, 1, 1.5, 2, 3, 5],
    \item $|\cos\theta^{*}|$: [0, 0.2, 0.4, 0.55, 0.7, 0.85, 1].
\end{itemize}
The normalized distributions of these observables for $\at = 0^\circ$ and $35^\circ$ are shown in~\cref{fig:smearings_1,fig:smearings_2}. The distributions are shown for each observable at the parton level and for the various experimental channels including the smearing factors discussed in~\cref{subsec:mostsensitive_detector}. The significances obtained for each preselected observable in a two-dimensional combination with $\pTx{H}$ using the optimized binning are shown in~\cref{fig:significances_optimized}. The same table considering combinations with $\Delta\phi_{t\bar{t}}$ is also displayed. 

We observe that the optimised binning using just 6 bins almost reproduces the significance based on the default binning after bin merging in the case of the 1D distributions of either $\pTx{H}$ or $\Delta\phi_{t\bar t}$. 
For the case of the 2D distributions, the optimised binning yields a very similar performance as the default binning, with an improvement typically at the few-percent level. 
The similarity of the significances obtained in the default and the optimised binning also indicates a robustness of the method of 2D distributions against the choice of the binning as long as one makes sure that each bin is populated by at least a few events.
Therefore we consider the optimised binning as validated and will from now on refer to significances computed from these exclusively.

%%% figure %%%
%%% figure %%%

%%% figure %%%
%%% figure %%%

%%% figure %%%
%%% figure %%%

%%%%%%%%%%%%%%%%%%%%%%%%%%%%%%%%%%
\clearpage
\subsection{Comparison to a multivariate analysis}
\label{subsec:BDT}

In this section, we compare the significance obtained from the two-dimensional distributions reported in the previous section to the one obtained from the output of a multivariate analysis algorithm trained on the full set of observables\footnote{We checked that providing additional low-level input variables leads to a negligible performance improvement (below the permille level).} introduced in~\cref{sec:obs}. We assume that such a multivariate analysis is able to make better use of the provided kinematic information of the observables and therefore outperforms the practical usage of only two observables. In the following, we use this comparison to a multivariate analysis to estimate the loss of sensitivity due to the usage of only two physically-motivated observables.

We implement a Boosted Decision Tree (BDT) based on the XGBoost algorithm~\cite{Chen:2016:XST:2939672.2939785}. The goal of the BDT is to distinguish a signal characterized by $\at \ne 0$ from the $\at=0$ hypothesis. Separate trainings are performed for the three $t\bar{t}H$ channels. The scaling factors and smearings described in~\cref{tab:yields} are applied to $t\bar{t}H$ events generated at the parton level to emulate detector effects and obtain realistic yields in each channel. The SM $t\bar{t}H$ sample is used as `background' and the $g_t=1,~\at = 90^\circ$ sample is used as `signal' in the training as this choice leads to the best performance in the full parameter space~\cite{Metodiev:2017vrx}.\footnote{All input variables are \cp-even such that the BDT is not sensitive to the interference term.} To leverage the statistics of the entire dataset, we employ a $k$-fold strategy~\cite{raschka:2020model} with $k= 5$. 80\% of the dataset is dedicated to training, while the remaining 20\% serves as testing subset. This operation is repeated five times to test the full dataset. The performance of the BDT is evaluated by computing the significance from the BDT output score based on~\cref{eq:significance_i2,eq:significance_uncertainty}.

\begin{figure}[!htpb]
    \centering
    \includegraphics[width=.98\textwidth]{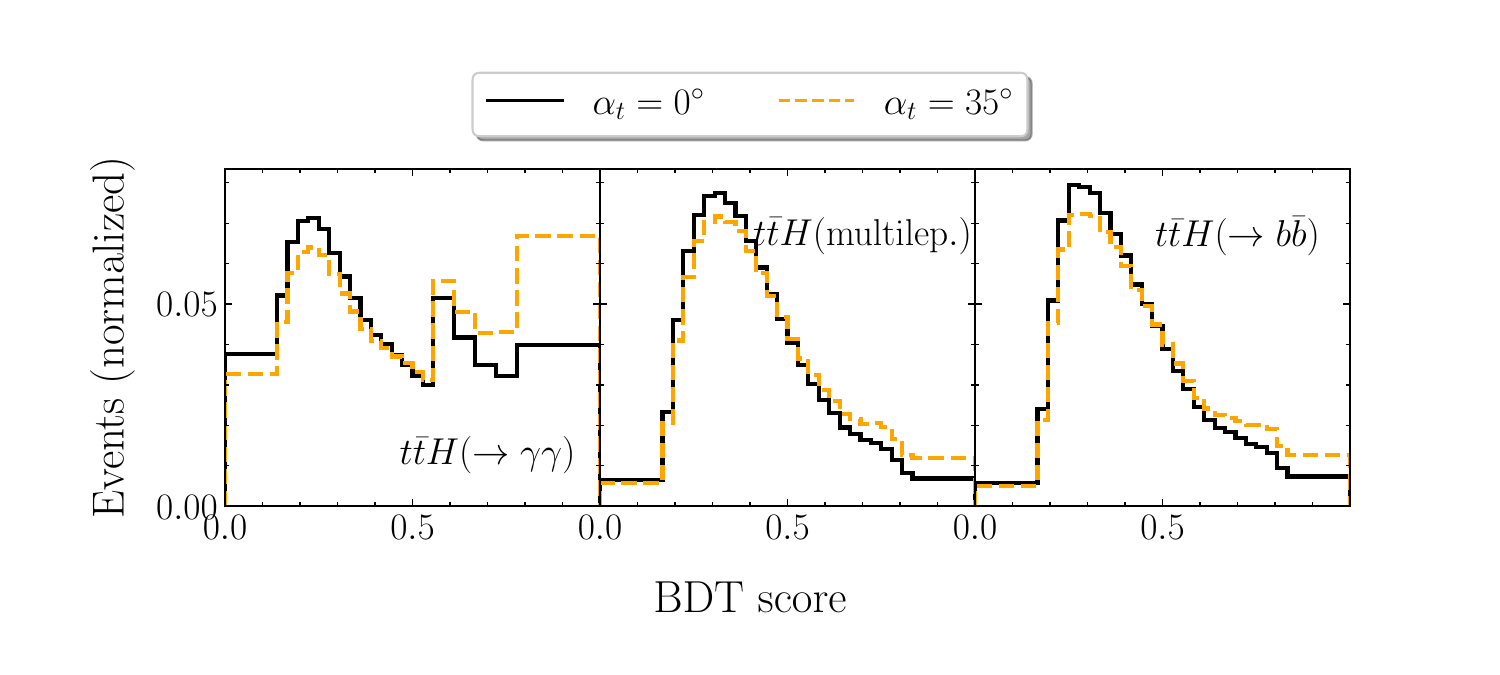}

    \caption{Distributions of the output BDT score for the (left) $t\bar{t}H(\to\gamma\gamma)$, (center) $t\bar{t}H$(multilep.), and (right) $t\bar{t}H(\to bb)$ channels, evaluated for the (black) SM and (orange) $g_t=1,~\at = 35^\circ$ sample.}
    \label{fig:bdt_scores}
\end{figure}

The output BDT score distributions for the SM scenario and the $g_t=1,~\at = 35^\circ$ signal sample in each $t\bar{t}H$ channel are shown in~\cref{fig:bdt_scores}. The bin merging method described in~\cref{subsec:sig_computation} is applied to guarantee at least two SM $t\bar{t}H$ events per bin. The significance values computed from these distributions are summarized and compared to the best significance values obtained from two-dimensional distributions in~\cref{tab:sigBDT}. As expected, the significances obtained from the BDT are higher, but only to a limited extent ($\sim 10\%$ for $H\to\gamma\gamma$, $\sim 17\%$ for $H\to \text{multilep.}$ and $\sim 25\%$ for $H\to b\bar b$), with a combined significance reaching $2.21$ compared to $1.91$ for the two-dimensional method. The improvement in sensitivity of the BDT is larger for the $t\bar{t}H(\mathrm{multilep.})$ and $t\bar{t}H(\to b\bar{b})$ channels, as these channels select more \ttH~events  (see~\cref{tab:yields}) and therefore allow better use of kinematic shape information. We conclude that the approach taken so far consisting of using two-dimensional distributions to probe \cp violation in the top-Yukawa coupling is sufficient to reach a close-to-optimal sensitivity with 300 fb$^{-1}$ of data, especially since the combined sensitivity is driven by the statistically limited $H\to\gamma\gamma$ channel.

%%% figure %%%

\begin{table}[!htpb]
    \centering
    \small
    \begin{tabular}{|c|c|c|}
    \hline
    Channel & Significance (BDT) & Significance (2D) \\
    \hline
    $t\bar{t}H(\to\gamma\gamma)$ & 1.75 & 1.57  \\
    $t\bar{t}H$(multilep.) & 1.17 & 0.94 \\
    $t\bar{t}H(\to bb)$ & 0.69 & 0.55 \\
    \hline
    Combined & 2.21 & 1.91 \\
    \hline
    \end{tabular}
    \caption{Significance obtained from the output BDT score distributions and best significance obtained from two-dimensional distributions in~\cref{subsec:mostsensitive_binning}.}
    \label{tab:sigBDT}
\end{table}

%%%%%%%%%%%%%%%%%%%%%%%%%%%%%%%%%%
\clearpage
\subsection{Expected exclusion limits}
\label{subsec:sensitivity}

\begin{figure}[!htbp]
    \centering
    \includegraphics[width=.54\textwidth]{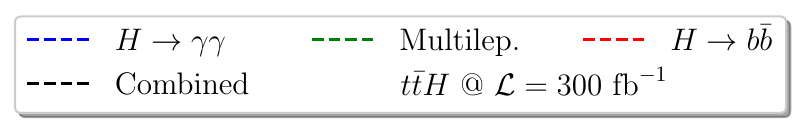} \\
    \includegraphics[trim=1.cm 0.25cm 1.cm 0.25cm, width=.32\textwidth]{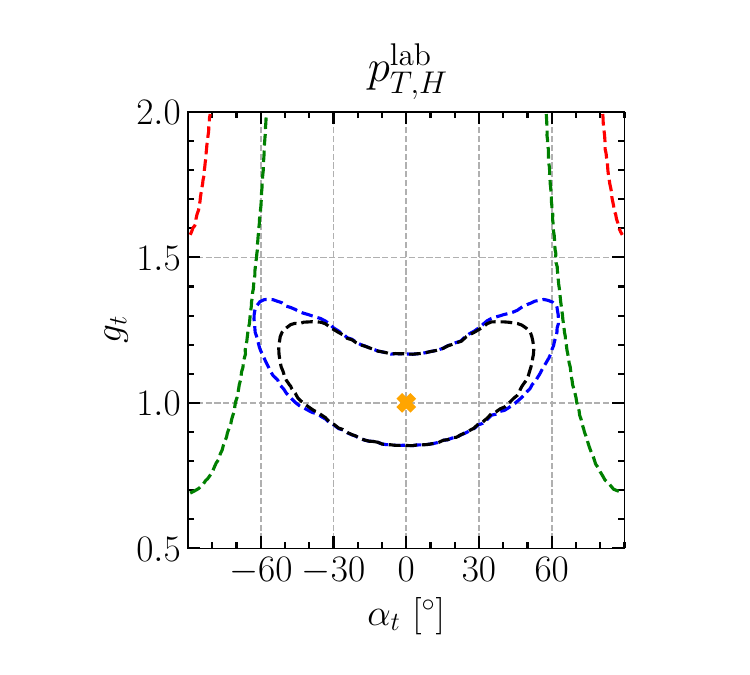}
    \includegraphics[trim=1.cm 0.25cm 1.cm 0.25cm, width=.32\textwidth]{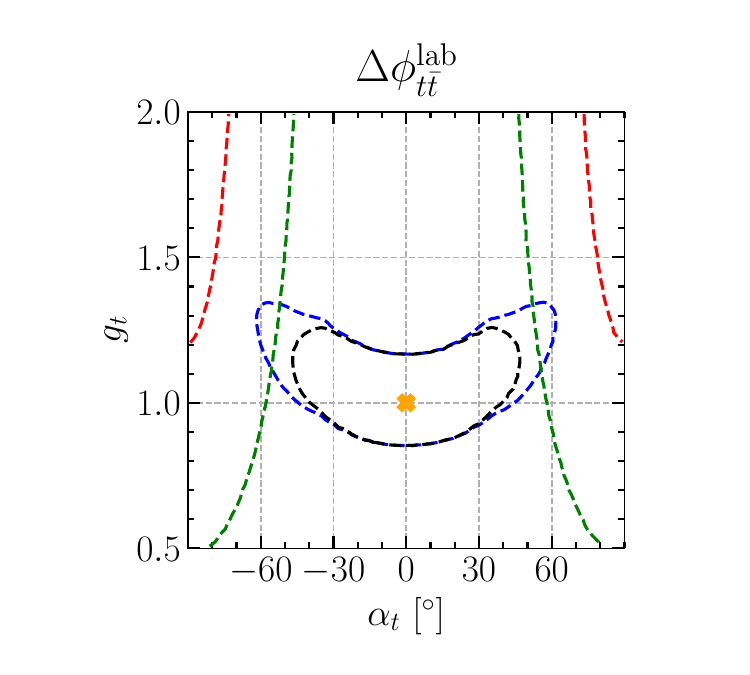} \\
    \includegraphics[trim=1.cm 0.25cm 1.cm 0.25cm, width=.32\textwidth]{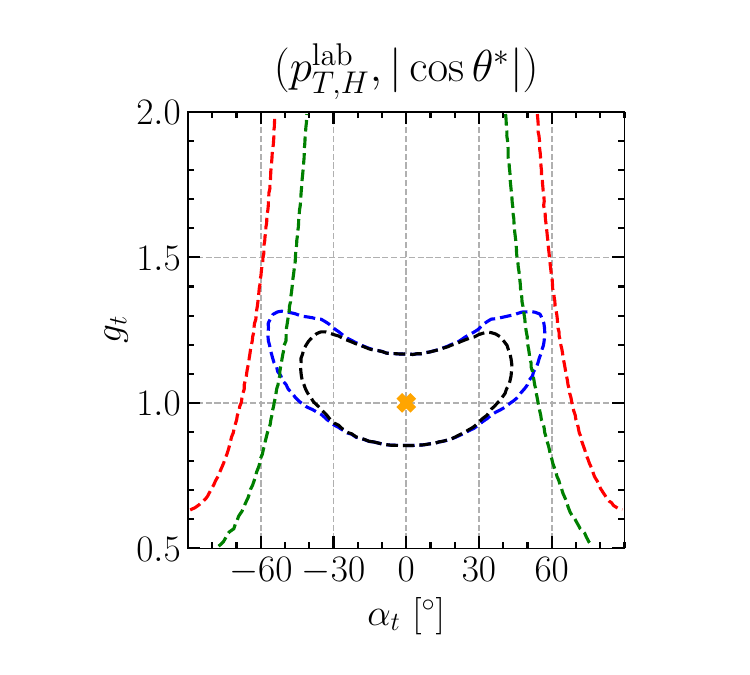}
    \includegraphics[trim=1.cm 0.25cm 1.cm 0.25cm, width=.32\textwidth]{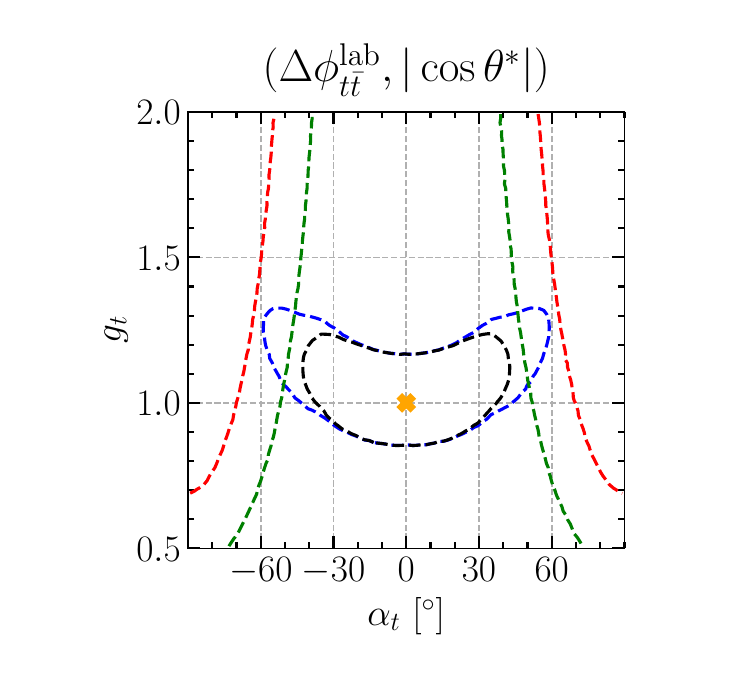}
    \includegraphics[trim=0.9cm 0.25cm 0.9cm 0.25cm, width=.32\textwidth]{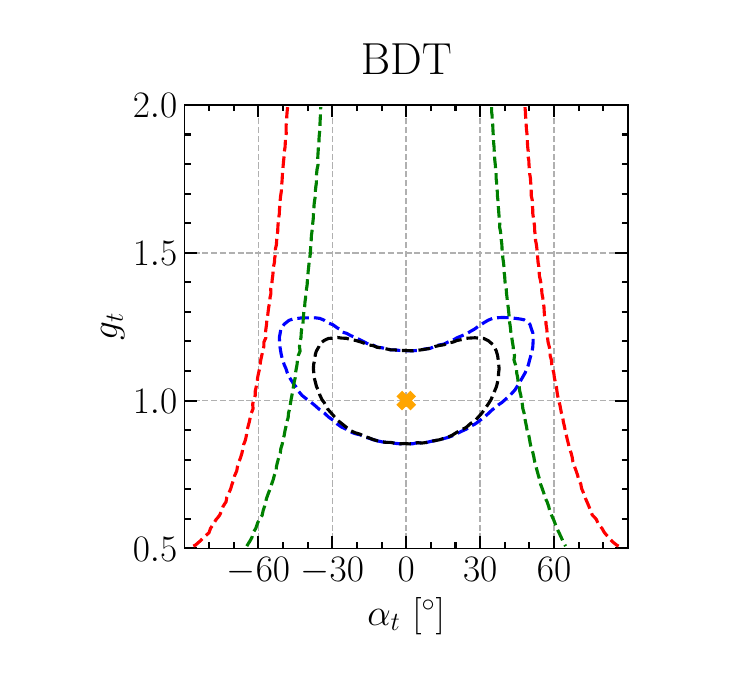}
    \caption{Expected limits at the 95\% confidence level in the $(\at,\gt)$ plane for ${\mathcal{L} = 300\invfb}$ when considering (upper left) the current $\pTx{H}$ STXS setup and (upper right) the $\Delta\phi_{t\bar t}^\text{lab}$ observable alone. On the second row, the results from the two-dimensional observables $(\pTx{H},|\cos\theta^*|)$ and $(\Delta\phi_{t\bar t}^\text{lab},|\cos\theta^*|)$ are shown. The expected limits obtained from the BDT output score described in~\cref{subsec:BDT} are also shown for comparison.}
    \label{fig:limits}
\end{figure}

Only results for $\at = 35^\circ$ and $\gt = 1$ have been presented so far. In this section, we show and discuss the expected exclusion limits at the 95\% confidence-level (using the Gaussian limit, i.e. $S=1.96$) in the full $(\at,\gt)$ plane for an integrated luminosity of $300\invfb$. The significance at any given point of the parameter space is computed based on the distribution of the observables under consideration for both the SM and BSM cases, where the BSM $t\bar{t}H$ yields are obtained from~\cref{eq:ttH_yield}. The resulting two-dimensional contours are shown for each $t\bar{t}H$ channel as well as their statistical combination in~\cref{fig:limits}. The contours for the $t\bar{t}H$(multilep.) and $t\bar{t}H(\to bb)$ channels are not closed at high values of $\gt$, which is due to the fact that any discrimination power arising from differences in the total rate between SM and BSM is neglected in these two channels (see~\cref{subsec:sig_computation}). 

In the current STXS $\pTx{H}$ setup (i.e., $\pTx{H}$ as the single discriminating variable with 6 bins), the discriminating power between SM and BSM hypotheses in the $t\bar{t}H(\to\gamma\gamma)$ channel originates mainly from differences in the total event rate (see~\cref{subsec:default_binning}). This channel almost entirely drives the sensitivity and allows to constrain $|\at| \lesssim 43^\circ$ for $\gt=1$, while the combination of the three channels improves the constraint marginally to $|\at| \lesssim 41^\circ$ for $\gt=1$. 
In this scenario, the $t\bar{t}H(\to bb)$ channel contribution to the combined sensitivity in the entire parameter space is rather small. The $t\bar{t}H$(multilep.) channel contributes to the combined sensitivity more strongly for $\gt>1$, where higher $t\bar{t}H$ yields are expected such that a higher discrimination power from the $\pTx{H}$ kinematic shape is obtained. In this region, an improvement up to $36$\% is observed in the combined limit on $|\at|$ due to $t\bar{t}H$(multilep.), obtained for $\gt=1.27$. 

If the current STXS $\pTx{H}$ setup is extended by splitting further $t\bar{t}H$ events according to their $|\cos\tstar|$ value, the combined limit at $\gt=1$ is improved by 12\% to reach $|\at| \lesssim 36^\circ$. The best improvement of the combined limit on $|\at|$ over the $\ttH(\to\gamma\gamma)$ standalone limit is $40$\% and is reached at $\gt=1.24$. The improvement originates mainly from the improved sensitivity of the $t\bar{t}H$(multilep.) channel, although the sensitivity of the $t\bar{t}H(\to bb)$ channel is also significantly improved and contributes to the combination. In this scenario, a higher sensitivity from the $t\bar{t}H$(multilep.) is expected with respect to the $t\bar{t}H(\to\gamma\gamma)$ channel for $1.10<\gt<1.31$. We obtain comparable results when combining $p_T^H$ with one of the other promising candidates identified in the previous sections, i.e. $\Delta\phi_{t\bar{t}}$, $b_1$ or $b_2$ in the lab frame.

\cref{fig:limits} also shows the exclusion contours for $\Delta\phi_{t\bar{t}}$ instead of $\pTx{H}$ as single variable, as well as for the two-dimensional binning of $\Delta\phi_{t\bar{t}}$ and $|\cos\tstar|$, which was among the best combinations not involving $\pTx{H}$ identified in~\cref{subsec:default_binning}. As expected from~\cref{fig:significances_optimized}, the $\Delta\phi_{t\bar{t}}$ observable alone performs better than $\pTx{H}$ alone mainly due to its higher level of kinematic shape discrimination, which leads to an increased sensitivity in the $t\bar{t}H$(multilep.) and $t\bar{t}H(\to bb)$ channels. Regarding the two-dimensional observable combination $(\Delta\phi_{t\bar{t}},|\cos\tstar|$), the picture is qualitatively very similar to the ($\pTx{H}$,$|\cos\tstar|$) case and compatible limits are found after statistical combinations of the three channels ($|\at| \lesssim 36^\circ$ for $\gt=1$). 

Finally we show in~\cref{fig:limits} the exclusion contours resulting from the significance computed from the output BDT score described in~\cref{subsec:BDT}. Since the BDT inputs include the full set of observables shown in~\cref{tab:obs}, we expect it to outperform our two-dimensional approach. Still, only a slight improvement over the limits obtained from simple two-dimensional variable combinations is observed overall. In particular, the expected limit of $|\at| \lesssim 33^\circ$ for $\gt = 1$ from the BDT is close to the corresponding result obtained from the ($\pTx{H}$,$|\cos\tstar|$) distribution. 

To summarize, we found comparable performance between the best $\pTx{H}$-based and $\Delta\phi_{t\bar{t}}$-based two-dimensional variable combinations and the output from a multivariate analysis based on the full set of observables shown in~\cref{tab:obs}.

%%% figure %%%
%%% figure %%%
 
\clearpage
%%%%%%%%%%%%%%%%%%%%%%%%%%%%%%%%%%
%%%%%%%%%%%%%%%%%%%%%%%%%%%%%%%%%%

\section{Extending the STXS framework}
\label{sec:mostsensitive_STXS}

The additional data from the ongoing Run 3 of the LHC is expected to increase the size of the proton-proton dataset of the ATLAS and CMS collaborations from $140\invfb$ to about $300\invfb$. The increased statistics will open the possibility to extract more precise information from the $t\bar{t}H$ kinematic shapes. As of now, the sensitivity to the \cp nature of the top-Yukawa coupling is largely driven by the $t\bar{t}H(\to\gamma\gamma)$ channel. However, based on our results, we expect a significant increase in sensitivity of the $t\bar{t}H$(multilep.) and $t\bar{t}H(\to bb)$ channels by the end of Run 3. If confirmed, the statistical combination of the various $t\bar{t}H$ channels will become key to constrain further the \cp\ nature of the top-Yukawa coupling over the next few years. We propose in~\cref{subsec:bkg} a qualitative discussion regarding the validity of our background assumptions used to evaluate the significance for the best candidate observables. This allows us to reduce further our list of candidates and to formulate our final proposal for a \cp-sensitive STXS extension in~\cref{subsec:stxs_proposal}.

%%%%%%%%%%%%%%%%%%%%%%%%%%%%%%%%%%

\subsection{Further background considerations}
\label{subsec:bkg}

The significance computation used to evaluate the sensitivity of each observable combination relies on some assumptions which were discussed in~\cref{subsec:sig_computation}. In particular, a constant signal ($t\bar{t}H$) over background (non-$t\bar{t}H$) ratio is assumed throughout all the bins considered in our study and $S/B = 1$, $0.4$, and $0.1$ are assumed in the $t\bar{t}H(\to\gamma\gamma)$, $t\bar{t}H$(multilep.), and $t\bar{t}H(\to bb)$ channels, respectively. These values are based on typical values observed in the corresponding ATLAS and CMS analyses. Qualitatively, this assumption should be conservative under the condition that the background shape is not peaking around the highest significance bins, which we propose to verify here.

%%%%%%

\subsubsection*{Higgs background processes}

We start with a short discussion of Higgs background processes, i.e.~mainly \tH and \tWH events in which a second top quark could be wrongly reconstructed. In the SM, the cross-sections of the \tH and \tWH processes are comparably low with respect to \ttH , with $\sigma_{\SM}^{\tH} / \sigma_{\SM}^{\ttH} \approx 0.15$ and $\sigma_{\SM}^{\tWH} / \sigma_{\SM}^{\ttH} \approx 0.03$. Both Higgs background processes become more important for higher fractions of \cp-odd top-Yukawa coupling, however, at $(\gt, \at) = (1, 35^\circ)$, we still find $\sigma^{\tH} / \sigma^{\ttH} \approx 0.24$ and $\sigma^{\tWH} / \sigma^{\ttH} \approx 0.05$. Given the low cross-sections in the parameter space of interest and the additional yield reduction expected from the analysis selection, we conclude that the $\tH$ and $\tWH$ processes can only play a minor role in the measurement of the \ttH kinematic distributions and therefore satisfy the assumptions made in~\cref{subsec:sig_computation}. 

%%%%%%

\subsubsection*{Non-Higgs background processes}

\begin{figure}[!htbp]
    \centering
    \includegraphics[width=.66\textwidth]{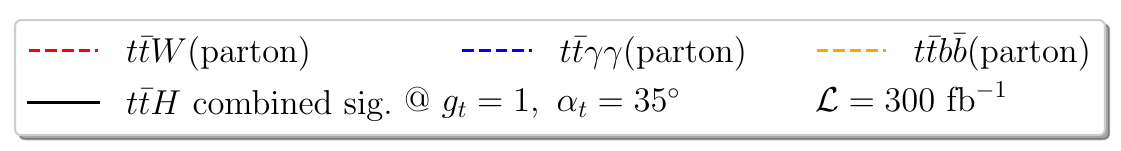}
    \includegraphics[width=.4\textwidth]{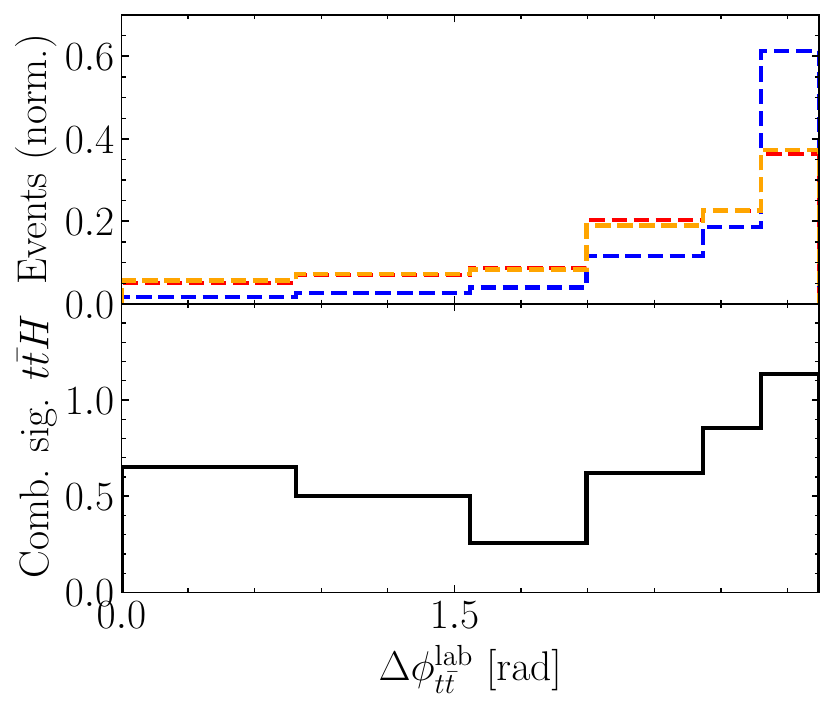}
    \includegraphics[width=.4\textwidth]{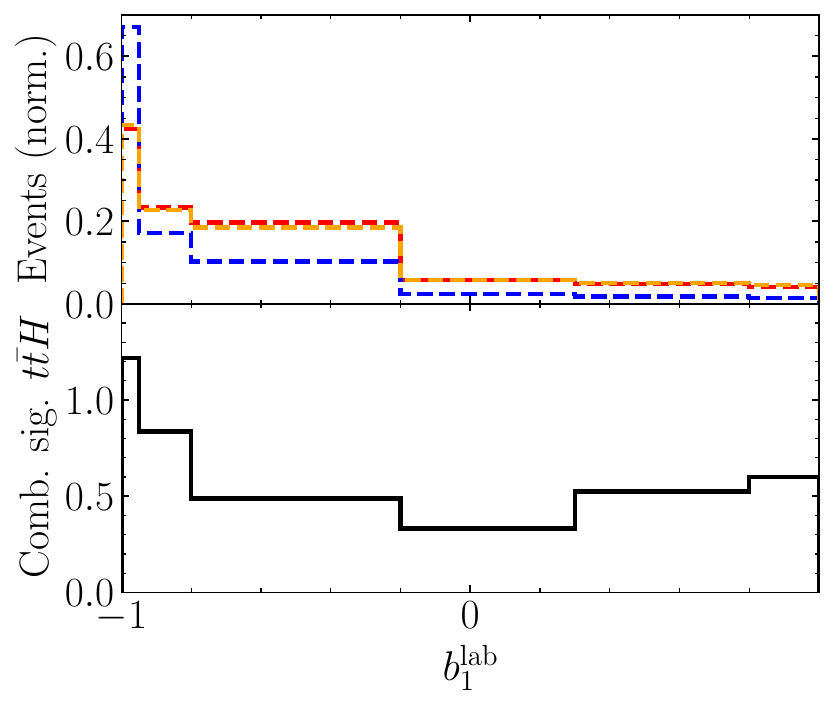}
    \includegraphics[width=.4\textwidth]{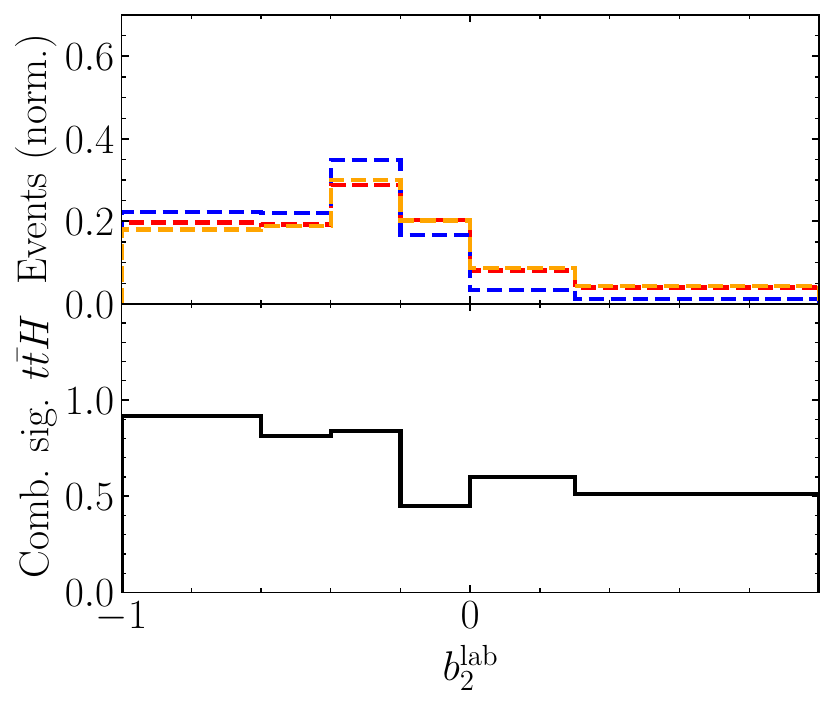}
    \includegraphics[width=.4\textwidth]{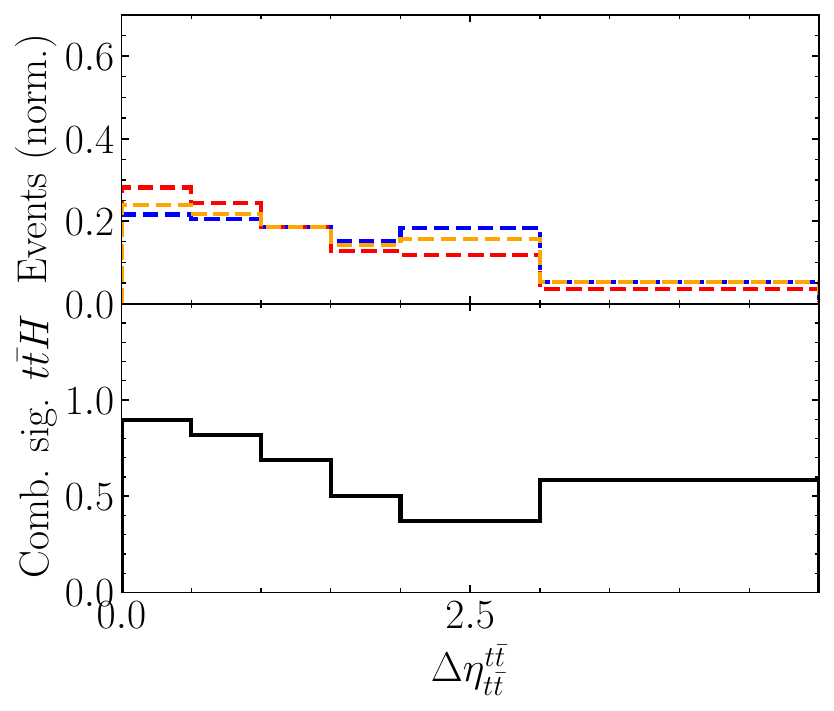}
    \includegraphics[width=.4\textwidth]{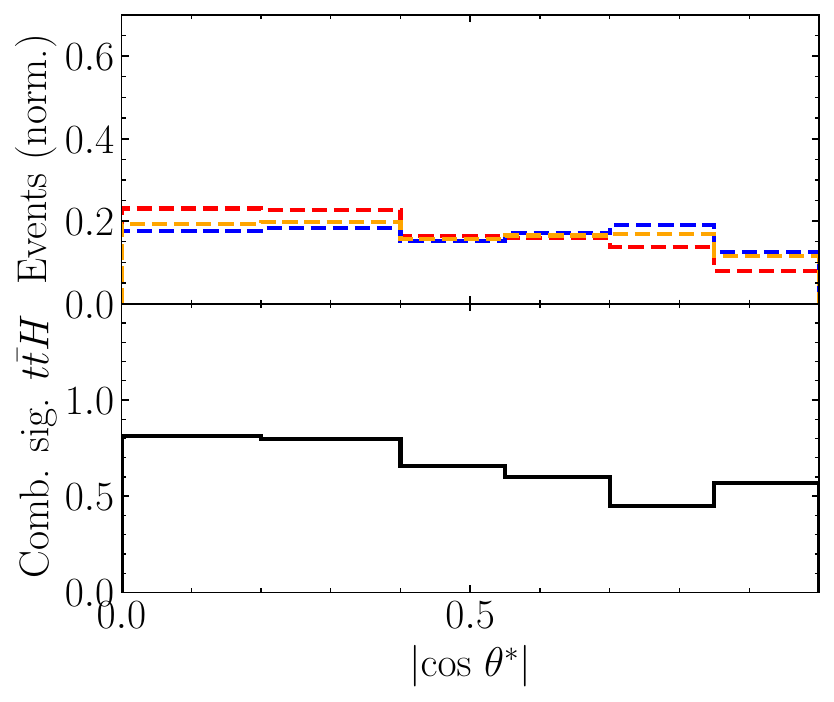}
    \caption{Distributions of the (top, left) azimuthal angle difference between the $t\bar{t}$ pair, (top, right) $b_1$ observable, and (center, left) $b_2$ observable in the lab frame, and distribution of the (center, right) pseudorapidity difference between the $t\bar{t}$ pair and (bottom) $|\cos\tstar|$ observable in the $t\bar{t}$ frame. The distributions for SM $t\bar{t}\gamma\gamma$, $t\bar tW$ and $t\bar{t} b\bar{b}$ events are shown in each upper panel plot. The observables are reconstructed by setting $p_H = p_{\gamma\gamma}$ for $t\bar{t}\gamma\gamma$, $p_H = p_W$ for $t\bar t W$, and $p_H = p_{b\bar b}$ for $t\bar t b \bar b$. The events are generated at the parton-level following the setup described in~\cref{subsec:event_generation}. All distributions are normalised to unity. We also show in each case the per-bin combined significance obtained for the \ttH sample at $g_t=1$, $\at = 35^\circ$ in a lower panel plot.}
    \label{fig:backgrounds}
\end{figure}

We discuss here the most relevant non-Higgs background process in each channel: $t\bar{t}\gamma\gamma$ for the $H\to\gamma\gamma$ channel, $t\bar tW$ for the multi-lepton channel, and $t\bar{t} b\bar{b}$ for the $H\to b\bar b$ channel, for which samples were generated at the parton-level based on the setup described in~\cref{subsec:event_generation}. For each background process, we assume that the non-top-quark particles are misidentified as a Higgs boson to compute the observables introduced in~\cref{sec:obs}. In practice, we implement this by setting $p_H = p_{\gamma\gamma}$ for $t\bar{t}\gamma\gamma$, $p_H = p_W$ for $t\bar t W$, and $p_H = p_{b\bar b}$ for $t\bar t b \bar b$. Following these assignments, we show in~\cref{fig:backgrounds} the $t\bar{t}\gamma\gamma$, $t\bar tW$ and $t\bar{t} b\bar{b}$ distributions for the various observables selected in~\cref{sec:results}. We also show in the same Figure the combined significance obtained in each bin for the $g_t=1$, $\at = 35^\circ$ sample. This gives us a qualitative estimate of whether our $S/B$ assumptions about the background uncertainties from~\cref{subsec:sig_computation} are conservative or not. If the non-Higgs backgrounds for a specific variable peaks in the bin which mainly drives the sensitivity, we consider this variable's sensitivity to be possibly inflated in~\cref{sec:results} and discard it in the final proposal. The parton-level non-Higgs background distributions were cross-checked against distributions obtained from the same samples but after parton shower (using \texttt{Pythia8}~\cite{Sjostrand:2014zea}) and while minor shape differences are present, the conclusions below remain unchanged. 

The $\Delta\phi_{t\bar{t}}$ and $b_1$ distributions in the lab frame show the same behaviour, with a clear peak at one end of the distribution in both the background distributions and the significance. The peak is particularly pronounced for the $t\bar{t}\gamma\gamma$ background and has the same location as the peak in the significance. Consequently, the sensitivity of these observables may be overestimated in our results shown in~\cref{sec:results} and they are excluded from the final shortlist. The distribution of $b_2$ in the lab frame has a peak of the background in the third bin, which is, however, not the bin dominating the total significance. For the $\Delta\eta_{t\bar{t}}$ and $|\cos\theta^*|$ distributions in the $t\bar{t}$ frame, the sensitivity as well as the background distributions are relatively flat across the full range. Therefore we expect our sensitivity estimates to be more robust for these three last observables, which are the final ones selected.

%%%%%%%%%%%%%%%%%%%%%%%%%%%%%%%%%%
\clearpage
\subsection{Proposal for \texorpdfstring{$t\bar t H$}{ttH} STXS extension}
\label{subsec:stxs_proposal}

The findings of our analysis are the following, considering as a benchmark the $g_t=1$, $\at = 35^\circ$ sample with 300 fb$^{-1}$ of data:
\begin{itemize}
    \item The combination of $\pTx{H}$ with $\Delta\phi_{t\bar{t}}^\text{lab}$, $b_1^\text{lab}$, $b_2^\text{lab}$, $\Delta\eta_{t\bar{t}}^{\ttbar}$, or $|\cos\theta^*|$ feature compatible sensitivity within $5$\% of the best combination. We favour combinations with \pTx{H} due to the existing STXS binning.
    \item When using the existing STXS binning in $\pTx{H}$ (6 bins) adding any second observable composed of 6 bins with optimized boundaries, the two-dimensional combinations including $\pTx{H}$ listed above yield compatible sensitivity.
    \item If we compare these results with the sensitivity that we obtain from a multivariate analysis trained on all the observables introduced in~\cref{sec:obs}, a limited gain of $16$\% is found when using the multivariate analysis (after combining all channels).
    \item A qualitative comparison of the main non-Higgs background shapes with respect to the significance in each of the tested bins shows that $\Delta\phi_{t\bar{t}}^\text{lab}$ and $b_1^\text{lab}$ are disfavored. This leads to the exclusion of the respective combinations with $\pTx{H}$.
    \item The final shortlist of candidate observables to be combined with $\pTx{H}$ include $b_2^\text{lab}$, $\Delta\eta_{t\bar{t}}^{\ttbar}$, and $|\cos\theta^*|$. Our analysis provides no support to favour one over another.
\end{itemize}

\begin{figure}
    \centering
    \includegraphics[width=.65\textwidth]{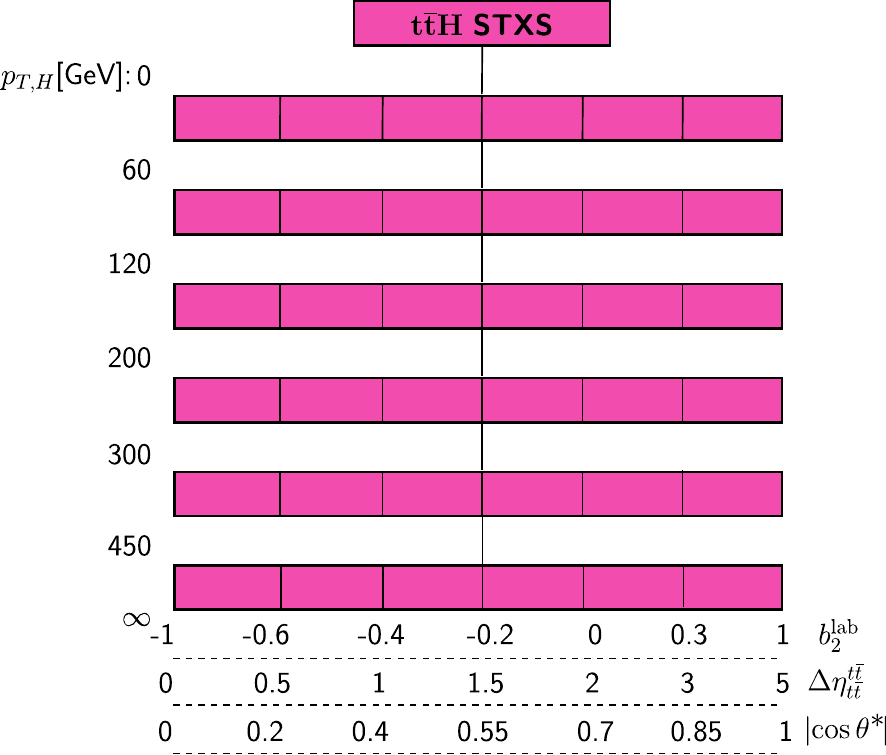}
    \caption{Proposal for an extension of the current STXS binning for the $\ttH$ production mode. Each bin in $\pTx{H}$ is further split in bins of either $b_2^{\mathrm{lab}}$$, \Delta\eta_{t\bar{t}}^{\ttbar}$, or $|\cos\theta^*|$.}
    \label{fig:stxs_proposal}
\end{figure}

Based on these findings, our final proposal is to extend the current STXS binning by a second dimension that could either be $b_2^\mathrm{lab}$,$\Delta\eta_{t\bar{t}}^{\ttbar}$, or $|\cos\theta^*|$, as illustrated in \cref{fig:stxs_proposal}.

%%% figure %%%
%%% figure %%%

\begin{figure}[!htbp]
    \centering
    \includegraphics[width=.92\textwidth]{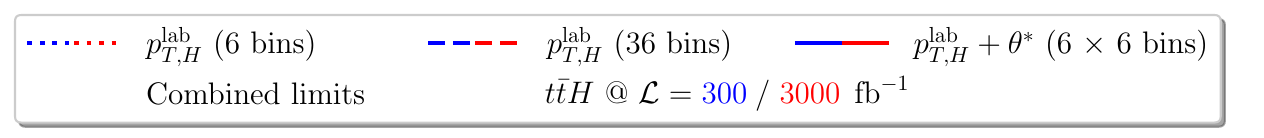}
    \includegraphics[width=.48\textwidth]{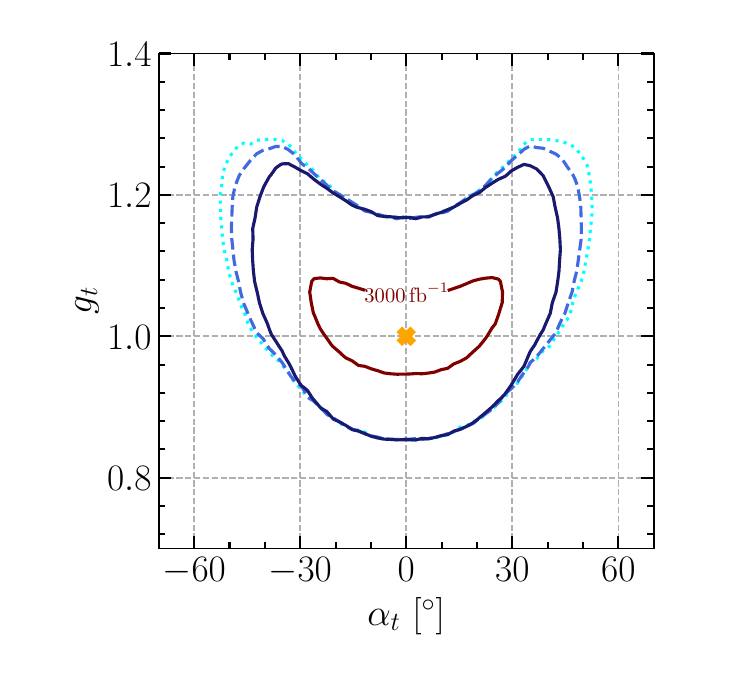}
    \includegraphics[width=.48\textwidth]{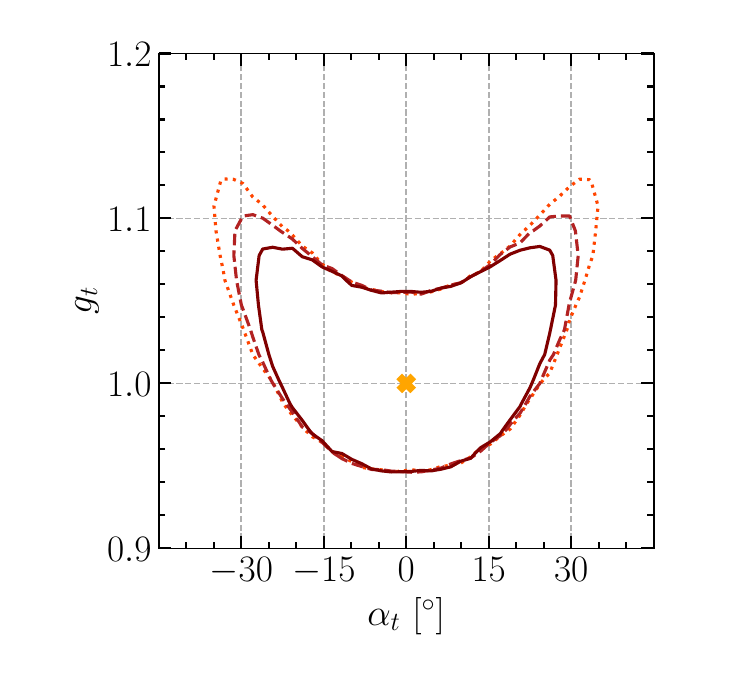}
    \caption{Constraints in the $(\gt,\at)$ plane for (blue) $\mathcal{L} = 300\invfb$ and (red) ${\mathcal{L} = 3000\invfb}$  at the $95~\%~\mathrm{CL}$ using the one-dimensional $\pTx{H}$ distribution --- evaluated using 6 (dotted line) and 36 (dashed line) bins --- as well as the two-dimensional $(\pTx{H},|\cos\theta^*|)$ distributions (solid line, $6\times 6$ bins). The contour for $\mathcal{L} = 3000\invfb$ is shown alone on the right-hand side and together with the $\mathcal{L} = 300\invfb$ contour on the left-hand side for better comparison.
    }
    \label{fig:limits_HL}
\end{figure}

To further support our proposal, we show in \cref{fig:limits_HL} a comparison of the expected two-dimensional limits in the ($g_t$, $\at$) parameter space that we obtain from the current STXS binning for \ttH~production, a finer one-dimensional binning including 36 bins, and our proposal taking the example of $|\cos\theta^*|$ as the second observable. This comparison is shown for $300\invfb$ and $3000\invfb$ (without any other modification in the analysis). Although the $3000\invfb$ scenario may be conservative as the moderate increase of center-of-mass energy from $\sqrt{s}=13\,$TeV to $14\,$TeV, improved acceptance and potential systematic uncertainty reduction are not taken into account, it still shows that our proposal allows to improve the sensitivity of the STXS measurement in the longer term. The sensitivity improvement from the introduction of a second observable due to the improved kinematic shape discrimination with higher number of \ttH\ events goes up to +$33$\% (+$28$\%), reached at $g_t=1.24~(1.08)$ for $300\invfb$ ($3000\invfb$) of data. On the other hand, the introduction of more $\pTx{H}$ bins lead to lower sensitivity improvements, going up to +$24$\% (+$15$\%), reached at $g_t=1.26~(1.10)$ for $300\invfb$ ($3000\invfb$) of data.\footnote{The sensitivity improvements due to the introduction of more bins are slightly reduced at $3000\invfb$ with respect to $300\invfb$. This effect originates from the fact that the discrimination power from the differences in the total rate between SM and BSM is enhanced due to the larger number of $t\bar{t}H(\to\gamma\gamma)$ events, and therefore it reduces the sensitivity boost due to additional kinematic shape information.}  

%%% figure %%%
%%% figure %%%

Our proposal for the extension of the STXS framework assumes a common definition of the top quark. This definition should guarantee the possibility to calculate higher-order loop corrections in a well-defined way. A pseudo-top-quark definition has already been worked out in the context of the LHC Top Working Group~\cite{ParticleLevelTopDefinitions}. We leave a study of the applicability of this definition to the $\ttH$ process for future work.

%%%%%%%%%%%%%%%%%%%%%%%%%%%%%%%%%%
%%%%%%%%%%%%%%%%%%%%%%%%%%%%%%%%%%

\section{Conclusion}
\label{sec:conclusion}

\cp violation in the Higgs sector is a possibility with far-reaching phenomenological and cosmological consequences such as for electroweak baryogenesis, which should be thoroughly explored. In this work, we focused on constraining the \cp structure of the top-Yukawa coupling via top-associated Higgs production. While multivariate analysis techniques allow to best exploit the experimental data, measurements of one- or two-dimensional contributions allow for a clearer reinterpretation of the data and an easier combination of different Higgs decay channels and across experiments. With this motivation, we have studied in this paper which combinations of observables are most sensitive to the \cp nature of the top-Yukawa coupling.

Our study considers eleven observables (in various reference frames) already identified as being \cp-sensitive in the literature. For the comparison of their performance, we investigated the three most important experimental channels measured at the LHC: the channel when the Higgs boson decays into two photons, the channel where the Higgs boson decays into a bottom quark and antiquark, and the \ttH events leading to multi-lepton final states. For each of these channels, we included detection effects and selection efficiencies to mimic the data efficiency and resolution from the experimental analysis results. Based on the resulting distributions, we evaluated the sensitivity of each (combination of) observable(s) to distinguish between a mixed \cp state of the top-Yukawa coupling and the SM case via the computation of significances based on a benchmark signal corresponding to $g_t=1$ and $\alpha_t=35^{\circ}$ and $300\invfb$ of data. The best sensitivity is found for combinations of observables with the azimuthal angle difference between the top quarks, $\Delta\phi_{t\bar{t}}$, while combinations with the Higgs transverse momentum, $\pTx{H}$, show comparable performance. 

With an extension of the current STXS v1.2 binning in $\pTx{H}$ in mind, we selected five observables which have the highest sensitivity in combination with $\pTx{H}$: $\Delta\phi_{t\bar t}$, $b_1$ and $b_2$ in the laboratory frame, and $\Delta\eta_{t\bar t}$ and $|\cos\theta^*|$ in the $t\bar{t}$ frame (see~\cref{sec:obs} for exact definitions). For these observables, we optimized the binning for the final sensitivity comparison with six bins per observable to match the number of bins used in the current STXS $\pTx{H}$ binning. We found that all five observables show very similar performance when used in combination with $\pTx{H}$. As a comparison, we also trained a boosted decision tree with all variables considered in our study.
Overall, we found that for $300\invfb$ of data, the exclusion limits are only slightly improved with the multivariate approach over our best two-dimensional combinations of observables. 

For the final selection of an observable combination, we also qualitatively investigated the behaviour of the various non-Higgs backgrounds. For $\Delta\phi_{t\bar t}$ and $b_1$, we found enhanced backgrounds in the bins which are most sensitive to the \cp nature of the top-Yukawa interactions, which make these observables less favorable. 
This is not the case for $\Delta\eta_{t\bar t}$, $|\cos\theta^*|$, and $b_2$. Their two-dimensional combinations with $\pTx{H}$ constitute our final proposal for a \cp-sensitive STXS extension. Their sensitivities are very similar and the best choice likely depends on more specific analysis details not considered in this work. Finally, we compared the expected limits in the $(g_t,\alpha_t)$ plane obtained with the current STXS binning, an extended one-dimensional $\pTx{H}$ binning, and our proposal. Both with $300\invfb$ and $3000\invfb$ of data, our proposal of two-dimensional binning exceeds the sensitivity of the other alternative STXS binning scenarios.

It is important to note that all the proposed STXS extensions in this work require the reconstruction of both top quarks per event. Therefore, a thorough discussion within the community is needed to agree beforehand on a common top-quark definition.

%%%%%%%%%%%%%%%%%%%%%%%%%%%%%%%%%%
%%%%%%%%%%%%%%%%%%%%%%%%%%%%%%%%%%

\section*{Acknowledgements}

\sloppy{
We thank Sarah Heim and Frank Tackmann for helpful discussions. This work was supported by ANR PIA funding: ANR-20-IDEES-0002. H.B.\ acknowledges support from the Alexander von Humboldt foundation. E.F. and M.M. acknowledge funding via Germany’s Excellence Strategy – EXC-2123 QuantumFrontiers – 390837967. E.F. also thanks the CERN TH Department for support during the initial phase of this project. 
}

%%%%%%%%%%%%%%%%%%%%%%%%%%%%%%%%%%
%%%%%%%%%%%%%%%%%%%%%%%%%%%%%%%%%%
%%%%%%%%%%%%%%%%%%%%%%%%%%%%%%%%%%

\vfill
\clearpage
\newpage

\appendix

%%%%%%%%%%%%%%%%%%%%%%%%%%%%%%%%%%
%%%%%%%%%%%%%%%%%%%%%%%%%%%%%%%%%%
\clearpage
\section{Supplementary parton-level distributions for \texorpdfstring{$t\bar t H$}{ttH}}
\label{app:distributions}

%%%%%%%%%%%%%%%%%%%%%%%%%%%%%%%%%%

\subsection{Higgs boson rest frame}

\begin{figure}[!htpb]
    \centering
    \includegraphics[trim=0.cm 0.5cm 0.cm 0.5cm, width=.8\textwidth]{figs/example_distributions/legend.pdf}
    \includegraphics[width=.4\textwidth]{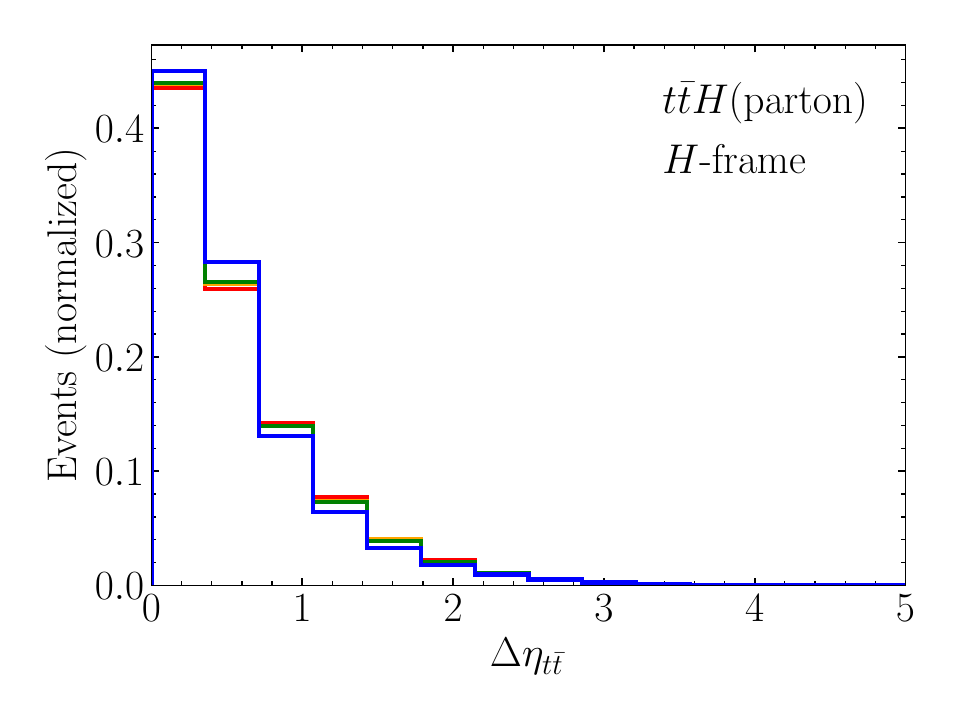}
    \includegraphics[width=.4\textwidth]{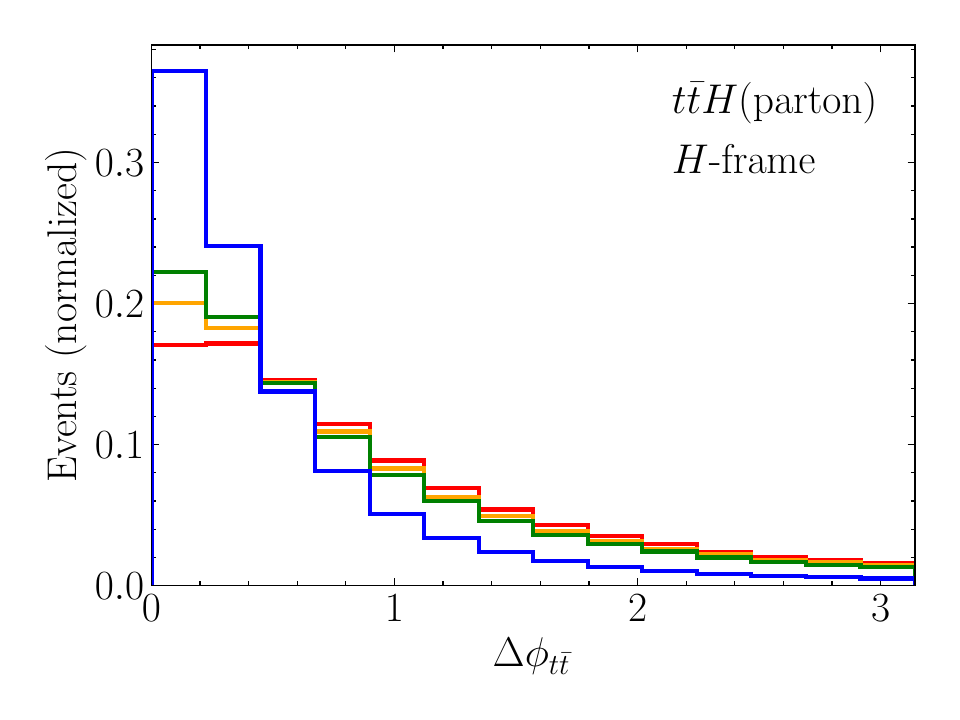}
    \includegraphics[width=.4\textwidth]{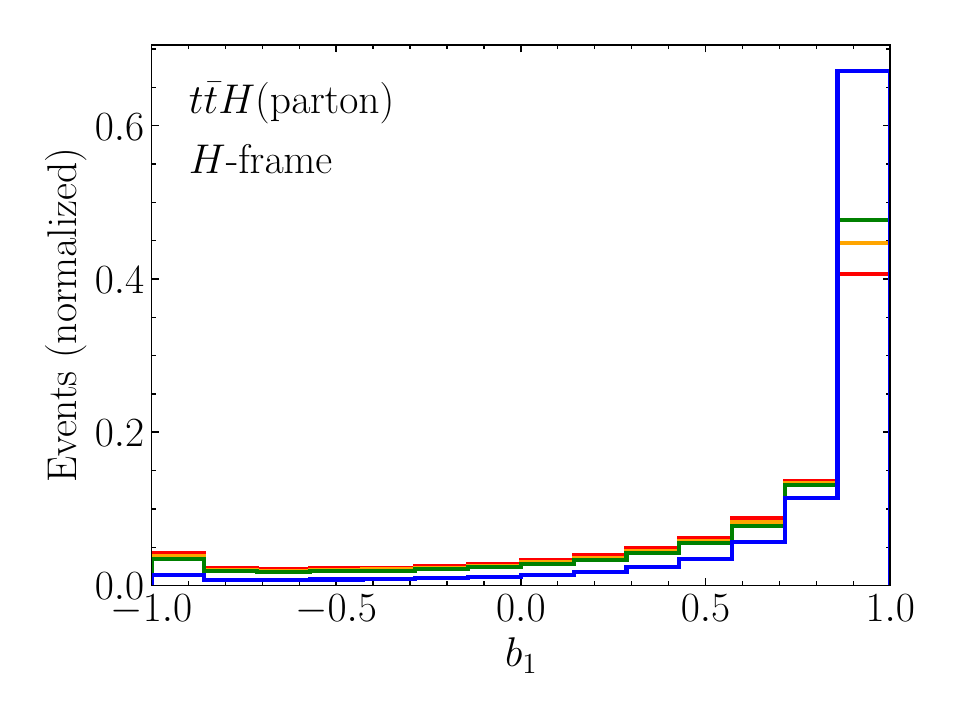}
    \includegraphics[width=.4\textwidth]{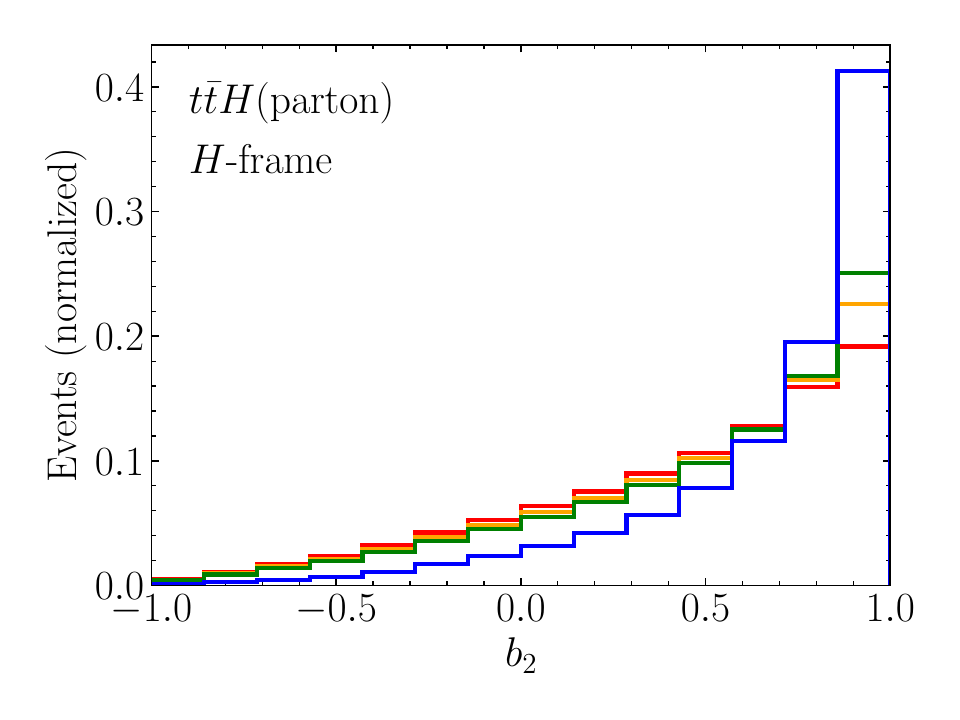}
    \includegraphics[width=.4\textwidth]{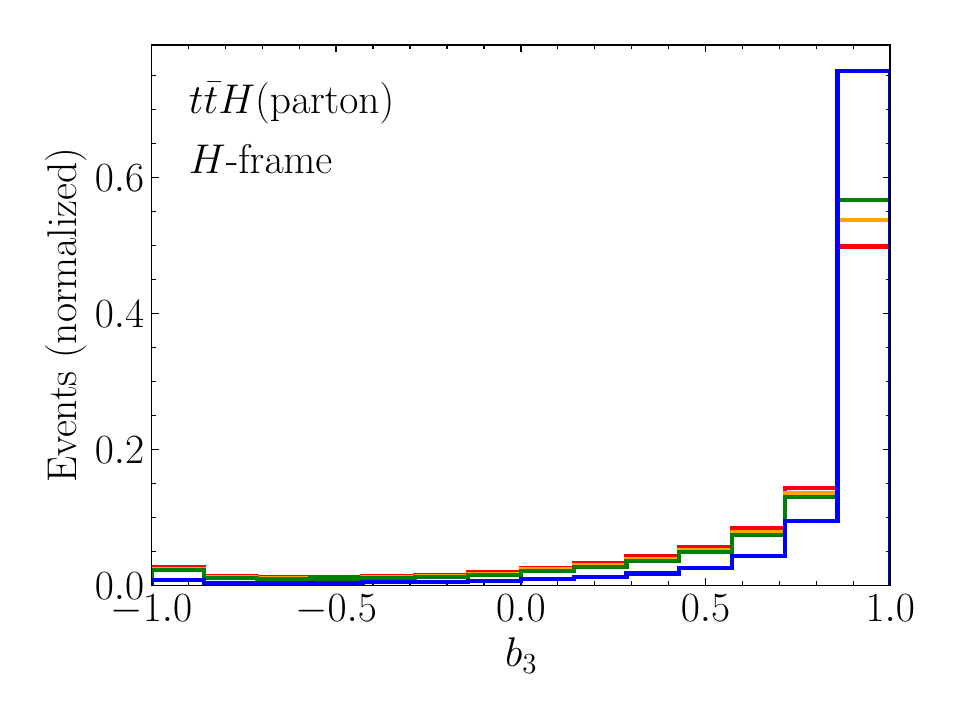}
    \includegraphics[width=.4\textwidth]{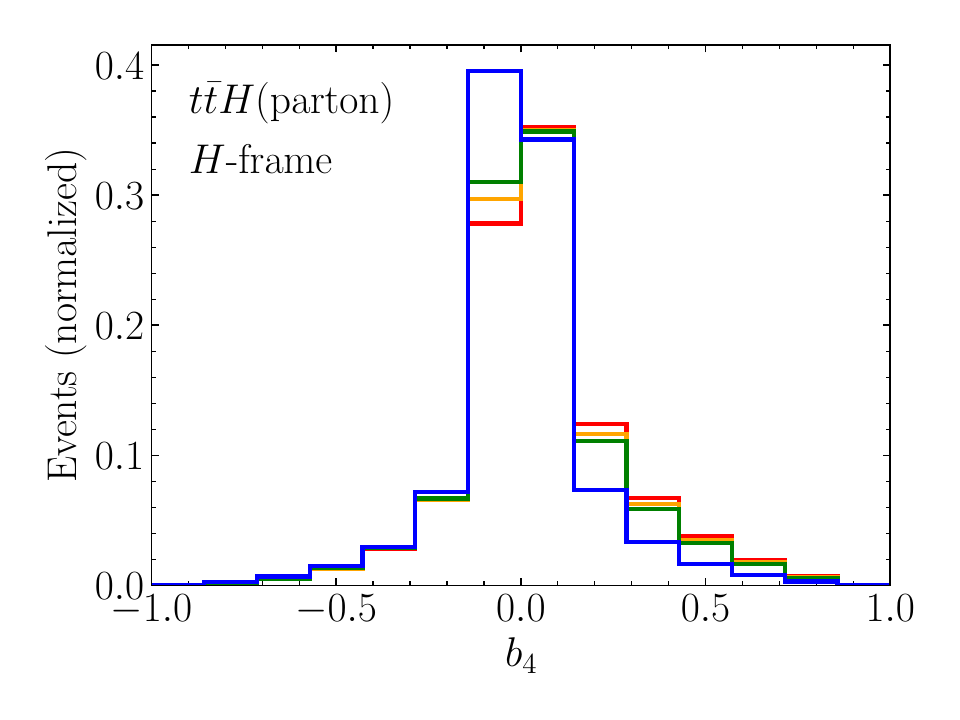}
    \caption{Distributions of the observables defined in~\cref{sec:obs} in the Higgs boson rest frame. $t\bar{t}H$ events generated at parton-level with $\gt=1$ and various values of \cp phase \at are considered following the event generation described in~\cref{subsec:event_generation}. All distributions are normalised to unity.}
    \label{fig:suppl_Higgs}
\end{figure}

The supplementary kinematic distributions for observables defined in the Higgs boson rest frame are shown in \cref{fig:suppl_Higgs}.

%%% figure %%%
%%% figure %%%

%%%%%%%%%%%%%%%%%%%%%%%%%%%%%%%%%%

\clearpage
\subsection{\texorpdfstring{\ttbar}{tt} rest frame}

\begin{figure}[!htpb]
    \centering
    \includegraphics[trim=0.cm 0.5cm 0.cm 0.5cm, width=.8\textwidth]{figs/example_distributions/legend.pdf}
    \includegraphics[width=.4\textwidth]{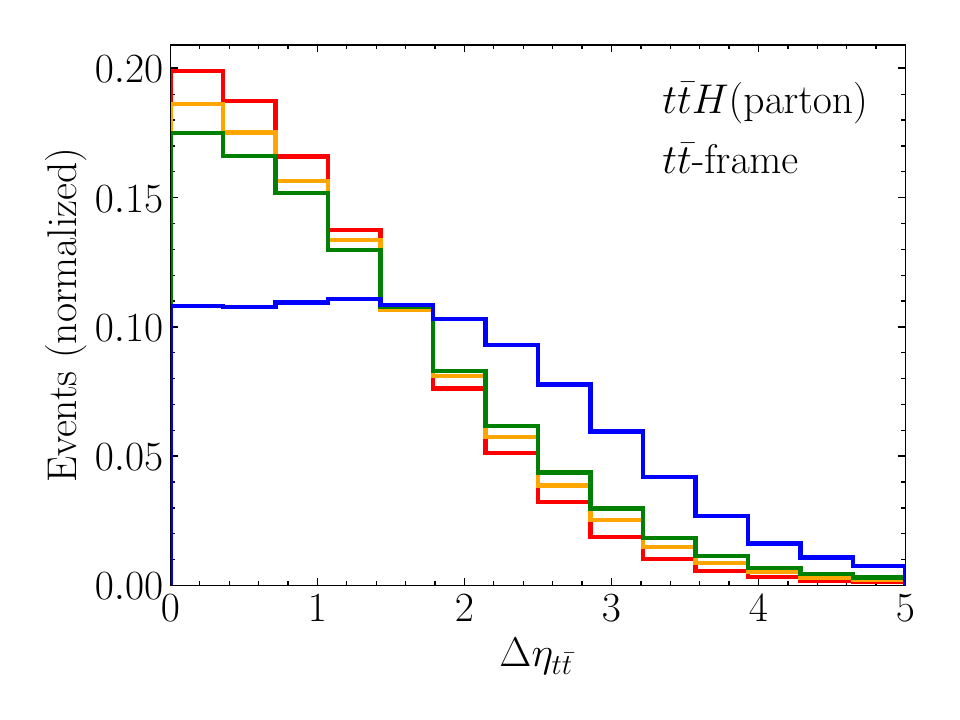}
    \includegraphics[width=.4\textwidth]{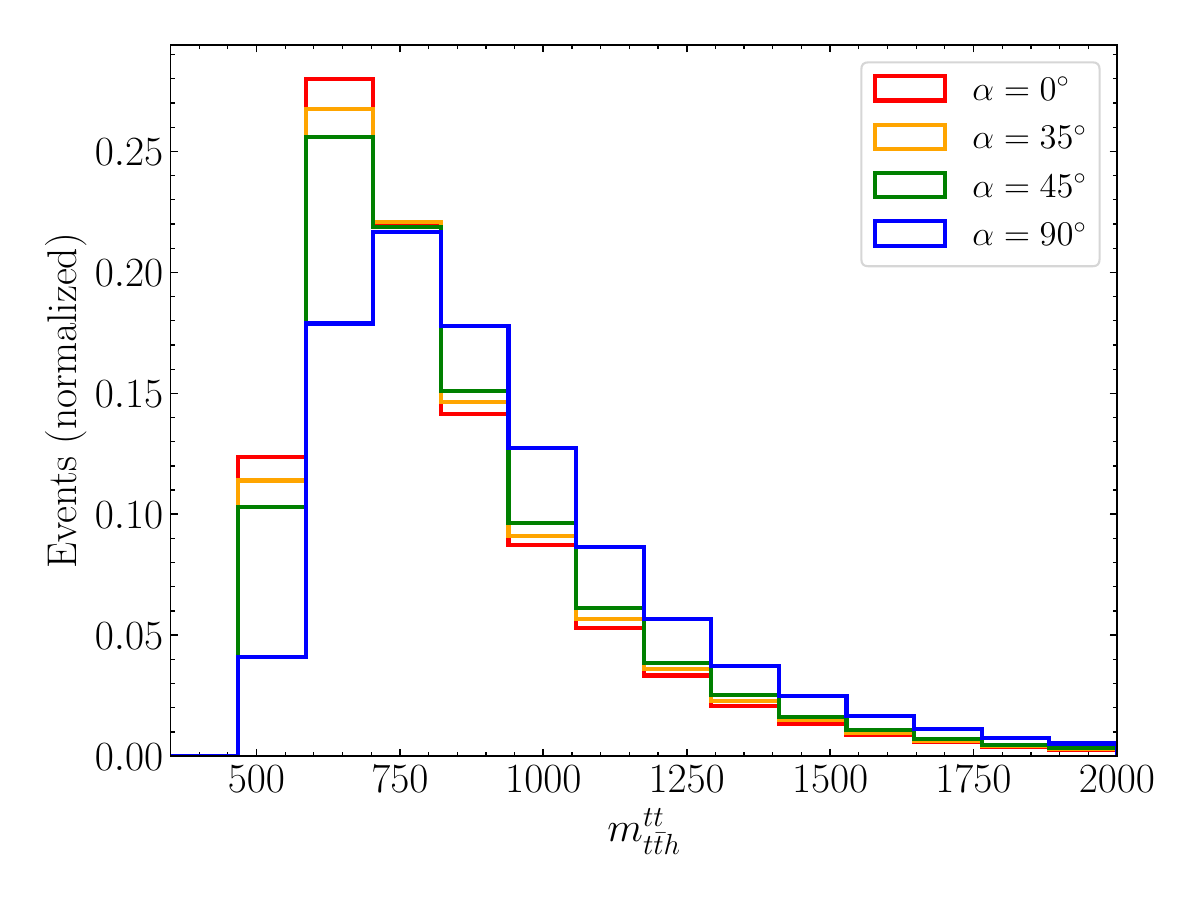}
    \includegraphics[width=.4\textwidth]{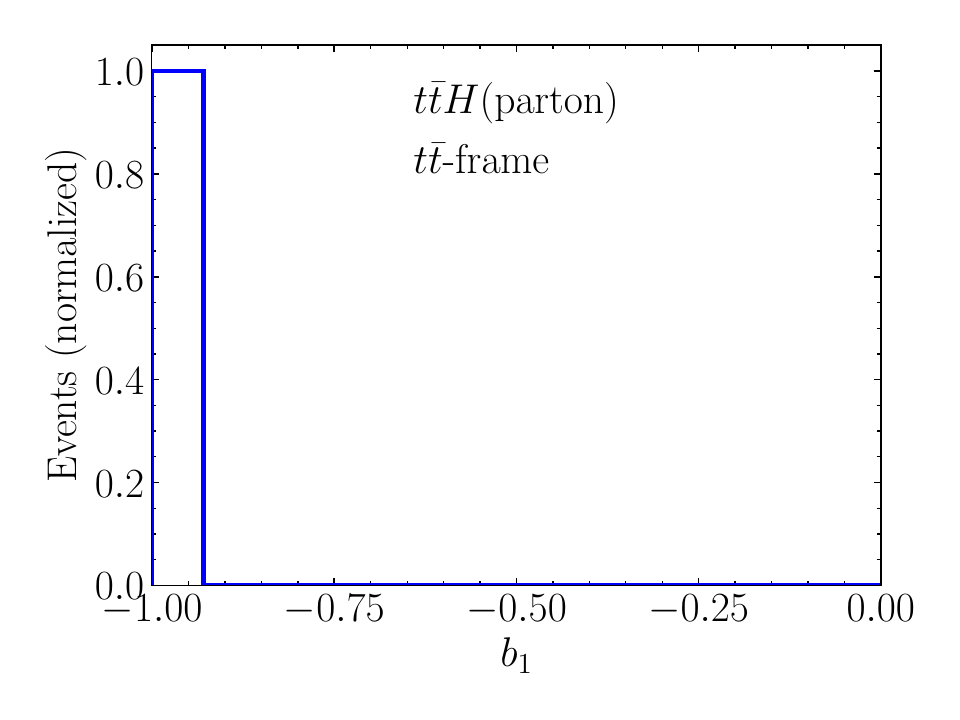}
    \includegraphics[width=.4\textwidth]{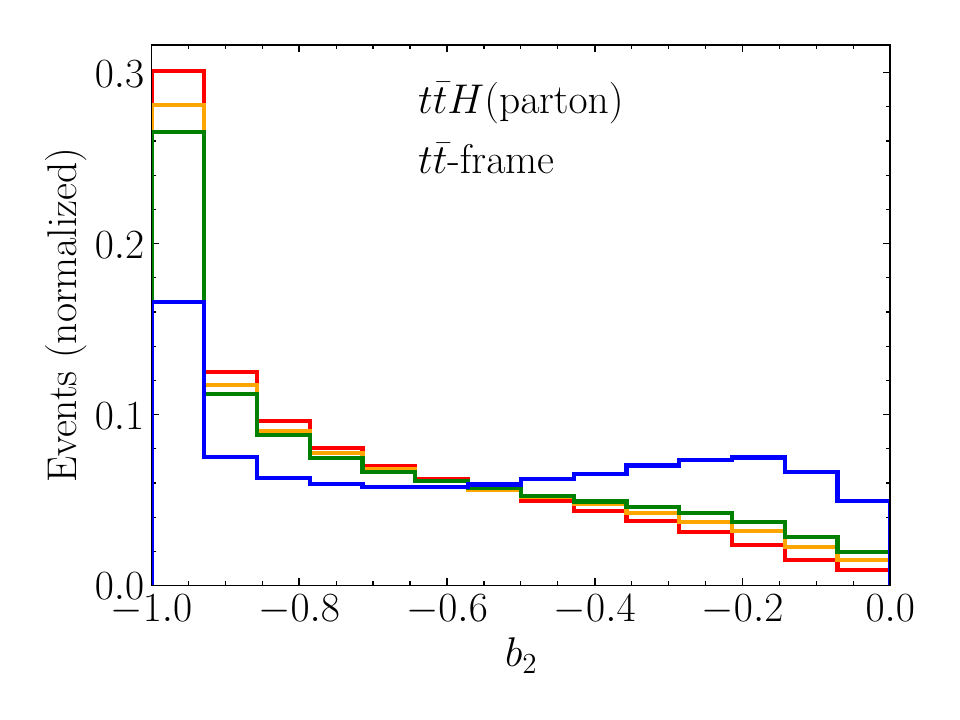}
    \includegraphics[width=.4\textwidth]{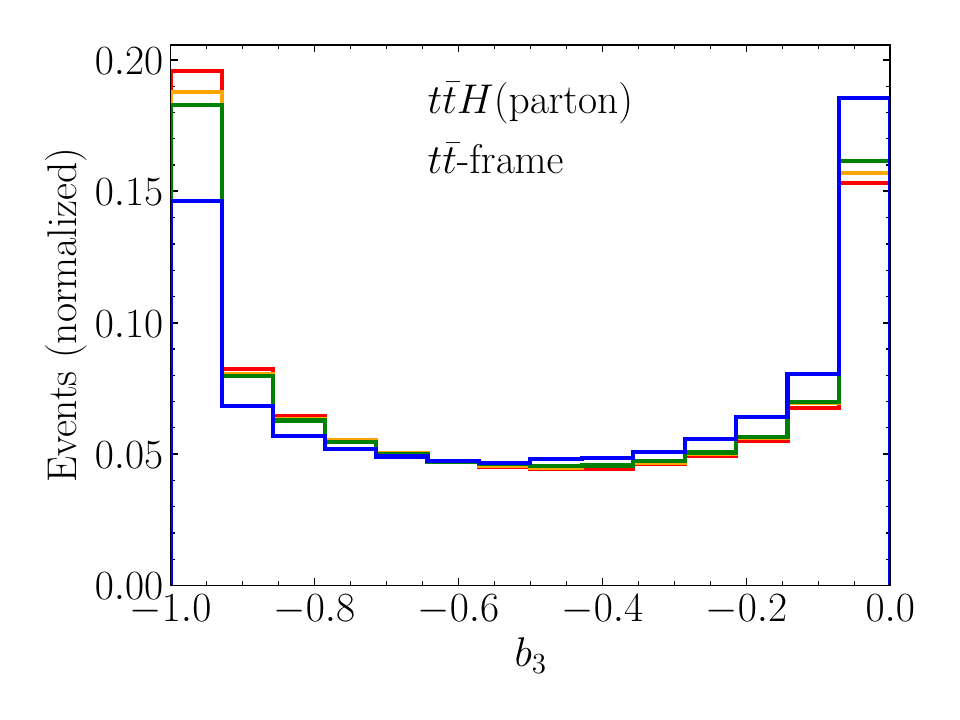}
    \includegraphics[width=.4\textwidth]{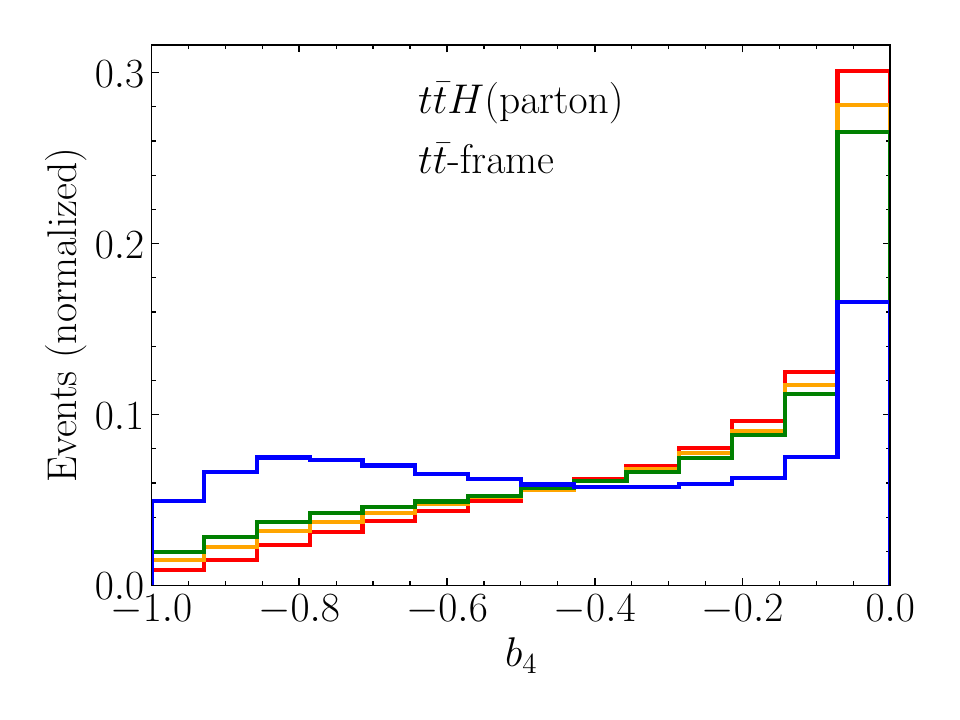}
    \caption{Distributions of the observables defined in~\cref{sec:obs} in the $t\bar{t}$ rest frame. $t\bar{t}H$ events generated at parton-level with $\gt=1$ and various values of \cp phase \at are considered following the event generation described in~\cref{subsec:event_generation}. All distributions are normalised to unity.}
    \label{fig:suppl_tt}
\end{figure}

The supplementary kinematic distributions for observables defined in the $t\bar t$ rest frame are shown in \cref{fig:suppl_tt}.

%%% figure %%%
%%% figure %%%

%%%%%%%%%%%%%%%%%%%%%%%%%%%%%%%%%%
\clearpage
\subsection{\texorpdfstring{\ttH}{ttH} rest frame}

\begin{figure}[!htpb]
    \centering
    \includegraphics[trim=0.cm 0.5cm 0.cm 0.5cm, width=.8\textwidth]{figs/example_distributions/legend.pdf}
    \includegraphics[width=.4\textwidth]{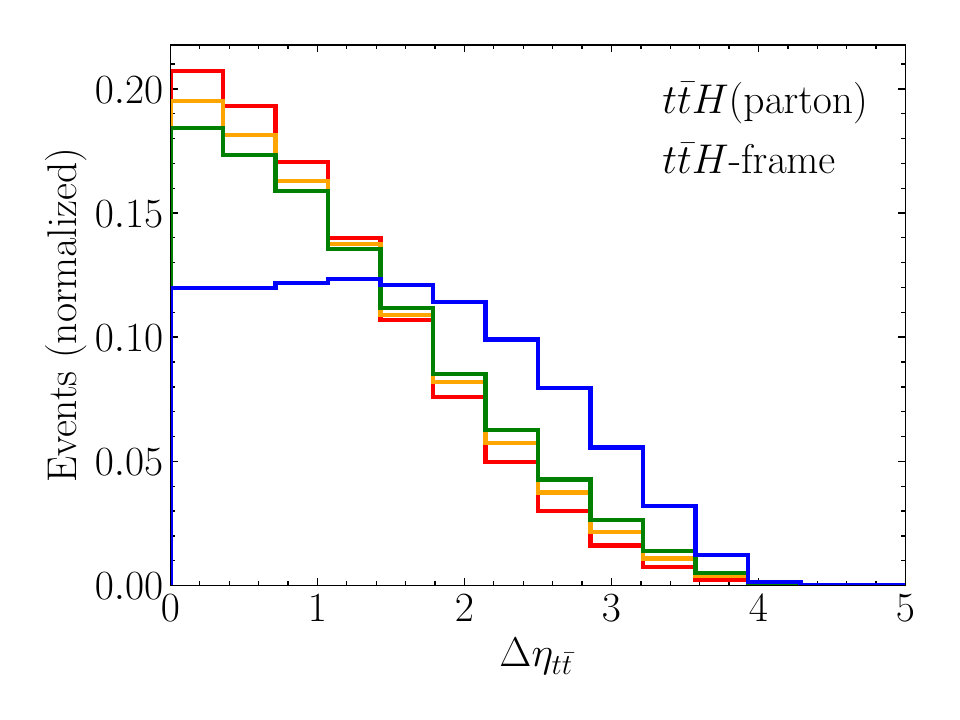}
    \includegraphics[width=.4\textwidth]{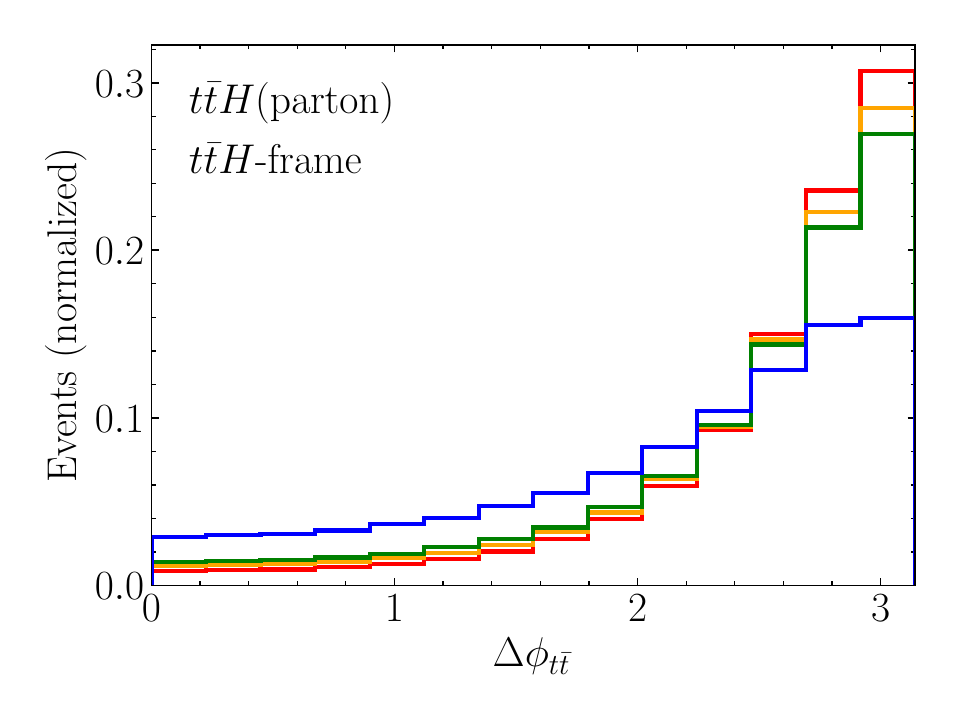}
    \includegraphics[width=.4\textwidth]{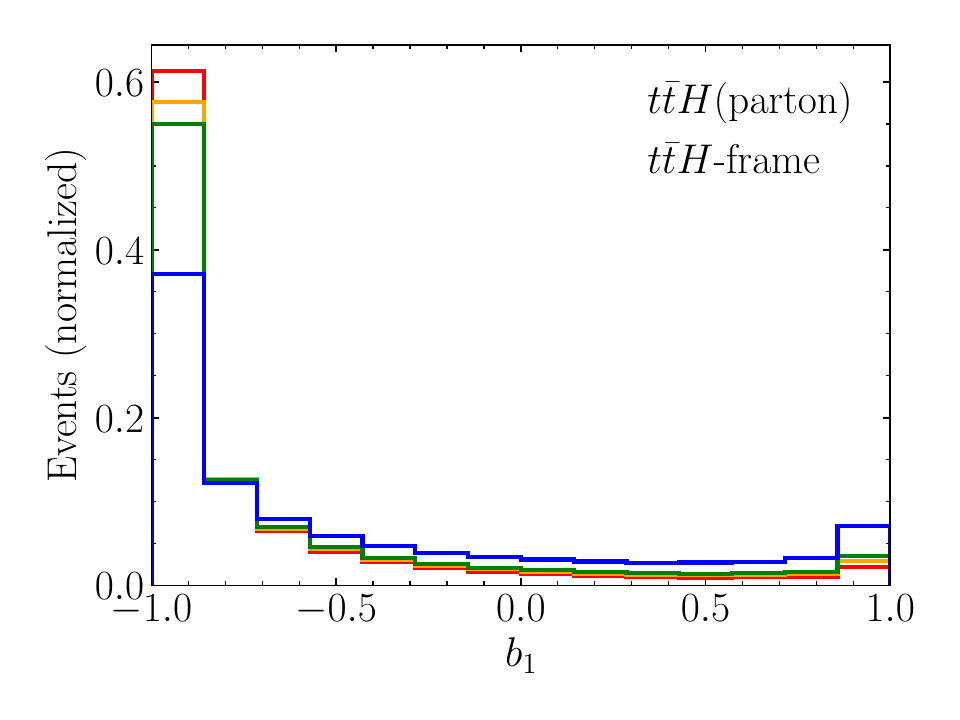}
    \includegraphics[width=.4\textwidth]{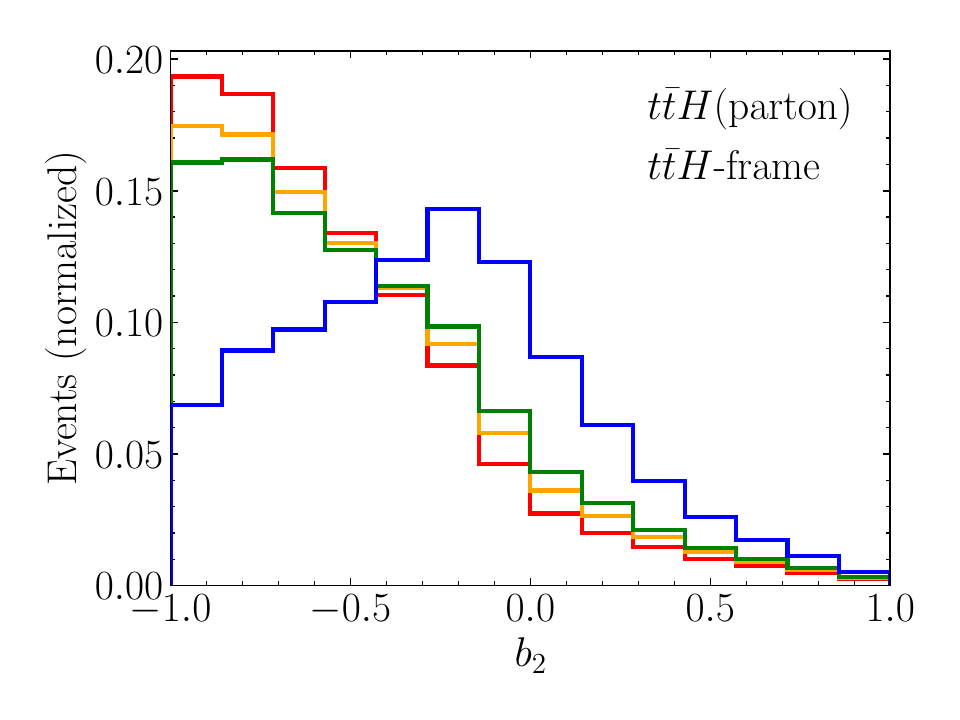}
    \includegraphics[width=.4\textwidth]{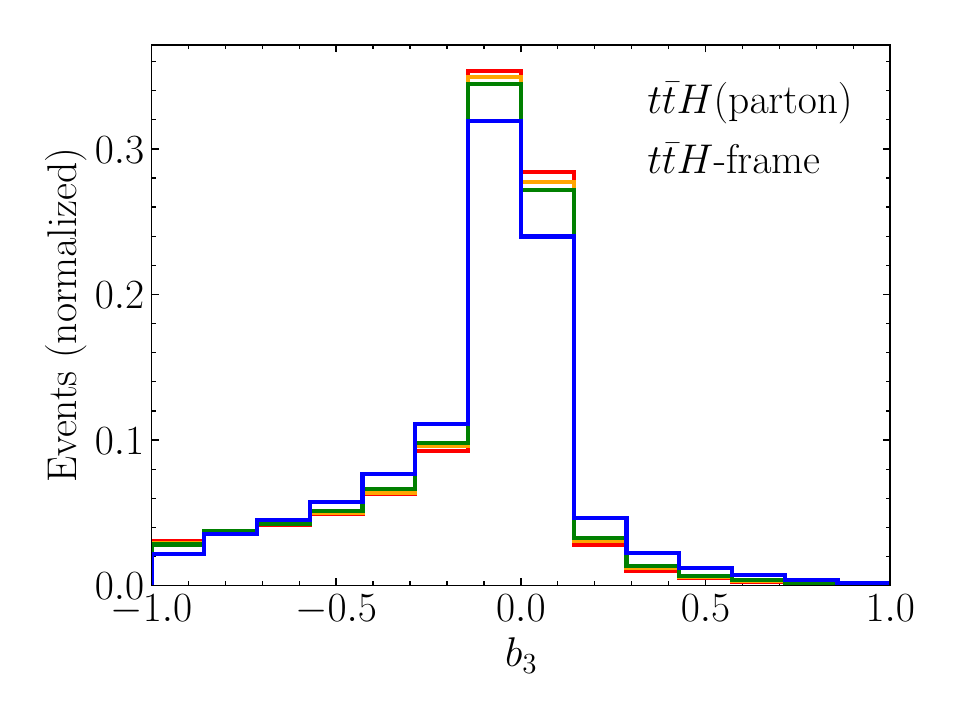}
    \includegraphics[width=.4\textwidth]{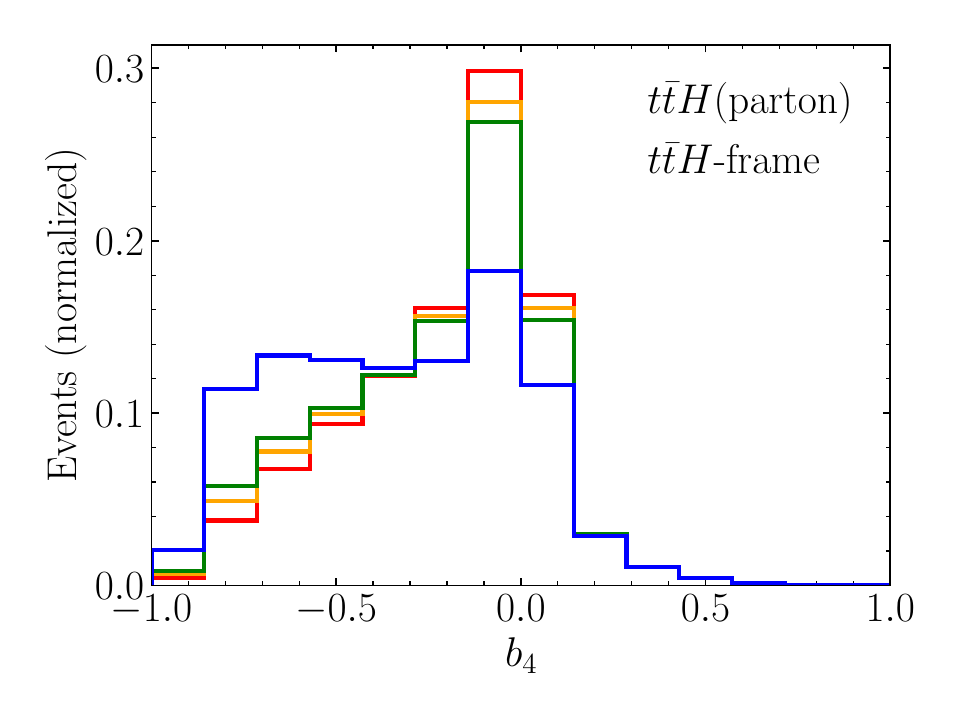}
    \caption{Distributions of the observables defined in~\cref{sec:obs} in the $t\bar{t}H$ rest frame. $t\bar{t}H$ events generated at parton-level with $\gt=1$ and various values of \cp phase \at are considered following the event generation described in~\cref{subsec:event_generation}. All distributions are normalised to unity.}
    \label{fig:suppl_ttH}
\end{figure}

The supplementary kinematic distributions for observables defined in the $t\bar t H$ rest frame are shown in \cref{fig:suppl_ttH}.

%%% figure %%%
%%% figure %%%

%%%%%%%%%%%%%%%%%%%%%%%%%%%%%%%%%%
\clearpage
\subsection{Impact of the \texorpdfstring{$\eta$}{eta} cut} 
\label{app:eta_cut}

%%% figure %%%
\begin{figure}[!htpb]
    \centering
    \includegraphics[trim=0.cm 0.5cm 0.cm 0.5cm, width=.8\textwidth]{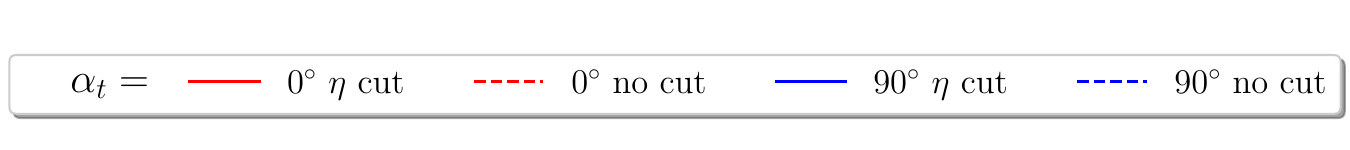}
    \includegraphics[width=.4\textwidth]{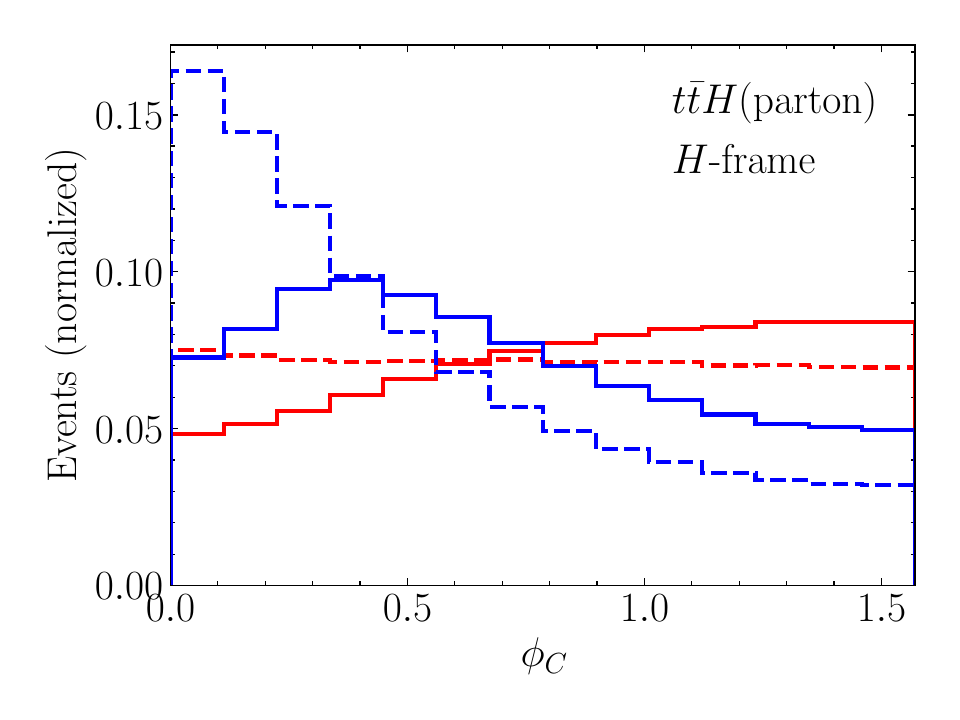}
    \includegraphics[width=.4\textwidth]{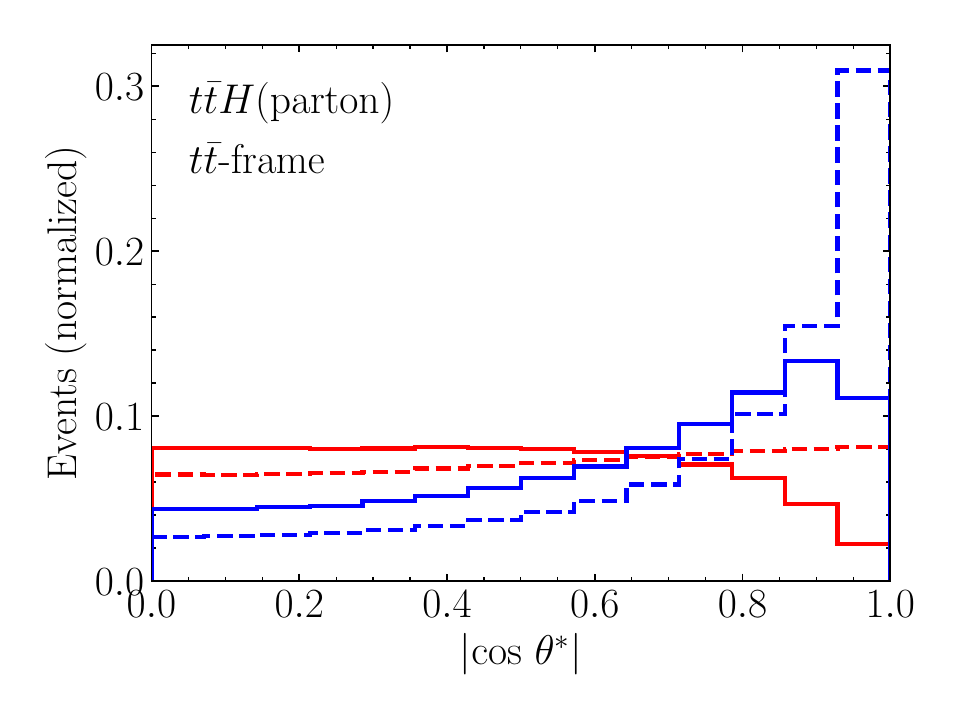}
    \caption{Distributions of (left) the $\phi_C$ and (right) the $|\cos\tstar|$ observables in the Higgs and $t\bar{t}$ frame, respectively. The distributions without any cut (solid line) and requiring $|\eta| < 2$ (dashed line) are shown for the $\alpha_t = 0^\circ$ and $~90^\circ$ samples. All distributions are normalised to unity.} 
    \label{fig:eta_cut}
\end{figure}
%%% figure %%%

% %%%%%%%%%%%%%%%%%%%%%%%%%%%%%%%%%%
% %%%%%%%%%%%%%%%%%%%%%%%%%%%%%%%%%%
\clearpage
\section{Full two-dimensional significance tables}
\label{app:significance_tables}

\begin{figure}[!htpb]
\centering
\includegraphics[width=16cm]{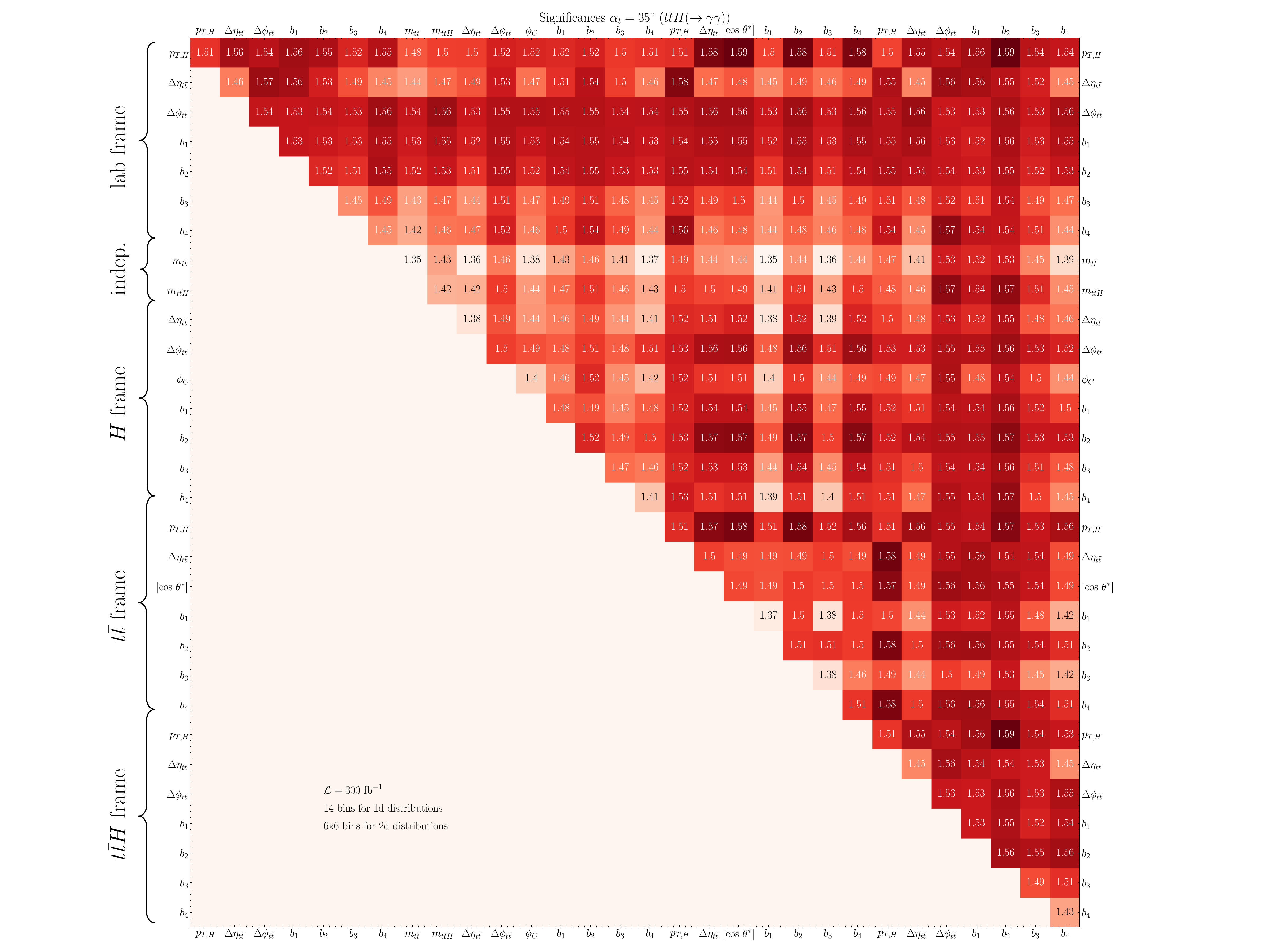}
\caption{Significances of the $g_t=1$, $\at=35^\circ$ signal for the $t\bar{t}H(\to\gamma\gamma)$ channel. The diagonal matrix elements show the significances obtained from single observables with default binning, while the off-diagonal matrix elements show the signifiances obtained from the combination of two observables. Only half of the histogram is filled to avoid repeating the same information.}
\label{fig:significances_gaga}
\end{figure}

\begin{figure}[!htpb]
\centering
\includegraphics[width=16cm]{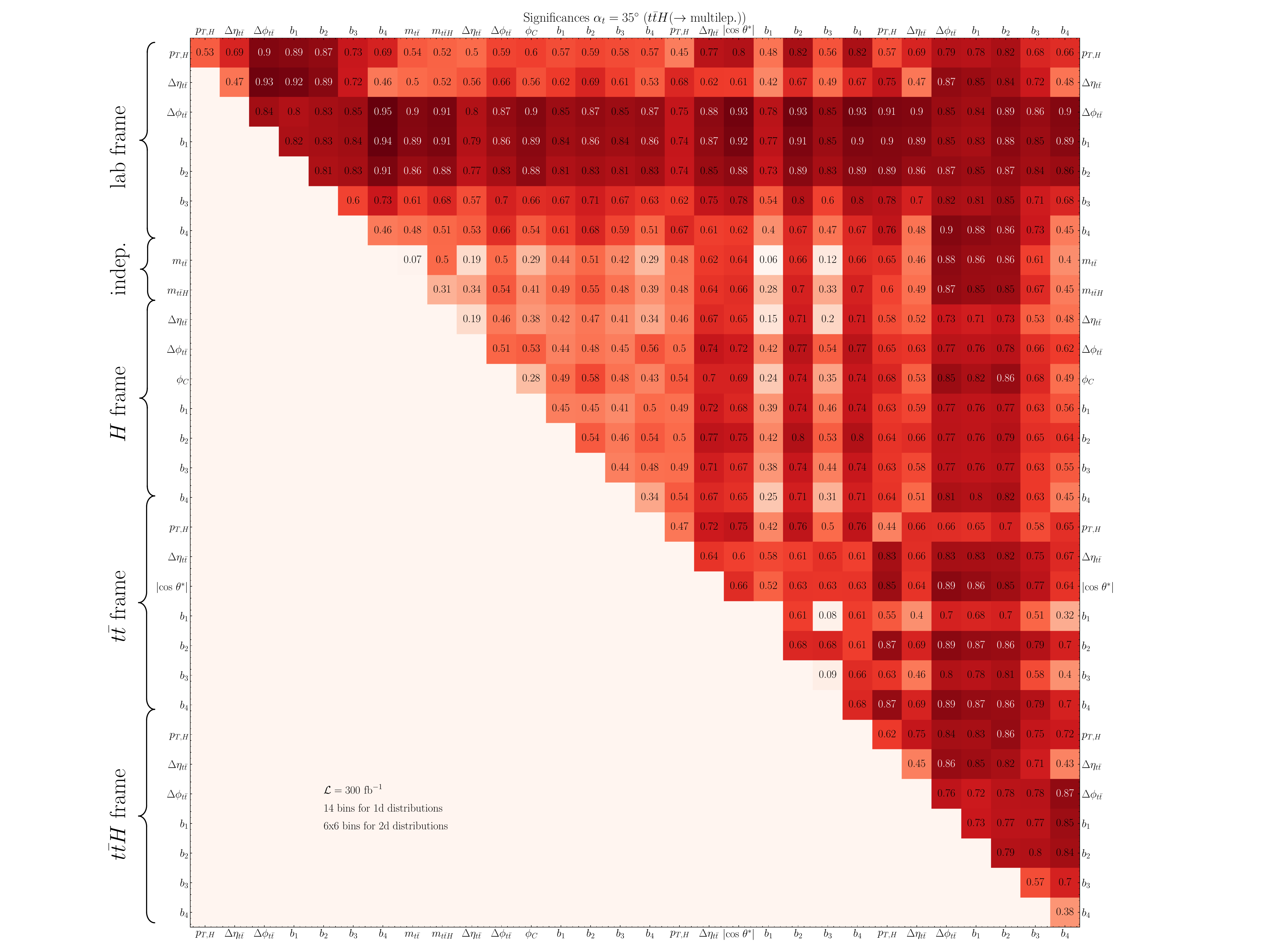}
\caption{Significances of the $g_t=1$, $\at=35^\circ$ signal for the $t\bar{t}H$(multilep.) channel. The diagonal matrix elements show the significances obtained from single observables with default binning, while the off-diagonal matrix elements show the signifiances obtained from the combination of two observables. Only half of the histogram is filled to avoid repeating the same information.}
\label{fig:significances_lep}
\end{figure}

\begin{figure}[!htpb]
\centering
\includegraphics[width=16cm]{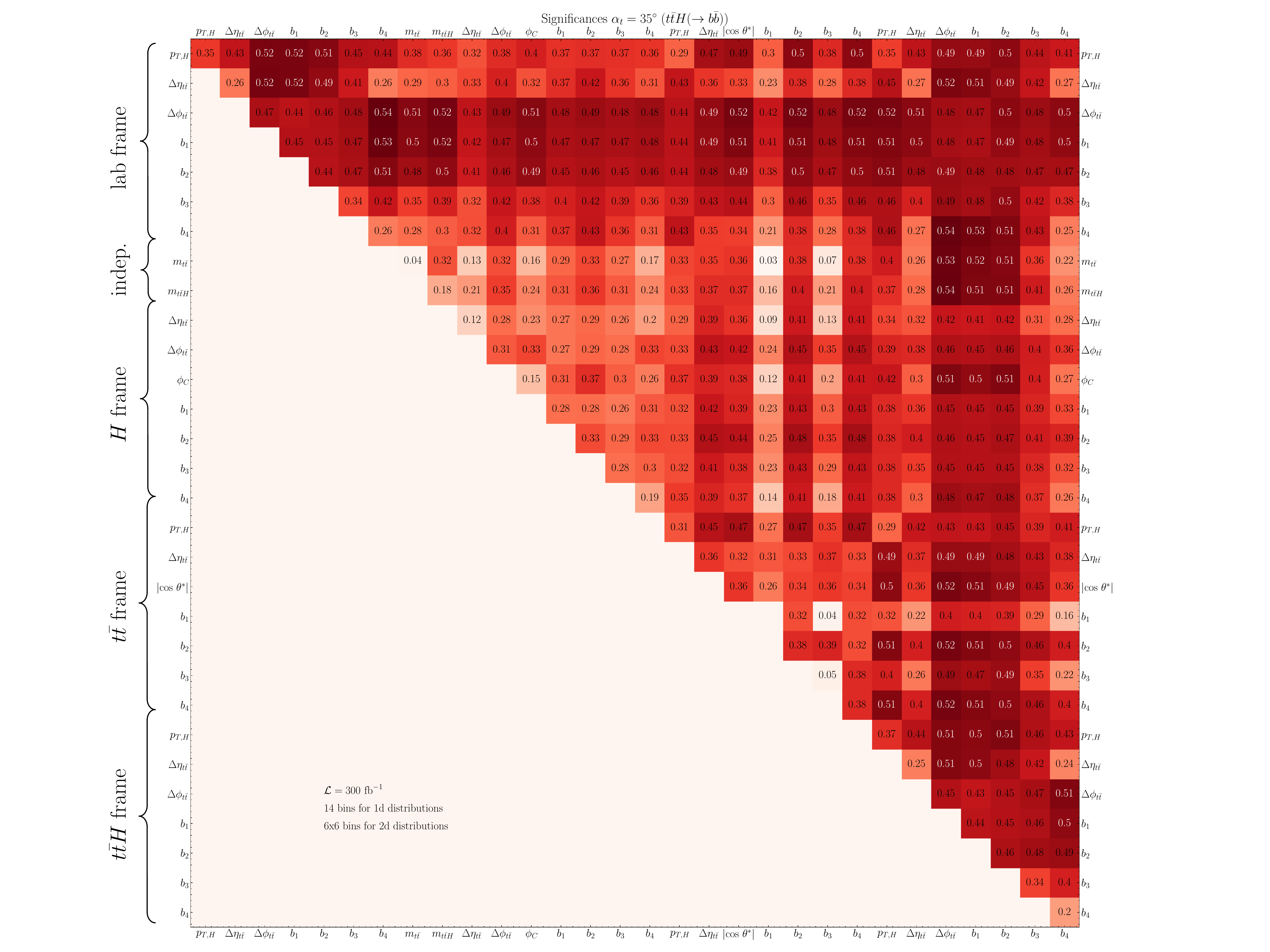}
\caption{Significances of the $g_t=1$, $\at=35^\circ$ signal for the $t\bar{t}H(\to bb)$ channel. The diagonal matrix elements show the significances obtained from single observables with default binning, while the off-diagonal matrix elements show the signifiances obtained from the combination of two observables. Only half of the histogram is filled to avoid repeating the same information.}
\label{fig:significances_bb}
\end{figure}

\begin{figure}[!htpb]
\centering
\includegraphics[width=16cm]{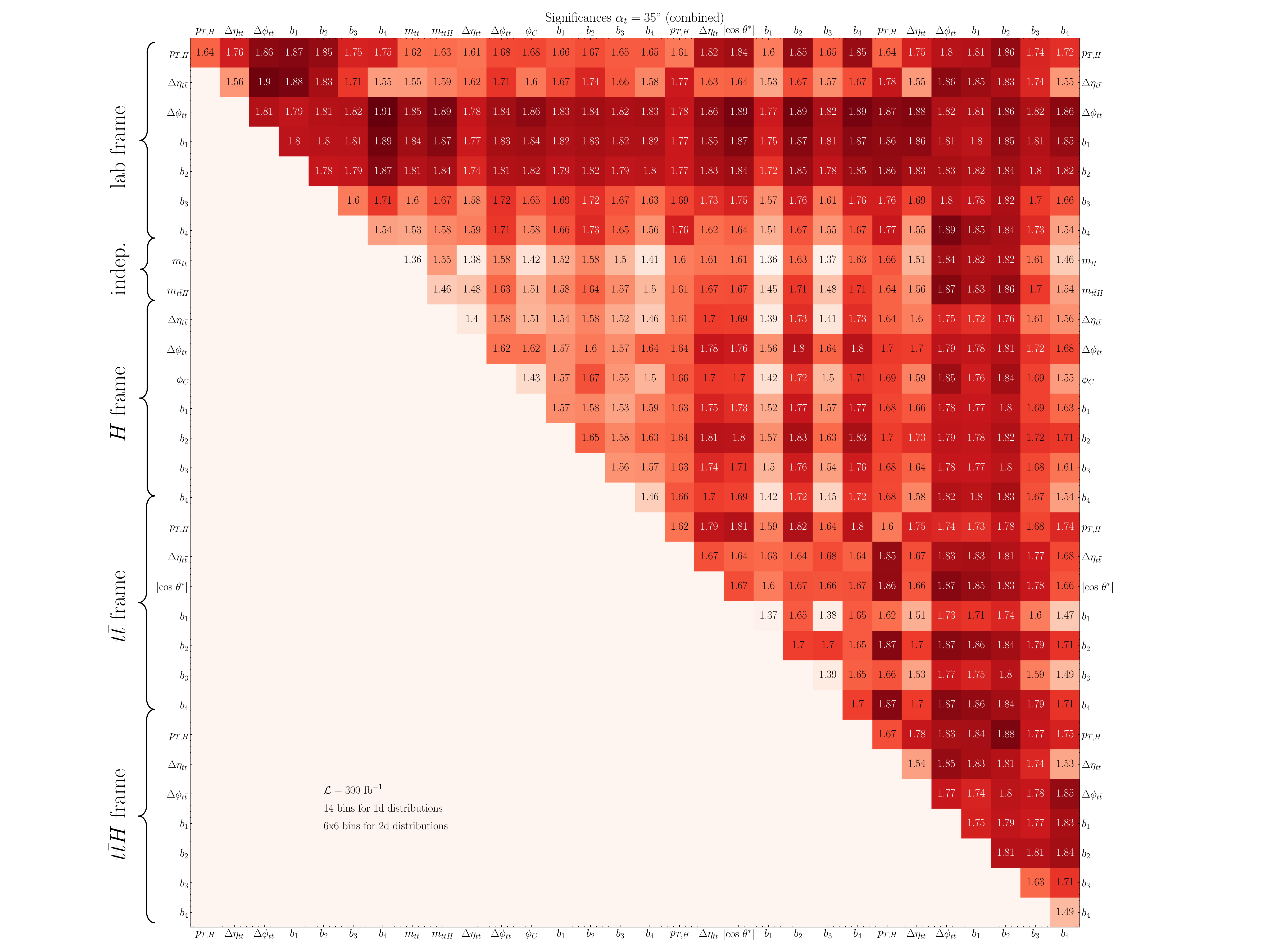}
\caption{Combined significances of the $g_t=1$, $\at=35^\circ$ signal including the $t\bar{t}H(\to\gamma\gamma)$, $t\bar{t}H$(multilep.) and ${t\bar{t}H(\to bb)}$ channels. The diagonal matrix elements show the significances obtained from single observables with default binning, while the off-diagonal matrix elements show the signifiances obtained from the combination of two observables. Only half of the histogram is filled to avoid repeating the same information.}
\label{fig:significances_combined}
\end{figure}

The full two-dimensional significance tables for the $g_t=1$, $\at = 35^{\circ}$ sample are shown in \cref{fig:significances_gaga} for the $t\bar{t}H(\to\gamma\gamma)$ channel, in \cref{fig:significances_lep} for the $t\bar{t}H$(multilep.) channel, and in \cref{fig:significances_bb} for the $t\bar{t}H(\to bb)$ channel. The combined significances including the three channels are depicted in \cref{fig:significances_combined}.

%%% figure %%%
%%% figure %%%

%%% figure %%%
%%% figure %%%

%%% figure %%%
%%% figure %%%

%%% figure %%%
%%% figure %%%

%%%%%%%%%%%%%%%%%%%%%%%%%%%%%%%%%%
%%%%%%%%%%%%%%%%%%%%%%%%%%%%%%%%%%
%%%%%%%%%%%%%%%%%%%%%%%%%%%%%%%%%%
\clearpage
\printbibliography

%%%%%%%%%%%%%%%%%%%%%%%%%%%%%%%%%%
%%%%%%%%%%%%%%%%%%%%%%%%%%%%%%%%%%
%%%%%%%%%%%%%%%%%%%%%%%%%%%%%%%%%%

\end{document}